\documentclass{aa}  

\usepackage{graphicx}
\usepackage{txfonts}
\usepackage{tabu}
\usepackage{amsmath,amsfonts,bm}
\begin{document}

   \title{Dynamical tide in stellar radiative zones}
   \subtitle{General formalism and evolution for low-mass stars}

   \author{J. Ahuir,
          \inst{1}
          S.Mathis
          \inst{1}
          \and
          L. Amard\inst{2}
          }
   \institute{D\'epartement d'Astrophysique-AIM, CEA/DRF/IRFU, CNRS/INSU, Universit\'e Paris-Saclay, Universit\'e Paris-Diderot, Universit\'e de Paris, F-91191 Gif-sur-Yvette, France\\
              \email{jeremy.ahuir@cea.fr}
         \and    
         University of Exeter, Department of Physics \& Astronomy, Stoker Road, Devon, Exeter EX4 4QL, UK
             }

   \date{Received XXX ; accepted YYY}

  \abstract
   {Most exoplanets detected so far are close-in planets, which are likely to be affected by tidal dissipation in their host star. To get a complete picture of the evolution of star-planet systems one needs to consider the impact of tides within both stellar radiative and convective zones.}
   {We aim to provide a general formalism allowing us to assess tidal dissipation in stellar radiative zones for late-type stars and early-type stars, including stellar structure with both a convective core and envelope like in F-type stars, allowing for the study of the dynamics of a given system throughout stellar evolution. On this basis, we investigate the influence of stellar structure and evolution on tidal dissipation in the radiative core of low-mass stars.}
   {We develop a general theoretical formalism to evaluate tidal dissipation in stellar radiative zones applicable to early-type and late-type stars. From the study of adiabatic oscillations throughout the star, we compute the energy flux transported by progressive internal gravity waves and the induced tidal torque. By relying on grids of stellar models we study the influence of stellar structure and evolution on tidal dissipation of F, G and K-type stars from the pre-main sequence (PMS) to the red giant branch (RGB).}
   {For a given star-planet system, tidal dissipation reaches a maximum value on the pre-main sequence for all stellar masses. On the main sequence (MS), it decreases to become almost constant. The dissipation is then several orders of magnitude smaller for F-type stars than for G and K-type stars. During the Sub-Giant phase and the RGB, tidal dissipation increases by several orders of magnitude, along with the expansion of the stellar envelope. We show that the dissipation of the dynamical tide in the convective zone dominates the evolution of the system during most of the PMS and the beginning of the main sequence, as the star rotates rapidly. Tidal dissipation in the radiative zone then becomes the strongest contribution during the Sub-Giant phase and the RGB, as the density at the convective-radiative interface increases. For similar reasons, we also find that the dissipation of a metal-poor star is stronger than the dissipation of a metal-rich star during the PMS, the Sub-Giant phase and the RGB. During the MS the opposite trend is observed. Finally, we show that the contribution of a convective core for the most massive solar-type stars is negligible compared to that of the envelope, as the mass distribution of the core is unfavorable to the dissipation of tidal gravity waves.}
   {}

   \maketitle
%

\section{Introduction}
Around 46 \% of observed exoplanets are located within 20 times the radius of their host star (using the database exoplanet.eu\footnote{http://www.exoplanet.eu, the database was consulted on December 18, 2020 to provide this estimate.}, e.g. \citealt{schneider}). Such a configuration leads to important star-planet interactions, affecting the dynamics of these compact systems \citep{cuntz}. The study of these interactions is therefore crucial to understand the population of currently observed planetary systems and their evolution. The study of the orbital architecture of exosystems then improves our understanding of such processes and constrains the evolution models we rely on. The secular evolution of a star-planet system is essentially driven by stellar tides \citep{leconte}, unless star-planet magnetic interactions develop due to the motion of the planet in the ambient magnetized stellar wind, which may have a significant role in the evolution of the system \citep{strugarek14,strugarek15,strugarek17}. In particular, the dissipation of tides in the host star, by ensuring angular momentum exchanges between the star rotation and the planetary orbit, is thought to play a major role in the secular evolution of orbital architecture \citep{bolmont16,benbakoura}, and star-planet obliquity \citep{lai,damiani}. Furthermore, as the reservoir of angular momentum of the planet is less important than the one in its orbit, the planet tends to be synchronized within a few thousands of years. 

The gravitational response of the star to the planet leads to two kinds of flows: the non-wave-like equilibrium tide \citep{zahn66,remus, ogilvie13}, which consists in the displacement induced by the hydrostatic adjustment of the stellar structure, and the dynamical tide, corresponding to tidally-forced internal waves. In stellar convective zones, the dynamical tide is constituted by inertial waves, restored by the Coriolis
force, and get dissipated by the turbulent friction applied by the convection on tidal waves \citep{ogilvie04,ogilvie07}. The induced tidal dissipation may vary over several orders of magnitude with tidal frequency, stellar mass, age, rotation, and metallicity \citep{mathis15,gallet17,bolmont17}. When inertial waves are likely to be excited, i.e. when the tidal frequency ranges between \([-2\Omega_\star, 2\Omega_\star]\), \(\Omega_\star\) being the stellar rotation angular velocity, such a dissipation is several orders of magnitude higher than the dissipation of the equilibrium tide \citep{ogilvie07}.

To get a complete picture of tidal dissipation in stars, one needs also to consider the dynamical tide in stellar radiative zones (we refer the reader to \citealt{ogilvie14} and \citealt{mathis19} for extensive reviews), which may compete with the dissipation in convective layers \citep{ivanov}. \citet{zahn70,zahn75} first highlighted this process as a key dissipation mechanism in early-type stars, to account for the circularization of massive close binaries \citep{zahn77, savonije83,savonije84,savonije97,papaloizou85,papaloizou97,savonijealberts95}. Such a process happens for \(a/R_\star < 4\), \(a\) being the orbital semi-major axis and \(R_\star\) the stellar radius \citep{north}. For those stars, which have a convective core and a radiative envelope, the dissipation of gravity waves is more efficient than the dissipation of the equilibrium tide \citep{zahn77}. \citet{goodman} and \citet{terquem} adopted a similar approach in the case of solar-type stars and showed that the resonant excitation of g-modes can compete with the dissipation of the equilibrium tide in the envelope. Such an effect has also been obtained in the case of gravito-inertial waves for uniformly rotating stars \citep{ogilvie07, chernov,ivanov}.

\citet{goldreich} first proposed a physical interpretation of the dynamical tide in radiative zones. Internal gravity waves are excited near the convective-radiative interfaces by the tidal potential, where the buoyancy frequency matches the tidal frequency. They propagate then into the radiative zone, where they are damped by radiative diffusion \citep{zahn75,zahn77}, critical layers \citep{alvan13} or non-linear breaking \citep{goodman, barker10, barker11,guillot}. In addition, the evanescent tail of the waves is subject to the friction applied by turbulent convection, that is commonly modelled with an eddy viscosity \citep{terquem}. There, they deposit their angular momentum, then altering the dynamics of the system considered. In particular, if we consider an early-type star with a radial differential rotation in its radiative zone, such a process leads to a synchronisation of the star starting from its surface \citep{goldreich}.

The main dependencies of the torque induced by the dissipation of tidal gravity waves are well understood. In particular, the location and stellar properties at the interface between the convective and radiative zones are critical parameters to estimate the amplitude of the dissipation. 
However, the effective contribution of the tidal forcing to the gravity waves torque requires an intensive investigation of forced oscillations. As already pointed out by \citet{kushnir}, such a factor has been studied independently for early-type stars in the context of binary systems \citep[e.g.][]{zahn75, goldreich}, as well as for late-type stars in the framework of the secular evolution of exoplanetary systems \citep[e.g. ][]{goodman,terquem}. In order to track the fate of the system from the birth of the star up to its death, one needs a unified formalism allowing us to take any changes in stellar structure during the evolution into account.

In this context, the influence of stellar structure and evolution, subject to complex variations over time \citep{kippenhahn}, on the tidal dissipation in radiative zones constitutes a key issue to be addressed. \citet{barker20} carried out a first study of tidal dissipation through gravity waves during stellar evolution. In a context of evolution of planetary systems, he compared the dissipation in the stellar radiative zone to the frequency averaged dissipation through inertial waves in the stellar convective zone, for a given stellar rotation period and orbital period. He finds that the dissipation of tidal gravity waves is the dominant mechanism for the migration of  close-in planets. This may be also an avenue to account for the survival of close-in exoplanets, depending on the host star properties \citep{barker10,guillot}.

Furthermore, tidal dissipation for a tri-layer structure, for example in the case of F-type stars as well as red giants in the red clump, should be studied. It can be very important for our understanding of binaries and planetary systems since in those configurations the dissipation of the dynamical tide in the convective zone is weak compared to the dissipation of the equilibrium tide \citep{mathis15, gallet17, beck18}. Some aspects of this question has been extensively studied by \citet{fuller17}, which assess the tidal dissipation through standing $g$-modes in the case of F-type and A-type stars by taking into account resonance locking.

The goal of our work is to provide a general formalism to assess tidal dissipation in stellar radiative zones which may be applicable to late-type stars and early-type, as well as tri-layer structures. On this basis, we investigate the influence of stellar structure and evolution on tidal dissipation for low-mass stars, and to study the contribution of a convective core for F-type stars. In Sect. 2, we study the forced adiabatic oscillations in a spherical geometry. From there we compute in Sect. 3 the energy flux carried by the waves and the induced tidal torque. In Sect. 4 we apply our formalism to massive stars, which matches with the \citet{zahn75} formulation, and low-mass stars. In Sect. 5 we investigate the influence of stellar structure and evolution on tidal dissipation in low-mass stars. For F-type stars we simultaneously evaluate both contributions of the convective core and envelope to the dynamical tide. All of those results are then summarized, discussed, and perspectives are discussed in Sect. 6.

\section{Forced adiabatic oscillations in a spherical geometry}
\subsection{Statement of the problem}

We aim to evaluate the tidal dissipation in stellar radiative zones driven by internal gravity waves (IGW). To this end we model the stellar interior as an inviscid fluid. For instance, the Sun has a Prandtl number of around \(3 \times 10^{-6}\) \citep{brun06}, meaning that thermal diffusion dominates viscosity effects.
 \begin{figure}[!h]
        \centering
        \includegraphics[scale=0.26]{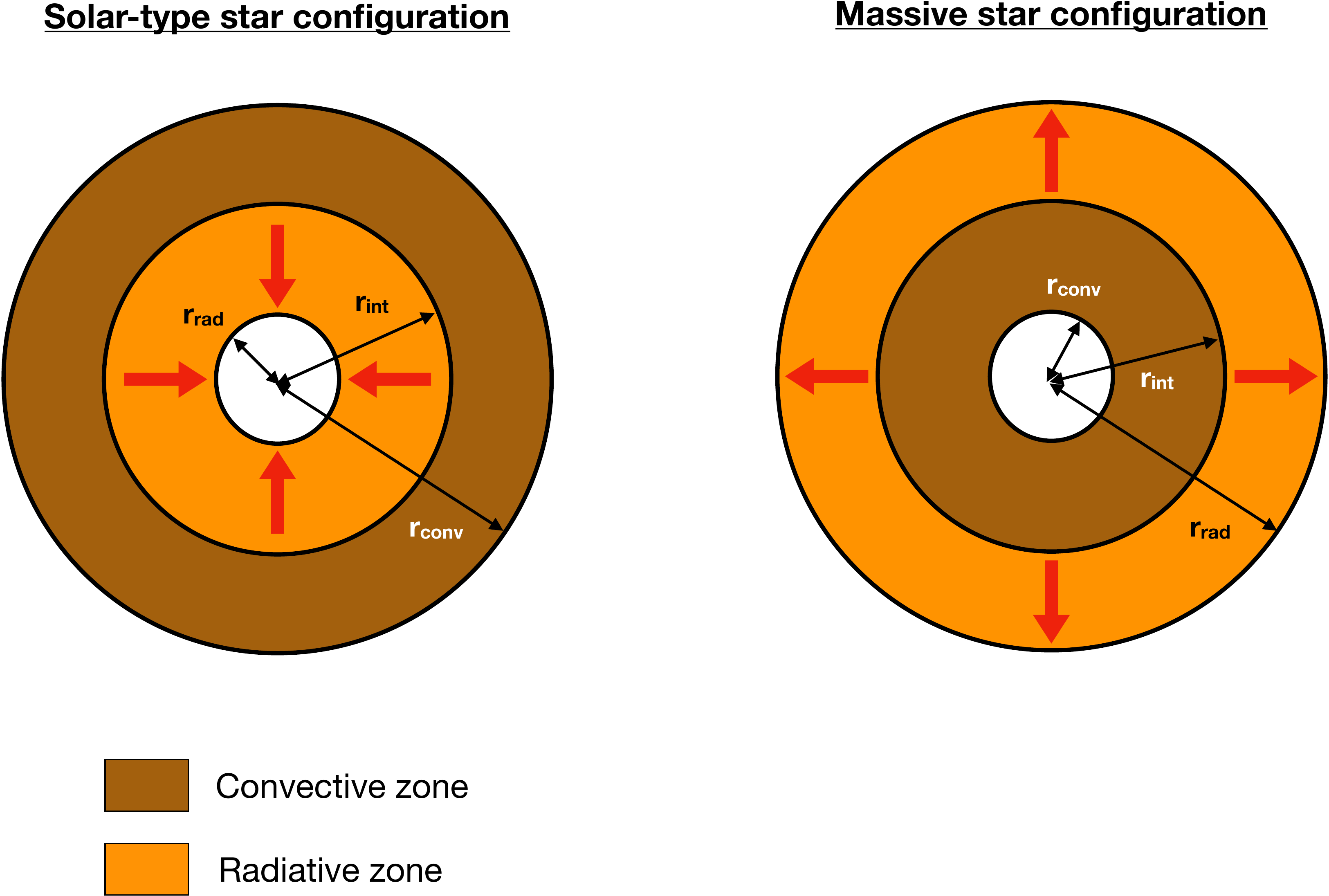}
  \caption{\label{shellular} Configurations of the radiative and convective spherical shells in our work. In brown: convective layer. In orange: radiative layer. The red arrows represent the energy flux carried by tidal gravity waves.}
\end{figure}

To account for the interior of both early-type and late-type stars, we consider a convective layer delimited by the radii \(r_\text{conv}\) and \(r_\text{int}\), where the interface convective zone-radiative zone is located. A radiative shell is then delimited by the radii \(r_\text{int}\) and \(r_\text{rad}\). In a solar-type star configuration, the waves are launched near the base of the convective envelope. Then energy propagates inwards to the center before their dissipation (see the left panel in Fig. \ref{shellular}) whereas in the massive star configuration energy is transported outwards to the stellar surface (see the right panel in Fig. \ref{shellular}). To take this changing direction of propagation into account we define \(\epsilon = \text{sign}(r_\text{conv} - r_\text{rad})\). This quantity is equal to 1 for an inward energy transport through gravity waves and equal to -1 otherwise. The values of \(r_\text{rad},\ r_\text{conv}\) and \(\epsilon\) for massive and solar-type stars are explicited in Table \ref{tab:values}.

\begin{table}[!h]
\centering 
      \caption{\label{tab:values} Values of \(r_\text{rad},\ r_\text{conv}\) and \(\epsilon\) for massive and solar-type stars.}
      \begin{tabu}{cccc}
            \hline
            \noalign{\smallskip}
            \text{Type of star} &$r_\text{rad}$&$r_\text{conv}$&$\epsilon$ \\
            \noalign{\smallskip}
            \hline
            \noalign{\smallskip}
            Massive & $R_\star$&0&-1\\
            Solar-type &0&$R_\star$&1\\
            \noalign{\smallskip}
            \hline
         \end{tabu}
   \end{table}

We focus in this work on the progressive low-frequency waves, that will be the most damped \citep[e.g.][]{press,zahn97,alvan15}. Indeed, \citet{terquem} have shown that the excitation of a fixed spectrum of g-modes, dissipated by radiative damping, does not affect the secular evolution of the system. In our model, the progressive gravity waves deposit angular momentum at the place they are damped, which drives spin-orbit angular momentum exchanges as well as the evolution of the system.\\ 
  
Following \citet{alvan15}, we first assess the cut-off frequency that separates progressive waves and g-modes. Along their propagation, gravity waves are subject to radiative damping, which in the quasi-adiabatic regime for low-frequency waves results in an amplitude of the wave damped by a factor \(\exp{(-\tau/2)}\) where
\begin{equation}
\tau (r, l ,\omega) = \epsilon\left[l(l+1)\right]^\frac{3}{2}\int_r^{r_\text{int}}K_T \frac{N^3}{\omega^4}\frac{dr_1}{r_1^3},
\end{equation}
with \(K_T\) the thermal diffusivity of the medium \citep[we refer the reader to][for more details]{zahn97}. In order to separate standing modes and progressive waves, we assume that a stationary wave can form if
\begin{equation}
\tau (r_\text{rad}, l ,\omega) \leq 1.
\end{equation}

In this case, the amplitude of the waves is sufficient at $r = r_\text{rad}$ for them to undergo reflection, leading to the formation of a standing g-mode. Such a condition defines the cut-off frequency \(\omega_c\) as
\begin{equation}\label{eqn:cutoff}
\omega \geq \left[l(l+1)\right]^\frac{3}{8}\left(\epsilon\int_{r_\text{rad}}^{r_\text{int}}K_T \frac{N^3}{r_1^3} dr_1\right)^\frac{1}{4} \equiv \omega_c, 
\end{equation}
 \begin{figure}[!h]
        \centering
        \includegraphics[scale=0.26]{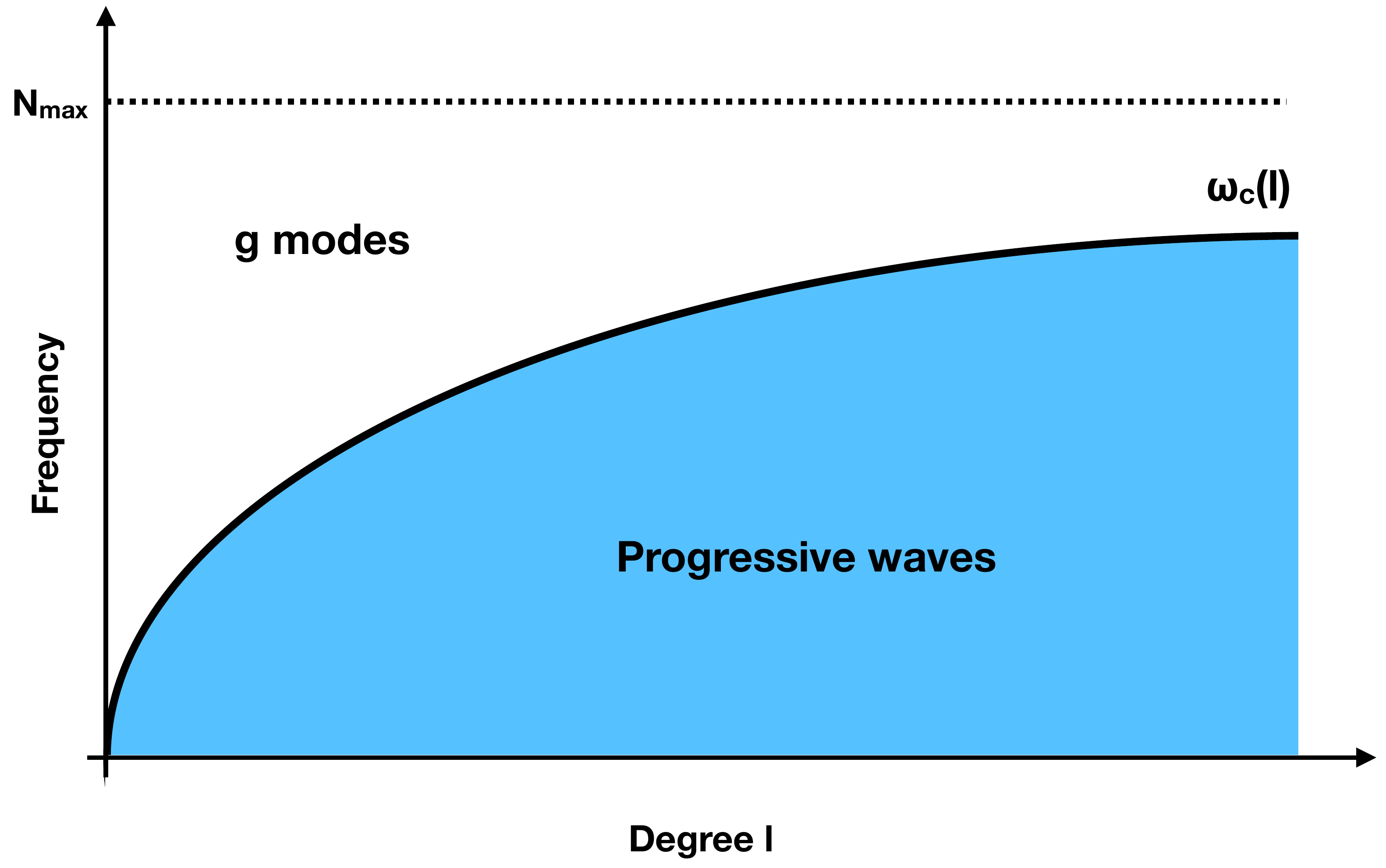}
  \caption{\label{cutoff} Nature of internal gravity waves as a function of degree \(l\) and frequency. The black dashed line corresponds to the maximal value of the Brunt-Väisälä frequency \(N_\text{max}\) in the radiative zone. The black line corresponds to the cut-off frequency \(\omega_c\) as a function of the degree, marking the separation between standing g-modes (in white, above) and progressive internal gravity waves (in blue, below).}
\end{figure}
above which standing modes form, as shown in Fig. \ref{cutoff}. Indeed, for frequencies higher than \(\omega_c\), gravity waves carry enough energy to undergo a reflection in spite of radiative damping, allowing the generation of individual modes. Below this frontier (in blue in Fig. \ref{cutoff}), waves are sufficiently damped so that standing modes cannot form. The spectrum is then only composed of progressive waves. Hence, as propagative waves are damped on a distance lesser than the size of the radiative zone, all the energy carried by gravity waves is dissipated inside the star before any reflection. Such a configuration leads to the most efficient dissipation if no critical layer or non-linear effect is taken into account.
 
 However, a tidal gravity wave is likely to transfer its entire angular momentum to the star through other dissipation mechanisms. Indeed, in the case of a low-mass star, if the amplitude of the wave exceeds a critical value, it can break near the center of the star, thus transferring its angular momentum to the mean flow and bringing the central regions of the star into co-rotation with the tidal forcing \citep{barker10, barker11}. A similar process happens near the stellar surface in the case of intermediate-mass and massive stars \citep{rogers}. This forms a critical layer that acts as an absorbent barrier for the subsequent waves. Such a process happens if the planetary mass exceeds a critical value, which decreases sharply with stellar mass and stellar age \citep{barker20}, as the strength of the stratification increases. We provide a similar wave-braking criterion in Appendix D. For stellar masses larger than 0.9 \(M_\odot\), the minimal planetary mass required to induce wave braking may fall below 1 Jupiter mass at ages lesser than 10 Gyr. Moreover, the interaction of tidal gravity waves with the differential rotation of the surrounding fluid may lead to the formation of a critical layer when the frequency of excited waves is of the same order of magnitude as the angular velocity of the fluid. In the radiative zone of solar-type stars, the fluid remains stable in such a configuration and the amplitude of gravity waves is damped by the critical layer \citep{alvan13}. For a mode of azimuthal number $m$ at an orbital harmonic $N$, in an inertial reference frame, such a layer exists at a radius $r_\text{CL}$ if
\begin{equation}
\omega = m\Omega_\text{RZ}(r_\text{CL}),
\end{equation}
where $\Omega_\text{RZ}(r_\text{CL})$ is the angular velocity of the radiative zone at the radius $r_\text{CL}$. As $\omega = N n_\text{orb}$ in an inertial reference frame, with $n_\text{orb}$ the mean motion of the planetary orbit, it leads to $\Omega_\text{RZ}(r) = N/m\ n_\text{orb}$. One can assess the range of orbital periods leading to a potential interaction with a critical layer by considering a coplanar and circular planetary orbit ($N=m=2$). The radiative core of a solar-type star tends to synchronize its spin with that of the convective zone during the MS \citep[e.g.][]{galletbouvier15,benomar}. Therefore, we assume as a first approximation a weak differential rotation within the star. Thus, one can provide an upper bound of the orbital period required for the creation of a critical layer by focusing on the evolution of the surface rotation rate of the star (which is also the rotation rate of the convective envelope if we assume that the latter is in solid body rotation). Hence, an interaction between a tidal gravity wave and a critical layer may occur within a solar-type star if the orbital period of the planet is shorter than 10 days during the PMS, and 50 days during the MS \citep[we refer the reader to][for more details on the rotational evolution of the radiative zone of solar-type stars]{galletbouvier15,amard16}.

In order to account for all these cases, for which tidal dissipation is likely to be effective, we assume in our model that all the energy carried by gravity waves is dissipated inside the star before any reflection. This way, one may be able to provide an upper bound of tidal dissipation and to unravel the impact of the stellar internal structure and rotation. In this context, we consider forced adiabatic oscillations in the stellar interior, the tidal torque being directly inferred from the angular momentum flux carried by the internal gravity waves.
 
\subsection{Forced dynamics of internal gravity waves}

We assume that the star is in hydrostatic equilibrium, which leads to
\begin{equation}
\vec{\nabla}p_0 = -\rho_0\vec{g_0},\\
\end{equation}
where \(p_0\) is the pressure inside the star, \(\rho_0\) its local density and \(\vec{g_0}\) the gravity. The subscript 0 in the aforementioned quantities refers to the unperturbed background. Such a structure is then perturbed by the tidal potential \(U_T\) applied by the companion. In this approach, by introducing velocity (\(\vec{v_1}\)), pressure (\(p_1\)) and gravitational potential (\(\varphi_1\)) perturbations induced by the planet, one can linearize the equations of hydrodynamics around the equilibrium state by ignoring all dissipative mechanisms:
\begin{equation}
\begin{cases}
\partial_t\rho_1 +\vec{v_1}\cdot\vec{\nabla}\rho_0+\rho_0\vec{\nabla}\cdot\vec{v_1} = 0 \\
\displaystyle \partial_t\vec{v_1} = -\frac{1}{\rho_0}\vec{\nabla}p_1 + \frac{\rho_1}{{\rho_0}^2}\vec{\nabla}p_0  - \vec{\nabla}\varphi_1 - \vec{\nabla}U_T \\
\displaystyle  \frac{1}{p_0}\left[\partial_t p_1 + \left(\vec{v_1}\cdot\vec{\nabla}\right)p_0\right]-\frac{\Gamma_1}{\rho_0}\left[\partial_t \rho_1 + \left(\vec{v_1}\cdot\vec{\nabla}\right)\rho_0\right]=0,
\end{cases}
\end{equation}
with \(\Gamma_1 = (\partial \ln p_0/\partial \ln \rho_0)_S\) the adiabatic exponent of the fluid, \(S\) being the specific macroscopic entropy. We make use of the spherical coordinates \((r,\theta,\varphi)\), with \(r\) the radial coordinate, \(\theta\) the colatitude, and \(\varphi\) the longitude, as well as their corresponding unit-vector basis (\(\vec{e_r}, \vec{e_\theta}, \vec{e_\varphi}\)). Then, after introducing the Lagrangian displacement field \(\bm \xi\), we develop all the fluctuations on the spherical harmonics \({Y_l}^m(\theta,\varphi) \propto P_l^m(\cos\theta)e^{im\varphi}\) as follows:
\begin{equation}
\bm\xi (r,\theta,\varphi,t)= \sum_{l,m}\left[\xi_{r;l,m}(r){Y_l}^m(\theta,\varphi)\vec{e_r}+ {\xi_{h;l,m}}(r)\bm\nabla_h{Y_l}^m(\theta,\varphi)\right]e^{-i\omega t},
\end{equation}
\begin{equation}
\rho_1 (r,\theta,\varphi,t)=\sum_{l,m} \tilde{\rho}_{l,m}(r){Y_l}^m (\theta,\varphi) e^{-i\omega t},
\end{equation}
\begin{equation}
p_1 (r,\theta,\varphi,t)=\sum_{l,m} \tilde{p}_{l,m}(r){Y_l}^m (\theta,\varphi) e^{-i\omega t},
\end{equation}
\begin{equation}
\varphi_1 (r,\theta,\varphi,t)=\sum_{l,m} \tilde{\varphi}_{l,m}(r){Y_l}^m (\theta,\varphi) e^{-i\omega t},
\end{equation}
\begin{equation}
U_T (r,\theta,\varphi,t)=\sum_{l,m} \varphi_{T;l,m} (r) {Y_l}^m  (\theta,\varphi) e^{-i\omega t},
\end{equation}
with \(\bm\nabla_h = \partial_\theta \vec{e_\theta} + (1/\sin\theta) \partial_\varphi\vec{e_\varphi}\) the horizontal gradient. In the rest of this work, we focus on the behavior of a single mode, and we assume that \(l\) and \(m\) are fixed by the tidal potential (for instance, \(l = m = 2\) in a coplanar configuration, \(l=2\) and \(m=0\) for eccentricity tides, \(l=2\) and \(m=1\) for obliquity tides). Then, for the sake of simplicity, the degree and order dependencies of the components of each perturbed quantity will not be explicited anymore. For instance, we write \(\xi_r\) instead of \(\xi_{r;l,m}\). 

Furthermore, we adopt the Cowling approximation \citep{cowling}, where we neglect the fluctuations of the gravific potential of the wave. This approximation is well justified for low-frequency waves. This leads to \citep{press, auclair}
\begin{equation}\label{eqn:press}
\begin{cases}
\displaystyle \partial_r\left(r^2\xi_r\right)+\frac{\partial_r p_0}{\Gamma_1 p_0}\left(r^2\xi_r\right) = \left(\frac{l(l+1)}{\omega^2}-\frac{\rho_0 r^2}{\Gamma_1 p_0}\right)y+\frac{l(l+1)}{\omega^2}\varphi_T\\
\displaystyle \partial_r y - y\frac{N^2}{g_0} = \frac{1}{r^2}\left(\omega^2-N^2\right)\left(r^2\xi_r\right)-\partial_r\varphi_T,
\end{cases}
\end{equation}
where we have introduced the reduced pressure \(y=\tilde{p}/\rho_0\) and the Brunt-Väisälä frequency \(N^2 =g_0\left(\frac{\partial_r p_0}{\Gamma_1 p_0}-\frac{\partial_r\rho_0}{\rho_0}\right)\).

In order to reunite the \citet{zahn75}, \citet{goldreich}, \citet{goodman}, and \citet{barker10} prescriptions, among others, in a flexible framework allowing for the study of a given system's dynamics throughout stellar evolution, we carry out and present in this paper all the necessary derivations.

\subsection{Wave behavior in the radiative zone}
\subsubsection{Approximations in the radiative zone}
Following \citet{press}, in the stably stratified radiative zone, where \(N^2 > 0\), deriving the first equation of \eqref{eqn:press} with respect to $r$ and including the second equation gives
\begin{equation}
\partial_{rr}\left(r^2\xi_r\right)+\frac{\partial_r\rho_0}{\rho_0}\partial_r\left(r^2\xi_r\right)+\mathcal{K}r^2\xi_r=\mathcal{A}+\mathcal{F}_T,
\end{equation}
with 
\begin{equation}
\begin{split}
&\mathcal{K} = \left(\frac{N^2}{\omega^2}-1\right)\frac{l(l+1)}{r^2}+\partial_r\left(\frac{\partial_r p_0}{\Gamma_1 p_0}\right),\\
\\
&\mathcal{F}_T = -\frac{l(l+1)}{\omega^2}\left(\frac{l(l+1)}{\omega^2}-\frac{\rho_0 r^2}{\Gamma_1 p_0}\right)^{-1}\frac{N^2}{g_0}\varphi_T,\\
\\
&\mathcal{A}=\partial_r\left(r^2\xi_r\right)\frac{N^2}{g_0}\left[\frac{l(l+1)}{\omega^2}\left(\frac{l(l+1)}{\omega^2}-\frac{\rho_0 r^2}{\Gamma_1 p_0}\right)^{-1}-1\right]+\\
&\left(r^2\xi_r\right)\left[\frac{l(l+1)}{\omega^2}\left(\frac{l(l+1)}{\omega^2}-\frac{\rho_0 r^2}{\Gamma_1 p_0}\right)^{-1}\!\frac{N^2}{g_0}\frac{\partial_r p_0}{\Gamma_1 p_0}\right]-\partial_r\left[\frac{\rho_0 r^2}{\Gamma_1 p_0}y\right],
\end{split}
\end{equation}
where the last term in $\mathcal{A}$ can be replaced through Eq. \eqref{eqn:press}. By using the anelastic approximation to filter out acoustic waves, we can assume that 
\begin{equation}
\frac{\rho_0 r^2}{\Gamma_1 p_0} \ll \frac{l(l+1)}{\omega^2}.
\end{equation}
Therefore we can simplify the expression of \(\mathcal{A}\) as
\begin{equation}
\mathcal{A} \approx -\xi_r\frac{N^2 r^2}{c_s^2}-\partial_r\left[\frac{r^2 y}{c_s^2}\right],
\end{equation}
with \(c_s = \sqrt{\Gamma_1p_0/\rho_0}\) the speed of sound. In the anelastic approximation one neglects terms of order \(1/c_s^2\) \citep{spiegel}, which leads to the simplified relation:
\begin{equation}\label{eqn:xiequation}
\begin{split}
&\partial_{rr}\left(r^2\xi_r\right)+\partial_r\left(r^2\xi_r\right)\frac{\partial_r\rho_0}{\rho_0}+\mathcal{K}r^2\xi_r=-\frac{N^2}{g_0}\varphi_T.
\end{split}
\end{equation}
To obtain a Schrödinger-like equation in the radiative zone, we introduce a new function \(\psi(r)=\rho_0^{\frac{1}{2}}r^2\xi_r\) which leads to
\begin{equation}\label{eqn:psiequation}
\frac{d^2\psi}{dr^2}+\frac{l(l+1)}{r^2}\left(\frac{N^2}{\omega^2}-1\right)\psi=\frac{l(l+1)N^2}{\omega^2r^2}\left(-\rho_0^{\frac{1}{2}}r^2\frac{\varphi_T}{g_0}\right)+\mathcal{V},
\end{equation}
with  \(\displaystyle \mathcal{V}=\psi\left[\rho_0^{-\frac{1}{2}}\partial_{rr}(\rho_0^{\frac{1}{2}})-\partial_r\left(\frac{\partial_r p_0}{\Gamma_1 p_0}\right)\right] \sim L^{-2}\psi\),
given a characteristic length \(L\) of our system. Eq. \eqref{eqn:xiequation} and \eqref{eqn:psiequation} are equivalent to the \citet{zahn75}, \citet{savonije84} and \citet{goodman} formulations, whose main discrepancies come from the terms taken into account in \(\mathcal{V}\). Here, we assume that the characteristic length of variation of the background is large compared the wavelength of a gravity wave, i.e.
\begin{equation}
\left(\frac{N^2}{\omega^2}-1\right)\frac{l(l+1)}{r^2} \gg L^{-2}.
\end{equation}
Therefore, one can neglect the \(\mathcal{V}\) term and obtain the equation ruling the behavior of internal gravity waves in the radiative zone \citep{zahn75}:
\begin{equation}\label{eqn:rad}
\psi''+\frac{l(l+1)}{r^2}\left(\frac{N^2}{\omega^2}-1\right)\psi=\frac{l(l+1)N^2}{\omega^2r^2}\left(-\rho_0^{\frac{1}{2}}r^2\frac{\varphi_T}{g_0}\right),
\end{equation}
where, for an arbitrary quantity \(F\), \(F' = \frac{dF}{dr}\). Such a convention will be used in the rest of this work to alleviate the notations. 

\subsubsection{Solutions in the radiative zone}
We now have to solve Eq. \eqref{eqn:rad} in the radiative zone. Far from the interface we can assume that \(N^2 \gg \omega^2\), since we consider low-frequency waves. Then one can use the WKBJ approximation \citep{froman} and the solution becomes
\begin{equation}
\psi(r) = -\rho_0^{\frac{1}{2}}r^2\frac{\varphi_T}{g_0}+C_{W}\frac{1}{\sqrt{k_r}}e^{\epsilon i(\tau_W-\tau_0)},
\end{equation}
where \(k_r = \sqrt{\left(\frac{N^2}{\omega^2}-1\right)\frac{l(l+1)}{r^2}}\) is the radial wavenumber, \(C_{W}, \tau_0\) are constants and \(\tau_W = \epsilon\int_{r_\text{rad}}^r{k_r(r)dr}\). The factor \(\epsilon\) is equal to 1 for an inward transport of energy through gravity waves and equal to -1 otherwise.\\

Furthermore, a turning point occurs near the radiative-convective interface, where \(N \approx \omega\). The WKBJ approximation is not relevant anymore and we expand the square of the Brunt-Väisälä frequency around the interface \(r= r_\text{int}\) as 
\begin{equation}
N^2 = \omega^2 + \left|\frac{dN^2}{dr}\right|_\text{int} \epsilon(r_\text{int}-r). 
\end{equation}
Eq. \eqref{eqn:rad} now becomes an inhomogeneous Airy equation \citep{zahn75,goodman}:
\begin{equation}
\frac{d^2\psi}{d\eta^2}+v^2\eta\psi = \frac{l(l+1)N^2}{\omega^2r^2}\left(-\rho_0^{\frac{1}{2}}r^2\frac{\varphi_T}{g_0}\right),
\end{equation}
with 
\begin{equation}\label{eqn:v}
v^2 =\frac{l(l+1)}{r_\text{int}^2\omega^2}\left|\frac{dN^2}{dr}\right|_\text{int}, 
\end{equation}
\begin{equation}
\eta =\epsilon(r_\text{int}-r).
\end{equation}
The solution \(\psi_h\) of the corresponding homogenous equation can be written as a linear combination of the Airy functions Ai and Bi as follows \citep{abramowitz}:
\begin{equation}
\psi_h(\eta) = C_{A}\ \text{Ai}\left[v^\frac{2}{3}(-\eta)\right] + C_{B}\ \text{Bi}\left[v^\frac{2}{3}(-\eta)\right],
\end{equation}
where \(C_{A}\text{ and }C_{B}\) are two constants. Furthermore, the Airy functions can be linked to the Bessel functions \(J_\frac{1}{3}\) and \(J_{-\frac{1}{3}}\) as
\begin{equation}
\text{Ai}(-x) =\frac{\sqrt{x}}{3}\left[J_{\frac{1}{3}}\left(\frac{2}{3}x^\frac{3}{2}\right)+J_{-\frac{1}{3}}\left(\frac{2}{3}x^\frac{3}{2}\right)\right],
\end{equation}
\begin{equation}
\text{Bi}(-x) = \sqrt{\frac{x}{3}}\left[J_{-\frac{1}{3}}\left(\frac{2}{3}x^\frac{3}{2}\right)-J_{\frac{1}{3}}\left(\frac{2}{3}x^\frac{3}{2}\right)\right].
\end{equation}
We rely for the rest of this work on a formulation based on Bessel functions, which will allow us to link \citet{zahn70,zahn75} and \citet{ivanov} approaches, as well as to simplify the matching of the different solutions. The solution \(\psi_h\) therefore becomes
\begin{equation}
\psi_h(\tau) = \left(\frac{\tau}{2}\right)^{\frac{1}{3}}\left[\alpha_\text{rad} J_{\frac{1}{3}}(\tau)+\beta_\text{rad} J_{-\frac{1}{3}}(\tau)\right],
\end{equation}
with \(\tau = \frac{2}{3}v\eta^{\frac{3}{2}}, \alpha_\text{rad} = C_A.3^{-\frac{2}{3}}-C_B.3^{-\frac{1}{6}}\) and \(\beta_\text{rad} = C_A.3^{-\frac{2}{3}}+C_B.3^{-\frac{1}{6}} \). A particular solution \(\psi_p\) of the inhomogeneous Airy equation, vanishing at the interface, can be expressed as (see Appendix A)
\begin{equation}
\psi_\text{p}(\eta) = Z(\eta)+\left(\frac{\tau}{2}\right)^{\frac{1}{3}}\left[\alpha_\text{rad,p}J_{\frac{1}{3}}(\tau)+\beta_\text{rad,p}J_{-\frac{1}{3}}(\tau)\right],
\end{equation}
where \(Z(\eta)=-\rho_0^{\frac{1}{2}}r^2\frac{\varphi_T}{g_0}\) is the particular solution of Eq. \eqref{eqn:rad} in the WKBJ formulation associated with the equilibrium tide, \(\displaystyle\alpha_\text{rad,p} = -\frac{dZ}{d\eta}(0)\left(\frac{v}{3}\right)^{-\frac{2}{3}}\Gamma\left(\frac{4}{3}\right)\) and \(\beta_\text{rad,p} = -Z(0)\Gamma\left(\frac{2}{3}\right)\). Since \(\displaystyle J_{\pm\frac{1}{3}}(\tau) \underset{\tau \to +\infty}{\sim}\sqrt{\frac{2}{\pi\tau}}\cos(\tau\mp\frac{\pi}{6}-\frac{\pi}{4})\) far from the interface, the particular solution obtained with the Bessel formulation matches with the particular solution derived through the WKBJ approximation.

\subsection{Wave behavior in the convective zone}
\subsubsection{Approximations in the convective zone}

In the convective zone, since we consider adiabatic oscillations, we assume \(N^2 = 0\). Such an approximation holds in convective cores and in the regions of convective envelopes for which convection is efficient \citep[at radii ranging up to \(0.9\ R_\star\), below the superadiabatic layer ; we refer the reader to][]{lebreton}. In this configuration, Eqs. \eqref{eqn:press} become 
\begin{equation}
\begin{cases}
\displaystyle\partial_r\left(r^2\xi_r\right)=-\frac{\partial_r p_0}{\Gamma_1 p_0}\left(r^2\xi_r\right)+ \frac{\rho_0 r^2}{\Gamma_1 p_0}(\varphi_T-\chi)+\frac{l(l+1)}{\omega^2}\chi\\
\displaystyle\xi_r=\frac{\partial_r\chi}{\omega^2},
\end{cases}
\end{equation}
where we define \(\chi=y+\varphi_T\). Since \(N^2=0\), we have \(\frac{\partial_r p_0}{\Gamma_1 p_0}=\frac{\partial_r \rho_0}{\rho_0}\), which leads to
\begin{equation}
\begin{cases}
\displaystyle\partial_r\left(r^2\xi_r\right)=-\frac{\partial_r \rho_0}{\rho_0}\left(r^2\xi_r\right)+ \frac{\rho_0 r^2}{\Gamma_1 p_0}(\varphi_T-\chi)+\frac{l(l+1)}{\omega^2}\chi\\
\displaystyle\xi_r=\frac{\partial_r\chi}{\omega^2}
\end{cases}
\end{equation}
As we focus on low-frequency gravity waves, we rely on the anelastic approximation, which gives
\begin{equation}
\partial_r\left(r^2\xi_r\right)=-\frac{\partial_r \rho_0}{\rho_0}\left(r^2\xi_r\right)+ \frac{l(l+1)}{\omega^2}\chi+z\\
\end{equation}
with \(\displaystyle z = -\frac{\partial_r \rho_0}{\rho_0}\frac{r^2\varphi_T}{g_0}\). We now define the quantity \(X(r)=\rho_0 r^2\xi_r\), which  verifies the following relation:
\begin{equation}\label{eqn:eqconvinter}
X'= \frac{l(l+1)}{\omega^2}\rho_0\chi +\rho_0 z.
\end{equation}
By differentiating Eq. \eqref{eqn:eqconvinter} we derive the equation ruling the behavior of evanescent internal gravity waves in the convective zone \citep{zahn75,ivanov}:
\begin{equation}\label{eqn:conv}
X''-\frac{\partial_r\rho_0}{\rho_0}X'-\frac{l(l+1)}{r^2}X=\rho_0 z'.
\end{equation}

\subsubsection{Solutions in the convective zone}
We now need to solve equation \eqref{eqn:conv} in the convective zone. Following \citet{zahn75}, if we consider \(X_1,\ X_2\) two independent solutions of the corresponding homogeneous ordinary differential equation, the general solution of Eq. \eqref{eqn:conv} can be written as
\begin{equation}\label{eqn:solX}
X=\left[C_1\!-\!\int_{r_\text{conv}}^r{\!\!\!\!\Lambda^{-1}\rho_0 z' X_2 dr}\right]X_1+\left[C_2\!+\!\int_{r_\text{conv}}^r{\!\!\!\!\Lambda^{-1}\rho_0 z' X_1 dr}\right]X_2,
\end{equation}
where \(C_1\) and \(C_2\) are two constants of integration and \(\Lambda = X_2'X_1-X_1'X_2\) is their Wronskian.
If we consider the displacement functions \(\xi_1 = X_1/(\rho_0 r^2)\) and \(\xi_2 = X_2/(\rho_0 r^2)\) as well as their Wronskian \(\Lambda_\xi = \xi_1\xi_2'-\xi_2\xi_1'\), the particular solution of Eq. \eqref{eqn:conv} can be written in an alternative form, knowing that \(\Lambda \propto \rho_0\):
\begin{equation}\label{eqn:solpartconv}
\int_{r_\text{conv}}^r{\Lambda^{-1}\rho_0z'X_i dr}=\Lambda_\xi^{-1}(r_\text{int})r_\text{int}^{-2}\xi_i(r_\text{int})\int_{r_\text{conv}}^r{z'\frac{X_i}{X_i(r_\text{int})} dr},
\end{equation}
with \(i=1\) or 2. Furthermore, the integral appearing in Eq. \eqref{eqn:solpartconv} can be expressed as
\begin{equation}\label{eqn:ipps}
\begin{split}
\int_{r_\text{conv}}^r{z'X_i dr} =\mathcal{B}+\mathcal{F}_i,\\
\end{split}
\end{equation}
with
\begin{equation}
\begin{split}
&\mathcal{B} = \left[-\left\{\frac{\rho_0'}{\rho_0}\frac{r^2\varphi_T}{g_0}+\left(\frac{r^2\varphi_T}{g_0}\right)'\right\}X_i+\left(\frac{r^2\varphi_T}{g_0}\right)X_i'\right]_{r_\text{conv}}^r,\\
&\mathcal{F}_i = \int_{r_\text{conv}}^r{\left[\left(\frac{r^2\varphi_T}{g_0}\right)''-\frac{l(l+1)}{r^2}\left(\frac{r^2\varphi_T}{g_0}\right)\right]X_i dr}.
\end{split}
\end{equation}

\subsection{Matching of the solutions}
\subsubsection{Matching radiative-convective zone}
At the interface, we carry out the matching between the solutions in the radiative and the convective layers by taking tidal forcing into account. For the sake of simplicity we define \(S_+(\tau) = \left(\frac{\tau}{2}\right)^\frac{1}{3}J_\frac{1}{3}\left(\tau\right)\) and \(S_-(\tau) = \left(\frac{\tau}{2}\right)^\frac{1}{3}J_{-\frac{1}{3}}\left(\tau\right)\), which constitute a basis for the homogeneous solution in the radiative zone in a Bessel formulation. Near the interface, we can express a basis solution of the homogeneous equation in the convective zone \(\rho_0^{-\frac{1}{2}}X_i\), with \(i=1\) or 2, as a linear combination \(\alpha_i S_+ + \beta_i S_-\). This way we obtain
\begin{equation}
\rho_0^{-\frac{1}{2}}X(r_\text{int})=(\alpha_\text{conv}+\alpha_\text{conv, p})S_+(0)+(\beta_\text{conv}+\beta_\text{conv,p})S_-(0),
\end{equation}
where
\begin{equation}
\begin{cases}
\alpha_\text{conv} = \alpha_1 C_1 + \alpha_2 C_2\\
\beta_\text{conv} = \beta_1 C_1 + \beta_2 C_2\\
\alpha_\text{conv,p} = -\alpha_1\int_{r_\text{conv}}^{r_\text{int}}{\Lambda^{-1}\rho_0 z' X_2 dr}+\alpha_2 \int_{r_\text{conv}}^{r_\text{int}}{\Lambda^{-1}\rho_0 z' X_1 dr}\\ 
\beta_\text{conv,p} = -\beta_1\int_{r_\text{conv}}^{r_\text{int}}{\Lambda^{-1}\rho_0 z' X_2 dr}+\beta_2 \int_{r_\text{conv}}^{r_\text{int}}{\Lambda^{-1}\rho_0 z' X_1 dr}.
\end{cases}
\end{equation}
Furthermore, we can choose \(\rho_0^{-\frac{1}{2}}X_2\) to match the basis solution \(S_+\) at the interface. Then we have \(\alpha_2 = 1\) and \(\beta_2 = 0\). One can also assume without restricting the generality of the foregoing that \(\alpha_\text{conv} \ll \beta_\text{conv}\), which amounts to choose a solution \(X_1\) close to \(S_-\). We now aim to characterize all the solutions at the interface and to better constrain the integration constants. To this end, we write \(S_+\) and \(S_-\) as
\begin{equation}
S_\pm (\tau) = \frac{1}{2}\left[3^\frac{2}{3}\text{Ai}\left(v^\frac{2}{3}(-\eta)\right)\mp3^\frac{1}{6}\text{Bi}\left(v^\frac{2}{3}(-\eta)\right)\right],
\end{equation}
which at the interface leads to
\begin{equation}
\begin{cases}
\displaystyle S_+(0) = 0\\
\displaystyle S_-(0) = \frac{1}{\Gamma\left(\frac{2}{3}\right)}\\
\displaystyle \partial_r S_+(0) = \frac{-\epsilon}{\Gamma\left(\frac{4}{3}\right)}\left(\frac{v}{3}\right)^\frac{2}{3} \\
\displaystyle \partial_r S_-(0) = 0. \\
\end{cases}
\end{equation}
On the basis of those calculations, the general solution \(X_h = \alpha_\text{conv}S_+ + \beta_\text{conv}S_-\) of the homogeneous equation in the convective zone verifies the following condition \citep{zahn75}:
\begin{equation}\label{eqn:homrel}
\frac{\alpha_\text{conv}}{\beta_\text{conv}}\frac{\Gamma(\frac{2}{3})}{\Gamma(\frac{4}{3})}\left(\frac{v}{3}\right)^{\frac{2}{3}} = -\epsilon\frac{\frac{d}{dr}(\rho_0^{-\frac{1}{2}}X_h)_\text{int}}{(\rho_0^{-\frac{1}{2}}X_h)_\text{int}}.
\end{equation}
Moreover, in the same layer, the Wronskian \(\Lambda_\xi\) of the displacement functions corresponding to the basis solutions of the homogeneous equation can be expressed as
\begin{equation}
\begin{split}
\Lambda_\xi (r_\text{int}) &= \rho_0^{-2}(r_\text{int})r_\text{int}^{-4} \left[X_1(r_\text{int})X'_2(r_\text{int})-X_2(r_\text{int})X'_1(r_\text{int})\right]\\
&=-\epsilon\rho_0^{-1}(r_\text{int})r_\text{int}^{-4}\frac{\beta_1}{\Gamma\left(\frac{4}{3}\right)\Gamma\left(\frac{2}{3}\right)}\left(\frac{v}{3}\right)
^\frac{2}{3}.
\end{split}
\end{equation}
The values of the displacement functions themselves at the interface \(\xi_i(r_\text{int}) = \rho_0^{-1}(r_\text{int}) r_\text{int}^{-2} X_i(r_\text{int})\) become
\begin{equation}
\xi_i(r_\text{int}) = \begin{cases}
\displaystyle \rho_0^{-\frac{1}{2}}(r_\text{int}) r_\text{int}^{-2}\frac{\beta_1}{\Gamma\left(\frac{2}{3}\right)}\text{, if}\ i=1\\
0\text{, if}\ i=2.\\
\end{cases}
\end{equation}
Therefore from Eq. \eqref{eqn:solpartconv} we obtain:
\begin{equation}
\int_{r_\text{conv}}^{r_\text{int}}{\!\!\Lambda^{-1}\rho_0z'X_1 dr}=-\epsilon\rho_0^{\frac{1}{2}}(r_\text{int})\Gamma\left(\frac{4}{3}\right)\left(\frac{v}{3}\right)
^{-\frac{2}{3}}\!\!\!\int_{r_\text{conv}}^{r_\text{int}}{\! z'\frac{X_1}{X_1(r_\text{int})} dr}
\end{equation}
Then we have for the particular solution in the convective zone:
\begin{equation}
\begin{split}
\alpha_\text{conv,p} =&-\alpha_1\int_{r_\text{conv}}^{r_\text{int}}{\Lambda^{-1}\rho_0 z' X_2 dr}\\
&-\epsilon\rho_0^{\frac{1}{2}}(r_\text{int})\ \Gamma\left(\frac{4}{3}\right)\left(\frac{v}{3}\right)
^{-\frac{2}{3}}\!\!\!\int_{r_\text{conv}}^{r_\text{int}}{z'\frac{X_1}{X_1(r_\text{int})} dr}.
\end{split}
\end{equation}
The matching of the inhomogeneous solutions in the radiative and convective zones then gives
\begin{equation}
\begin{split}
&(\alpha_\text{rad}+\alpha_\text{rad,p})\begin{pmatrix}S_+(0)\\\partial_r S_+(0)\end{pmatrix}+(\beta_\text{rad}+\beta_\text{rad,p})\begin{pmatrix}S_-(0)\\\partial_r S_-(0)\end{pmatrix}\\ &=(\alpha_\text{conv}+\alpha_\text{conv,p})\begin{pmatrix}S_+(0)\\\partial_r S_+(0)\end{pmatrix}+(\beta_\text{conv}+\beta_\text{conv,p})\begin{pmatrix}S_-(0)\\\partial_r S_-(0)\end{pmatrix}.
\end{split}
\end{equation}
Since the particular solution in the radiative zone vanishes at the interface, we obtain
\begin{equation}
\alpha_\text{rad} = \alpha_\text{conv} + \alpha_\text{conv,p},
\end{equation}
\begin{equation}
\beta_\text{rad} = \beta_\text{conv} + \beta_\text{conv,p}.
\end{equation}

\subsubsection{Matching WKBJ-Bessel}
In the radiative zone, far from the interface, the functions \(J_\frac{1}{3} (\tau)\) and \(J_{-\frac{1}{3}}(\tau)\) for large values of \(\tau\) have the following asymptotic form:  \(\displaystyle J_{\pm\frac{1}{3}}(\tau)\sim\sqrt{\frac{2}{\pi\tau}}\cos(\tau\mp\frac{\pi}{6}-\frac{\pi}{4})\), which gives
\begin{equation}
\psi(\tau) = Z(\eta)+\frac{1}{\sqrt\pi}\!\left(\frac{2}{\tau}\right)^{\frac{1}{6}}\!\left[\mathcal{A}_+\!\cos\!\left(\tau-\frac{5\pi}{12}\right)+\mathcal{A}_-\!\cos\!\left(\tau-\frac{\pi}{12}\right)\right],
\end{equation}
with
\begin{equation}
\begin{split}
&\mathcal{A}_+ = \alpha_\text{conv}+\alpha_\text{conv, p}+\alpha_\text{rad,p},\\
&\mathcal{A}_- = \beta_\text{conv}+\beta_\text{conv, p}+\beta_\text{rad,p}.
\end{split}
\end{equation}
If we define \(\mathcal{C} = \frac{3^{\frac{1}{6}}}{\sqrt\pi}v^{\frac{1}{3}}\), knowing that \(k_r = v\eta^{\frac{1}{2}}\) near the interface, we obtain 
\begin{equation}
\psi(\tau) = Z(\eta)+\frac{\mathcal{C}}{\sqrt k_r}\left[\mathcal{A}_+\cos\!\left(\tau-\frac{5\pi}{12}\right)+\mathcal{A}_-\cos\!\left(\tau-\frac{\pi}{12}\right)\right].\\ 
\end{equation}
Furthermore, we can note that \(\int_{r}^{r_\text{int}}{k_r dr} =\epsilon\tau\). By introducing the constant \(\varphi = \tau + \tau_W - \tau_0  = \epsilon \int_{r_\text{rad}}^{r_\text{int}}{k_r dr} - \tau_0 \), the asymptotic solution becomes
\begin{equation}
\begin{split}
&\psi(\eta) = Z(\eta)\\
&+\frac{\mathcal{C}}{\sqrt k_r}\!\left[\mathcal{A}_+\cos\!\left(\varphi-\frac{5\pi}{12}\right)+\mathcal{A}_-\cos\!\left(\varphi-\frac{\pi}{12}\right)\right]\cos(\tau_W - \tau_0)\\
&+\frac{\mathcal{C}}{\sqrt k_r}\!\left[\mathcal{A}_+\sin\!\left(\varphi-\frac{5\pi}{12}\right)+\mathcal{A}_-\sin\!\left(\varphi-\frac{\pi}{12}\right)\right]\sin(\tau_W - \tau_0).
\end{split}
\end{equation}
The matching with the WKBJ solution \(\psi =Z+C_{W}\frac{1}{\sqrt{k_r}}e^{\epsilon i(\tau_W-\tau_0)}\) then leads to the following system:
\begin{equation}
\begin{cases}
\displaystyle\mathcal{A}_+\cos\left(\varphi-\frac{5\pi}{12}\right)+\mathcal{A}_-\cos\left(\varphi-\frac{\pi}{12}\right) = \frac{C_{W}}{\mathcal{C}}\\
\displaystyle\mathcal{A}_+\sin\left(\varphi-\frac{5\pi}{12}\right)+\mathcal{A}_-\sin\left(\varphi-\frac{\pi}{12}\right) = \epsilon i\frac{C_{W}}{\mathcal{C}},\\
\end{cases}
\end{equation}
which leads to \citep{ivanov}
\begin{equation}
\begin{cases}
\displaystyle\alpha_\text{conv} + \alpha_\text{conv,p} + \alpha_\text{rad,p} = \frac{C_{W}}{\mathcal{C}}\frac{\sin\left(\varphi-\frac{\pi}{12}\right)-\epsilon i\cos\left(\varphi-\frac{\pi}{12}\right)}{\sin\left(\frac{\pi}{3}\right)}\\
\displaystyle\beta_\text{conv}+\beta_\text{conv,p}+\beta_\text{rad,p} = -\frac{C_{W}}{\mathcal{C}}\frac{\sin\left(\varphi-\frac{5\pi}{12}\right)-\epsilon i\cos\left(\varphi-\frac{5\pi}{12}\right)}{\sin\left(\frac{\pi}{3}\right)}.
\end{cases}
\end{equation}
\\
Therefore, we can express the WKBJ amplitude \(C_{W}\) as a function of the particular solution coefficients as follows:
\begin{equation}
\begin{split}
\frac{\alpha_\text{conv}}{\beta_\text{conv}}(\beta_\text{conv,p}+\beta_\text{rad,p})\!-\!\alpha_\text{conv,p}\!-\!\alpha_\text{rad,p}= \frac{-2C_W}{\sqrt 3 \mathcal{C}} e^{\epsilon i\left(\varphi-\frac{\pi}{12}-\frac{\pi}{2}\right)},
\end{split}
\end{equation}
where we used the fact the \(\alpha_\text{conv}\ll\beta_\text{conv}\). We finally obtain the following expression for \(C_W\) \citep{zahn75}:
\begin{equation}
C_{W} = - K_0 e^{-\epsilon i\left(\varphi-\frac{\pi}{12}-\frac{\pi}{2}\right)},
\end{equation}
with \(K_0 = \frac{\sqrt 3}{2}\mathcal{C}\left[\displaystyle\frac{\alpha_\text{conv}}{\beta_\text{conv}}(\beta_\text{conv,p}+\beta_\text{rad,p})-\alpha_\text{conv,p}-\alpha_\text{rad,p}\right]\).

\subsubsection{Closure of the system}
In the previous sections, we have been able to characterize the particular solutions in the convective and the radiative zone thanks to the following coefficients:
\begin{equation}
\begin{cases}\label{eqn:coeffpart}
\alpha_\text{rad,p} =-\frac{dZ}{d\eta}(0)\left(\frac{v}{3}\right)^{-\frac{2}{3}}\Gamma_1\left(\frac{4}{3}\right)\\
\beta_\text{rad,p} =-Z(0)\Gamma\left(\frac{2}{3}\right)\\
\alpha_\text{conv,p} = -\alpha_1\int_{r_\text{conv}}^{r_\text{int}}{\Lambda^{-1}\rho_0 z' X_2 dr}\\
\qquad\qquad-\epsilon\rho_0^{\frac{1}{2}}(r_\text{int})r_\text{int}^2\ \Gamma\left(\frac{4}{3}\right)\left(\frac{v}{3}\right)
^{-\frac{2}{3}}r_\text{int}^{-2}\int_{r_\text{conv}}^{r_\text{int}}{z'\frac{X_1}{X_1(r_\text{int})} dr}\\
\beta_\text{conv,p} = -\beta_1\int_{r_\text{conv}}^{r_\text{int}}{\Lambda^{-1}\rho_0 z' X_2 dr}.
\end{cases}
\end{equation}
Then we can compute the quantity \(K_0\) as follows:
\begin{equation}
\begin{split}
&K_0=\mathcal{T}_0+\frac{3\Gamma\left(\frac{4}{3}\right)}{2\sqrt\pi}\left(\frac{v}{3}\right)^{-\frac{1}{3}}\!\!\!\!\rho_0^\frac{1}{2}(r_\text{int})r_\text{int}^2\times\\
&\left\{\left[-\epsilon\frac{\frac{d}{dr}(\rho_0^{-\frac{1}{2}}X_\text{h})_\text{int}}{(\rho_0^{-\frac{1}{2}}X_\text{h})_\text{int}}-\frac{\frac{dZ}{d\eta}(0)}{Z(0)}\right]\left(\frac{\varphi_T}{g_0}\right)_\text{int}\!\!\!+\epsilon r_\text{int}^{-2}\!\!\int_{r_\text{conv}}^{r_\text{int}}{\!\! z'\frac{X_1}{X_1(r_\text{int})} dr}\right\}
\end{split}
\end{equation}
where 
\begin{equation}
\mathcal{T}_0 = -\frac{\sqrt 3}{2}\mathcal{C}\left(\frac{\alpha_\text{conv}}{\beta_\text{conv}}\beta_1-\alpha_1\right)\int_{r_\text{conv}}^{r_\text{int}}{\Lambda^{-1}\rho_0 z' X_2 dr}.
\end{equation}
Furthermore, from Eq. \eqref{eqn:ipps} we have
\begin{equation}\label{eqn:intK0}
\begin{split}
&r_\text{int}^{-2}\int_{r_\text{conv}}^{r_\text{int}} z'\frac{X_1}{X_1(r_\text{int})}dr = r_\text{int}^{-2}\mathcal{F}_1+\mathcal{T}_1\\
&+\left(\frac{\varphi_T}{g_0}\right)_\text{int}\left\{-\frac{\rho_0'(r_\text{int})}{\rho_0(r_\text{int})}-\frac{\left(\rho_0^{-\frac{1}{2}}Z\right)_\text{int}'}{\left(\rho_0^{-\frac{1}{2}}Z\right)_\text{int}}+\frac{X_1'(r_\text{int})}{X_1(r_\text{int})}\right\},\\
\end{split}
\end{equation}
where
\begin{equation}
\begin{split}
\mathcal{F}_1 = &\int_{r_\text{conv}}^{r_\text{int}}{\left[\left(\frac{r^2\varphi_T}{g_0}\right)'' - \frac{l(l+1)}{r^2}\left(\frac{r^2\varphi_T}{g_0}\right)\right]\frac{X_1}{X_1(r_\text{int})}}dr,\\
\\
\mathcal{T}_1 = &\left(\frac{\varphi_T}{g_0}\right)_{r_\text{conv}}\!\left(\frac{r_\text{conv}}{r_\text{int}}\right)^2\times\\
&\left\{\left[\frac{\rho_0'(r_\text{conv})}{\rho_0(r_\text{conv})}+\frac{\left(\rho_0^{-\frac{1}{2}}Z\right)'_{r_\text{conv}}}{\left(\rho_0^{-\frac{1}{2}}Z\right)_{r_\text{conv}}}\right]\frac{X_1(r_\text{conv})}{X_1(r_\text{int})}-\frac{X_1'(r_\text{conv})}{X_1(r_\text{int})}\right\},
\end{split}
\end{equation}
From this we obtain
\begin{equation}\label{eqn:K0}
K_0 =\mathcal{T}_0+\frac{3\Gamma\left(\frac{4}{3}\right)}{2\sqrt\pi}\left(\frac{v}{3}\right)^{-\frac{1}{3}}\rho_0^\frac{1}{2}(r_\text{int})r_\text{int}^2 \left\{\epsilon r_\text{int}^{-2}\mathcal{F}_1+\epsilon\mathcal{T}_1+\mathcal{T}_2\right\},
\end{equation}
with
\begin{equation}
\mathcal{T}_2 = \left(\frac{\varphi_T}{g}\right)_{r_\text{int}}\left(\frac{v}{3}\right)^\frac{2}{3}\frac{\Gamma\left(\frac{2}{3}\right)}{\Gamma\left(\frac{4}{3}\right)}\frac{C_2}{\beta_1 C_1}.
\end{equation}
The constants of integration \(C_1\) and \(C_2\), defined in Eq. \eqref{eqn:solX}, correspond to the homogeneous solutions in the convective zone. This way, we are able to assess the WKBJ amplitude of the tidal gravity waves from stellar properties and the tidal potential. Such a quantity is of prime importance to estimate the energy and angular momentum fluxes carried by tidal internal gravity waves.

\section{Tidal dissipation in stellar radiative zones}
\subsection{Energy flux and luminosity}
Let us go back to the perturbed equations of hydrodynamics, assuming the Cowling approximation \citep{cowling}: 
\begin{equation}
\begin{cases}
\partial_t\rho_1 +\vec{v_1}\cdot\vec{\nabla}\rho_0+\rho_0\vec{\nabla}\cdot\vec{v_1} = 0 \\
\partial_t\vec{v_1} = -\frac{1}{\rho_0}\vec{\nabla}p_1 + \frac{\rho_1}{{\rho_0}^2}\vec{\nabla}p_0 - \vec{\nabla}U_T \\
\frac{1}{p_0}\left[\partial_t p_1 + \left(\vec{v_1}\cdot\vec{\nabla}\right)p_0\right]-\frac{\Gamma_1}{\rho_0}\left[\partial_t \rho_1 + \left(\vec{v_1}\cdot\vec{\nabla}\right)\rho_0\right]=0.
\end{cases}
\end{equation}
We introduce the buoyancy term \(\displaystyle\vec b = \frac{\rho_1}{{\rho_0}^2}\vec{\nabla}p_0 = -\frac{g\rho_1}{{\rho_0}}\vec{e_r}\), which verifies in the anelastic approximation
\begin{equation}
\partial_t b =-N^2\ \vec{v_1}\cdot\vec{e_r}.
\end{equation}
The momentum equation then gives
\begin{equation}
\partial_t\vec{v_1}\cdot\vec{v_1} = -\frac{1}{\rho_0}\vec{\nabla}p_1\cdot\vec{v_1} + \vec{b}\cdot\vec{v_1} - \vec{\nabla}U_T\cdot\vec{v_1},
\end{equation}
which leads to
\begin{equation}
\partial_t\left[\frac{1}{2}\rho_0 v_1^2+\rho_0\frac{b^2}{2N^2}\right]+\vec{\nabla}\!\cdot\!\left[\left(p_1+\rho_0 U_T\right)\vec{v_1}\right] = p_1\vec{\nabla}\cdot\vec{v_1}  - U_T\partial_t \rho_1\\
\end{equation}
From the anelastic approximation, one can simplify the continuity equation, as \(\partial_t \rho_1\) can be neglected. Furthermore, at low frequencies, the anelastic approximation reduces to the Boussinesq approximation, which means that we can assume that \(\nabla \cdot \vec{v_1} = 0\). In this framework, the conservation of energy can be expressed as follows:
\begin{equation}
\partial_t e+\vec{\nabla}\cdot\vec{F_E}=0, 
\end{equation}
where \(e = \frac{1}{2}\rho_0 v_1^2+\rho_0\frac{b^2}{2N^2}\) is the energy density of the fluid, sum of the kinetic energy density \(1/2\rho_0 v_1^2\) and of the potential energy density \(\rho_0 b^2/2N^2\), and \(\vec{F_E} = (p_1+\rho_0 U_T)\vec{v_1}\) is the energy flux carried by the tidal internal gravity waves.

If we decompose the relevant quantities on the spherical harmonics, the temporal mean of the energy flux along the direction of propagation becomes
\begin{equation}
\begin{split}
F_E(r,\theta,\varphi) &= \frac{1}{2}\Re\left\{\rho_0 (r)\left[y(r)+\varphi_T(r)\right]\left(-i\omega\xi_r (r)\right)^*\right\}|Y_l^m(\theta,\varphi)|^2\\
&=-\frac{\rho_0 \omega}{2}\Im\left\{\chi(r) \xi_r^*(r)\right\}|Y_l^m(\theta,\varphi)|^2.
\end{split}
\end{equation}
To assess the energy flux carried by a gravity wave travelling in the radiative zone far from the interface, we consider only the radial displacement associated with the dynamical tide in its WKBJ form \(\xi_r = C_{W}\ \rho_0^{-\frac{1}{2}}r^{-2}{k_r}^{-\frac{1}{2}}\exp\left[\epsilon i(\tau_W-\tau_0)\right]\). Furthermore, if we use the anelastic approximation on Eqs. \eqref{eqn:press} we obtain
\begin{equation}
\chi = \frac{\omega^2}{l(l+1)}\partial_r(r^2\xi_r).
\end{equation}
The mean energy flux carried by the waves then becomes:
\begin{equation}
F_E = -\epsilon \frac{\omega^3}{2l(l+1)}|C_{W}|^2r^{-2}|Y_l^m|^2\\
\end{equation}
From this expression, one can compute the energy luminosity \(L_E (r) = \int_{0}^{2\pi}\int_{0}^{\pi} F_E(r,\theta,\varphi)\ r^2\sin \theta\ d\theta d\varphi\) as
\begin{equation}\label{eqn:energylum}
L_E = -\epsilon\frac{1}{2}\frac{\omega^3}{l(l+1)}|C_{W}|^2,
\end{equation}
where the constant \(C_W\) can be computed though the expression of \(K_0\) in Eq. \eqref{eqn:K0}. As expected from adiabatic oscillations, the energy luminosity is conserved. Furthermore, in the case of an inward energy transport (for which \(\epsilon = 1\); see Table \ref{tab:values}), the energy luminosity is negative, while in the case of an outward energy transport (for which \(\epsilon = -1\); see Table \ref{tab:values}), \(L_E\) is positive.

\subsection{Computation of the tidal torque}
In the radiative zone, far from the interface, one can express the azimuthal displacement from the radial displacement in its WKBJ form as
\begin{equation}
\begin{split}
\xi_h &= \frac{\chi}{\omega^2 r}\\
&=\epsilon i\frac{r k_r}{l(l+1)}\xi_r.
\end{split}
\end{equation}
Such an expression then leads to the azimuthal velocity \citep{zahn97}
\begin{equation}
v_{1\varphi}=-\epsilon m\frac{rk_r}{l(l+1)}\frac{v_{1r}}{\sin \theta}.
\end{equation}
From those calculations, one can compute the temporal mean of angular momentum flux along the direction of propagation as
\begin{equation}\label{eqn:angmomflux}
\begin{split}
F_J(r,\theta,\varphi) &= \frac{1}{2}\rho_0\ r\sin \theta\  \Re\left\{ v_{1\varphi}(r,\theta,\varphi)v_{1r}^*(r,\theta,\varphi)\right\}\\
&=\frac{m}{\omega}F_E (r,\theta,\varphi).
\end{split}
\end{equation}
From this expression, one can compute the mean luminosity of angular momentum \(L_J (r) = \int_{0}^{2\pi}\int_{0}^{\pi} F_J(r,\theta,\varphi)\ r^2\sin \theta\ d\theta d\varphi\):
\begin{equation}\label{eqn:LJ_expr}
L_J = -\epsilon \frac{m}{2}\frac{\omega^2}{l(l+1)}|C_{W}|^2,
\end{equation}
which represents the angular momentum transported by tidal gravity waves per unit of time. Such a quantity is also conserved as we consider adiabatic oscillations. Furthermore, we find that prograde waves ($m > 0$) transport angular momentum towards the center of the star in the case of an inward energy transport ($\epsilon = 1$, corresponding to the configuration of solar-type stars), while they deposit angular momentum near the stellar surface in the case of an outward energy transport ($\epsilon = -1$, corresponding to the massive and intermediate-mass stars' configuration). Since we assume that tidal gravity waves are entirely dissipated in the radiative zone before any reflection (see \S 2.1), the star then undergoes a tidal torque \(T\) coming from the deposition of angular momentum by the excited waves. Hence, prograde waves transporting energy inwards in the solar-type configuration and outwards in the massive star configuration both lead to a spin-up of the radiative zone (we refer the reader to Appendix E for an explicit computation), which leads to
\begin{equation}\label{eqn:torque_general}
T = -\epsilon L_J. 
\end{equation}
Inserting Eq. \eqref{eqn:LJ_expr} in Eq. \eqref{eqn:torque_general} leads to
\begin{equation}
T = \frac{m}{2}\frac{\omega^2}{l(l+1)}|C_{W}|^2,
\end{equation}
which is independent of the direction of propagation of the waves. Such an expression has to be compared to the following formulation in the case of a coplanar and circular system \citep{murray}:
\begin{equation}\label{eqn:murray}
|T| = \frac{9}{4Q'}\frac{m_p^2}{M_\star}R_\star^2 \frac{n^4}{\omega_\text{dyn}^2},
\end{equation}
where \(M_\star\) is the stellar mass, \(R_\star\) is the stellar radius, \(m_p\) is the planet mass, \(n\) the mean motion, \(\displaystyle\omega_\text{dyn}^2 = \frac{GM_\star}{R_\star^3}\) and \(Q'\) is a modified quality factor, defined as the ratio of the total energy stored in the tidal velocity field divided by the energy dissipated over one planetary revolution \citep{goldreich63,kaula64,macdonald}. Such a quantity can be linked to the Love number \(k_2\), which is the ratio of the perturbed gravitational potential of the star divided by the tidal potential at the stellar surface, as 
\begin{equation}
Q' = \frac{3}{2|\Im(k_2)|}.
\end{equation}
From this formulation we can assess the tidal dissipation and the modified quality factor as
\begin{equation}
|\Im (k_2)| = \frac{1}{3}\frac{|m|}{l(l+1)}\frac{M_\star}{m_p^2 R_\star^2}\frac{\omega_\text{dyn}^2\omega^2}{n^4}|C_{W}|^2,
\end{equation}
\begin{equation}
Q' = \frac{9}{2}\frac{l(l+1)}{|m|}\frac{m_p^2 R_\star^2}{M_\star}\frac{n^4}{\omega_\text{dyn}^2\omega^2}\frac{1}{|C_{W}|^2}.
\end{equation}

\section{Application to stellar structures}
\subsection{Dynamical tide in intermediate-mass and massive stars: the \citet{zahn75} prescription}
In the case of a massive star, we consider a bi-layer structure with a convective core and a radiative envelope. The energy transported by a tidal gravity wave will therefore propagate outward, towards the stellar surface. This gives in our formulation \(r_\text{conv} = 0, \ r_\text{rad} = R_\star\) and \(\epsilon = -1\), as already presented in Table \ref{tab:values}. Furthermore, we assume that the radial displacement and the tidal potential vanish at the center due to spherical symmetry. Since Eq. \eqref{eqn:press} in the anelastic approximation leads to
\begin{equation}
\partial_r\left(r^2\xi_r\right)+\frac{\partial_r p_0}{\Gamma_1 p_0}\left(r^2\xi_r\right) = \frac{l(l+1)}{\omega^2}\varphi_T,
\end{equation}
the conditions at the center of the star can be reduced to
\begin{equation}
\xi_r(0) = 0,
\end{equation}
\begin{equation}
\partial_r\xi_r(0) = 0.
\end{equation}
Furthermore, as we consider here a convective core, we translate the aforementioned conditions by means of the function \(X = \rho_0r^2\xi_r\). We then obtain \(X(0) = X'(0) = 0\). If we assume that \(X_1\) is the solution regular at the center, then \(C_2=0\), i.e. \(\mathcal{T}_0 = \mathcal{T}_2 = 0\). Therefore, to keep a non zero solution we need to impose \(X_1(0) = 0\). Since \(r_\text{conv} = 0\) we have also \(\mathcal{T}_1 = 0\). This way, we can now express \(K_0\) as 
\begin{equation}
\begin{split}
K_0&=-\frac{3 \Gamma\left(\displaystyle\frac{4}{3}\right)}{2\sqrt\pi}\left(\frac{v}{3}\right)^{-\frac{1}{3}}\rho_0^\frac{1}{2}(r_\text{int})\mathcal{F},
\end{split}
\end{equation}
where 
\begin{equation}\label{eqn:Fmassive}
\mathcal{F} = \int_0^{r_\text{int}}{\left[\left(\frac{r^2\varphi_T}{g_0}\right)'' - \frac{l(l+1)}{r^2}\left(\frac{r^2\varphi_T}{g_0}\right)\right]\frac{X_1}{X_1(r_\text{int})}}dr 
\end{equation}
can be easily linked to the \(H_l\) parameter introduced by \citet{zahn70,zahn75}.
To compare with the results of \citet{zahn75}, we proceed as follows. From the expression of \(v\) in Eq. \eqref{eqn:v} we have
\begin{equation}
\begin{split}
K_0 &= -\left(\frac{l(l+1)}{\omega^2}\right)^\frac{1}{4}\frac{\sqrt 3 \Gamma\left(\displaystyle\frac{4}{3}\right)}{2\sqrt\pi}\left(\frac{v}{3}\right)^{-\frac{5}{6}}\rho_{0,c}^\frac{1}{2} \left(\frac{1}{r_\text{int}^2}\left|\frac{dN^2}{dr}\right|_\text{int}\right)^\frac{1}{4}\mathcal{F}\\
&=-\left(\frac{l(l+1)}{\omega^2}\right)^\frac{1}{4}K_{0,\text{Z75}},
\end{split}
\end{equation}
where \(K_{0,\text{Z75}}\) is the amplitude of the WKBJ solution as derived in \citet{zahn75}. Furthermore, in the present paper, the WKBJ solution is written as \(\displaystyle \xi_r = -\frac{\varphi_T}{g_0}+C_{W}\ \rho_0^{-\frac{1}{2}}r^{-2}\frac{1}{\sqrt{k_r}}e^{-i(\tau_W-\tau_0)}\),
while in \citet{zahn75} the following formulation is used:
\begin{equation}
\xi_r = -\frac{\varphi_T}{g_0}-C_\text{Z75}\ \rho_0^{-\frac{1}{2}}r^{-2}\left(\frac{N^2}{r^2}\right)^{-\frac{1}{4}}e^{-i(\tau_W-\tau_0)}.
\end{equation}
Therefore, in order to keep consistent formulations, we have
\begin{equation}
C_{W} =-\left(\frac{l(l+1)}{\omega^2}\right)^\frac{1}{4} C_{\text{Z75}}
\end{equation}
This change of convention then explains the different expressions obtained for the \(K_0\) term. From Eq. \eqref{eqn:energylum} the energy luminosity then becomes:
\begin{equation}
L_E = \frac{1}{2}\frac{\omega^2}{\sqrt{l(l+1)}}K_{0,\text{Z75}}^2,
\end{equation}
which is identical to the expression derived by \citet{zahn75}. From the expression of \(K_0\), the energy luminosity becomes
\begin{equation}
\begin{split}
L_E =-\frac{3^\frac{2}{3}\Gamma^2\left(\frac{1}{3}\right)}{8\pi}\omega^\frac{11}{3}\left[l(l+1)\right]^{-\frac{4}{3}}\rho_0(r_\text{int})r_\text{int}\left|\frac{dN^2}{d\ln r}\right|_{r_\text{int}}^{-\frac{1}{3}}\mathcal{F}^2,
\end{split}
\end{equation}
which is similar to the \citet{goodman} formulation in the case of low-mass stars. The tidal torque and the corresponding tidal dissipation can then be computed as follows:
\begin{equation}
|T| = \frac{|m|}{\omega}\frac{3^\frac{2}{3}\Gamma^2\left(\frac{1}{3}\right)}{8\pi}\omega^\frac{11}{3}\left[l(l+1)\right]^{-\frac{4}{3}}\rho_0(r_\text{int})r_\text{int}\left|\frac{dN^2}{d\ln r}\right|_{r_\text{int}}^{-\frac{1}{3}}\mathcal{F}^2,
\end{equation}
\begin{equation}
\begin{split}
|\Im (k_2)| = \frac{3^{-\frac{1}{3}}\Gamma^2\left(\frac{1}{3}\right)}{4\pi}|m|&\left[l(l+1)\right]^{-\frac{4}{3}}\times\\
&\frac{\omega^\frac{8}{3}}{n^4}\frac{GM_\star^2}{m_p^2R_\star^5}\rho_0(r_\text{int})r_\text{int}\left|\frac{dN^2}{d\ln r}\right|_{r_\text{int}}^{-\frac{1}{3}}\!\mathcal{F}^2.
\end{split}
\end{equation}

\subsection{Dynamical tide in low-mass stars}
\subsubsection{General computation of the tidal dissipation}
In the case of a solar-type star, we consider a bi-layer structure with a radiative core and a convective envelope. The energy transported by the wave will therefore propagate inward, towards the center. This gives in our formulation (see Table \ref{tab:values}) \(r_\text{conv} = R_\star,\ r_\text{rad} = 0\) and \(\epsilon = 1\). We assume that the density at the stellar surface vanishes. Such an hypothesis is valid for a polytropic model and allows us to fulfill automatically a stress-free boundary condition, according to which the Lagrangian pressure perturbation vanishes. Considering the same surface condition for a non-zero density will marginally affect our results. Indeed, a photospheric density slightly modifies the conditions of excitation of tidal gravity waves (for more details, we refer the reader to Appendix C). Boundary conditions at the stellar surface will be extensively investigated in a future work. For a zero density at the stellar surface we obtain
\begin{equation}
X(R_\star) = \rho_0(R_\star)R_\star^2\  \xi_r(R_\star) = 0.
\end{equation}
Furthermore, we choose a basis solution in the convective zone \(X_1\) such as \(X_1(R_\star) = 0\). Then, since \(X(R_\star) = C_1 X_1(R_\star)+C_2 X_2 (R_\star)\), we need \(C_2 = 0\) to fulfill the condition at the stellar surface, which leads to \(\mathcal{T}_0 = \mathcal{T}_2 = 0\). Furthermore, if we consider stellar matter as a polytrope near the stellar surface, we can define \(\theta,K,n\) as
\begin{equation}
P_0 = K\rho_0^{1+\frac{1}{n}},
\end{equation}
\begin{equation}
\rho_0 = \rho_c\theta^n,
\end{equation}
where \(\rho_c\) is the density at the center \citep{chandra,kippenhahn}. By defining a dimensionless radius \(\xi\), the stellar mass and the stellar radius can be defined from the solutions of the Lane-Emden equation as
\begin{equation}
R_\star = \left[\frac{(n+1)K\rho_c^{\frac{1}{n}-1}}{4\pi G}\right]^\frac{1}{2}\xi_1,
\end{equation}
\begin{equation}
M_\star = -4\pi\left[\frac{(n+1)K\rho_c^{\frac{1}{n}-1}}{4\pi G}\right]^\frac{3}{2}\rho_c^{\frac{3-n}{2n}}\xi_1^2\left(\frac{d\theta}{d\xi}\right)_{\xi_1},
\end{equation}
where \(\xi_1\) is the first zero of \(\theta (\xi)\). Since the mass and radius take finite values, we obtain
\begin{equation}
\frac{\rho_0'}{\rho_0} =\frac{n}{\theta}\frac{d\theta}{d\xi} \underset{r \to R_\star}\to -\infty.
\end{equation}
Therefore, near the surface, \(X_1\) is a solution of the following equation:
\begin{equation}
X''-\frac{\rho_0'}{\rho_0}X'=0,
\end{equation}
which leads to \(X_1' \propto \rho_0\). Therefore we obtain \(X_1'(R_\star) = 0\). From those considerations we now have \(\mathcal{T}_1 = 0\) and
\begin{equation}
K_0 =-\frac{3 \Gamma\left(\displaystyle\frac{4}{3}\right)}{2\sqrt\pi}\left(\frac{v}{3}\right)^{-\frac{1}{3}}\rho_0^\frac{1}{2}(r_\text{int})\mathcal{F},
\end{equation}
with
\begin{equation}\label{eqn:Flowmass}
\mathcal{F} = \int_{r_\text{int}}^{R_\star}{\left[\left(\frac{r^2\varphi_T}{g_0}\right)'' - \frac{l(l+1)}{r^2}\left(\frac{r^2\varphi_T}{g_0}\right)\right]\frac{X_1}{X_1(r_\text{int})}}dr.
\end{equation}
The energy luminosity then becomes
\begin{equation}
\begin{split}
L_E =-\frac{3^\frac{2}{3}\Gamma^2\left(\displaystyle\frac{1}{3}\right)}{8\pi}\omega^\frac{11}{3}\left[l(l+1)\right]^{-\frac{4}{3}}\rho_0(r_\text{int})r_\text{int}\left|\frac{dN^2}{d\ln r}\right|_{r_\text{int}}^{-\frac{1}{3}}\mathcal{F}^2.
\end{split}
\end{equation}
Such an expression is equivalent to the \citet{goodman} prescription if
\begin{equation}
\left|\partial_r \xi_r^\text{dyn}\right|_{r_\text{int}} = r_\text{int}^{-2}\mathcal{F},
\end{equation}
where \(\xi_r^\text{dyn}\) is the radial displacement linked to the dynamical tide. One can make sure that such a condition is fulfilled from straightforward calculations by expressing the radial displacement \(\xi_r\) in the \((S_+, S_-)\) basis, knowing that \(C_2 = \mathcal{T}_1 = 0\) (we refer the reader to Appendix B for more details). The tidal torque \(T\) can be computed as
\begin{equation}
|T| = \frac{|m|}{\omega}\frac{3^\frac{2}{3}\Gamma^2\left(\displaystyle\frac{1}{3}\right)}{8\pi}\omega^\frac{11}{3}\left[l(l+1)\right]^{-\frac{4}{3}}\rho_0(r_\text{int})r_\text{int}\left|\frac{dN^2}{d\ln r}\right|_{r_\text{int}}^{-\frac{1}{3}}\mathcal{F}^2.
\end{equation}
By comparing such an expression to the \citet{murray} formulation in Eq. \eqref{eqn:murray}, one can assess the tidal dissipation as
\begin{equation}
\begin{split}
|\Im (k_2)| = \frac{3^{-\frac{1}{3}}\Gamma^2\left(\displaystyle\frac{1}{3}\right)}{4\pi}|m|&\left[l(l+1)\right]^{-\frac{4}{3}}\times\\
&\frac{\omega^\frac{8}{3}}{n^4}\frac{GM_\star^2}{m_p^2R_\star^5}\rho_0(r_\text{int})r_\text{int}\left|\frac{dN^2}{d\ln r}\right|_{r_\text{int}}^{-\frac{1}{3}}\!\mathcal{F}^2.
\end{split}
\end{equation}

\subsubsection{Alternative formulation of the forcing term}
We now aim to simplify the forcing term \(\mathcal{F}\) in the case of a solar-type star for a better understanding of this contribution to tidal dissipation. The behavior of the tidal potential is ruled by the Poisson equation
\begin{equation}
\partial_r\left(r^2\partial_r\varphi_T\right) = l(l+1)\varphi_T.
\end{equation}
The gravitational acceleration can be computed from the mean density \(\bar{\rho}\) inside a sphere of radius \(r\) through Gauss's law
\begin{equation}
g_0 = \frac{4}{3}\pi G \bar\rho r,
\end{equation}
with \(\bar \rho = \frac{3}{4\pi r^3}\int_0^r{4\pi r_1^2}\rho(r_1)dr_1\). From those expressions we obtain
\begin{equation}
\begin{cases}
\partial_r\bar\rho = \frac{3}{r}\left(\rho(r)-\bar\rho\right)\\
\partial_r g_0 = \frac{g}{r}+4\pi G\left(\rho(r)-\bar\rho\right).
\end{cases}
\end{equation}
If we define \(Y=\frac{r^2\varphi_T}{g_0}\) we have
\begin{equation}
Y'=\left[1+3\left(1-\frac{\rho}{\bar\rho}\right)\right]\frac{Y}{r}+\frac{r^2\varphi_T}{g_0}
\end{equation}
which leads, by assuming slow variations of the stellar density, to \citep{zahn70,zahn75}:
\begin{equation}
Y''-6\left(1-\frac{\rho}{\bar\rho}\right)\frac{Y'}{r}-\left[l(l+1)-12\left(1-\frac{\rho}{\bar\rho}\right)\right]\frac{Y}{r^2}=0.
\end{equation}
The forcing term \(\mathcal{F}\) therefore becomes \citep{kushnir}:
\begin{equation}\label{eqn:Finhom}
\mathcal{F} =\int_{r_\text{int}}^{R_\star}{6\left(1-\frac{\rho}{\bar\rho}\right)\left[\frac{1}{r}\left(\frac{r^2\varphi_T}{g_0}\right)'-\frac{2}{r^2}\left(\frac{r^2\varphi_T}{g_0}\right)\right]\frac{X_1}{X_1(r_\text{int})}}dr
\end{equation}

\subsubsection{Case of a simplified bi-layer structure for a solar-type star}

We assume in this section a bi-layer structure for a given solar-type star, for which both the radiative core and the convective envelope are assumed to be homogeneous with respective constant densities \(\rho_r\) and \(\rho_c \ll \rho_r\). We also assume that the depth of the convective layer is small compared to the stellar radius. We introduce the aspect ratios
\begin{equation}
\alpha = \frac{R_r}{R_\star},
\end{equation}
\begin{equation}
\beta = \frac{M_r}{M_\star},
\end{equation}
\begin{equation}
\gamma = \frac{\rho_c}{\rho_r} = \frac{\alpha^3(1-\beta)}{\beta(1-\alpha^3)},
\end{equation}
where \(R_r\) is the radius of the radiative zone and \(M_r\) its mass. With this configuration we have
\begin{equation}
\bar{\rho} \approx \rho_r = \frac{3M_\star}{4\pi R_\star^3}\frac{\beta}{\alpha^3},
\end{equation}
\begin{equation}
\begin{split}
g_0 = \frac{\beta}{\alpha^3} \omega_\text{dyn}^2 r,
\end{split}
\end{equation}
with \(\omega_\text{dyn}^2 = \frac{GM_\star}{R_\star^3}\). Therefore, the forcing term \(\mathcal{F}\), in its alternative form, can be expressed as
\begin{equation}
\begin{split}
\mathcal{F} = 6\left(1-\gamma\right)\!\int_{r_\text{int}}^{R_\star}{\left[\frac{1}{r}\left(\frac{r^2\varphi_T}{g_0}\right)'-\frac{2}{r^2}\left(\frac{r^2\varphi_T}{g_0}\right)\right]\frac{X_1}{X_1(r_\text{int})}}dr.\\
\end{split}
\end{equation}
Furthermore, in the case of a solar-type star, we showed that
\begin{equation}
X_1'(r) \propto \rho_0(r)\text{, with } 1-\frac{r}{R_\star}\ll 1.
\end{equation}
From this we obtain
\begin{equation}
\frac{X_1(r)}{X_1(r_\text{int})} = \frac{\rho_c (R_\star - r)}{\rho_c (R_\star - r_\text{int})} = \frac{1}{1-\alpha}\left(1-\frac{r}{R_\star}\right).
\end{equation}
Then, the forcing term becomes
\begin{equation}
\mathcal{F} = 6\frac{1-\gamma}{1-\alpha}\int_{r_\text{int}}^{R_\star}{\left[\frac{1}{r}\left(\frac{r^2\varphi_T}{g}\right)'-\frac{2}{r^2}\left(\frac{r^2\varphi_T}{g}\right)\right]\left(1-\frac{r}{R_\star}\right)}\ dr.
\end{equation}
By introducing the quantity \(\Psi = \frac{\varphi_T}{r^2}\), independent of the stellar radius,  we can write \(Y = \frac{r^2\varphi_T}{g}\) as
\begin{equation}
Y = \frac{\alpha^3}{\beta}\frac{\Psi}{\omega_\text{dyn}^2}r^3.
\end{equation}
From this we obtain
\begin{equation}\label{eqn:Fscaling}
\mathcal{F} =3\frac{1-\gamma}{1-\alpha}\frac{\alpha^5}{\beta}\left(\frac{2\alpha}{3}-1\right) R_\star^2\frac{\Psi}{\omega_\text{dyn}^2}.
\end{equation}
By comparing with the \citet{goodman} prescription, we obtain
\begin{equation}
\partial_r \xi_r^\text{dyn} (r_\text{int}) = 3\frac{1-\gamma}{1-\alpha}\frac{\alpha^3}{\beta}\left(\frac{2\alpha}{3}-1\right) \frac{\Psi}{\omega_\text{dyn}^2}.
\end{equation}
Such a prescription justifies analytically the parametrizations of \citet{goodman} and \citet{barker10}. This leads to the following tidal dissipation:
\begin{equation}
\begin{split}
\frac{3}{2Q'} = \frac{3^{\frac{8}{3}}\Gamma^2\left(\displaystyle\frac{1}{3}\right)}{16\pi^2}|m| &\left[l(l+1)\right]^{-\frac{4}{3}}\times\\
&\frac{\omega^\frac{8}{3}}{n^4}\frac{M_\star R_\star^3}{G}\mathcal{E}(\alpha,\beta)\left|\frac{dN^2}{d\ln r}\right|_{r_\text{int}}^{-\frac{1}{3}}\frac{\Psi^2}{m_p^2},
\end{split}
\end{equation}
where \(\displaystyle \mathcal{E}(\alpha,\beta) = \frac{\alpha^{11}(1-\beta)(1-\gamma)^2}{\beta^2(1-\alpha^3)(1-\alpha)^2}\left(\frac{2\alpha}{3}-1\right)^2\).
If we consider the amplitude of the largest tide for a coplanar and circular orbit (Ogilvie 2004; Barker \& Ogilvie 2010; Ogilvie 2014):
\begin{equation}\label{eqn:psi22}
\Psi = \sqrt{\frac{6\pi}{5}}\frac{m_p}{M_\star}n^2,
\end{equation}
we obtain a simplified expression for the tidal dissipation:
\begin{equation}
\frac{3}{2Q'} = \frac{3^{\frac{11}{3}}\Gamma^2\left(\displaystyle\frac{1}{3}\right)}{40\pi}|m|\left[l(l+1)\right]^{-\frac{4}{3}}\!\omega^\frac{8}{3}\frac{R_\star^3}{GM_\star}\left|\frac{dN^2}{d\ln r}\right|_{r_\text{int}}^{-\frac{1}{3}}\!\!\mathcal{E}(\alpha,\beta).
\end{equation}

\subsection{Tidal dissipation for a tri-layer structure}
In main-sequence F-type stars, a convective core is present in addition to the envelope \citep{kippenhahn}. Indeed, the dominating CNO cycle, extremely sensitive to the temperature, concentrates the energy production at low radii. Therefore, in this configuration, convection is the most efficient process in the central region to transport energy towards the center. A convective core also exists in evolved stars, for which the dissipation of g-modes is important \citep{schlaufman,essick,weinberg}. Thus, more complex stellar structures may have to be investigated with our formalism. 

As tidally excited modes are observed in this type of stars for example in heartbeat stars  (we refer the reader to \citealt{fuller17} for an extensive study), the formation of standing $g$-modes is generally not prevented. For instance, following \citet{barker10}, wave braking may happen in the central part of the star if the companion is massive enough (we refer the reader to \citealt{barker20} and Appendix D for a derivation of this criterion). However, such a configuration is more likely to happen for stars older than 2 Gyr \citep{barker20}, for which the convective core has already disappeared. The only remaining possibility to prevent the formation of a standing mode is a potential interaction with a critical layer \citep{alvan13}. Indeed, at the beginning of the MS, the rotation period of the stellar radiative zone can be comparable to the orbital period of the companion and short enough to ensure efficient tidal interaction \citep[e.g.][]{galletbouvier15,amard19}. We present the general formalism for this potential (rare) event. In any case, our prescription may not provide an upper bound of tidal dissipation anymore for this type of stars. Indeed, tidal gravity modes may be resonantly excited and can produce more dissipation than progressive waves.

 \begin{figure}[!h]
        \centering
        \includegraphics[scale=0.25]{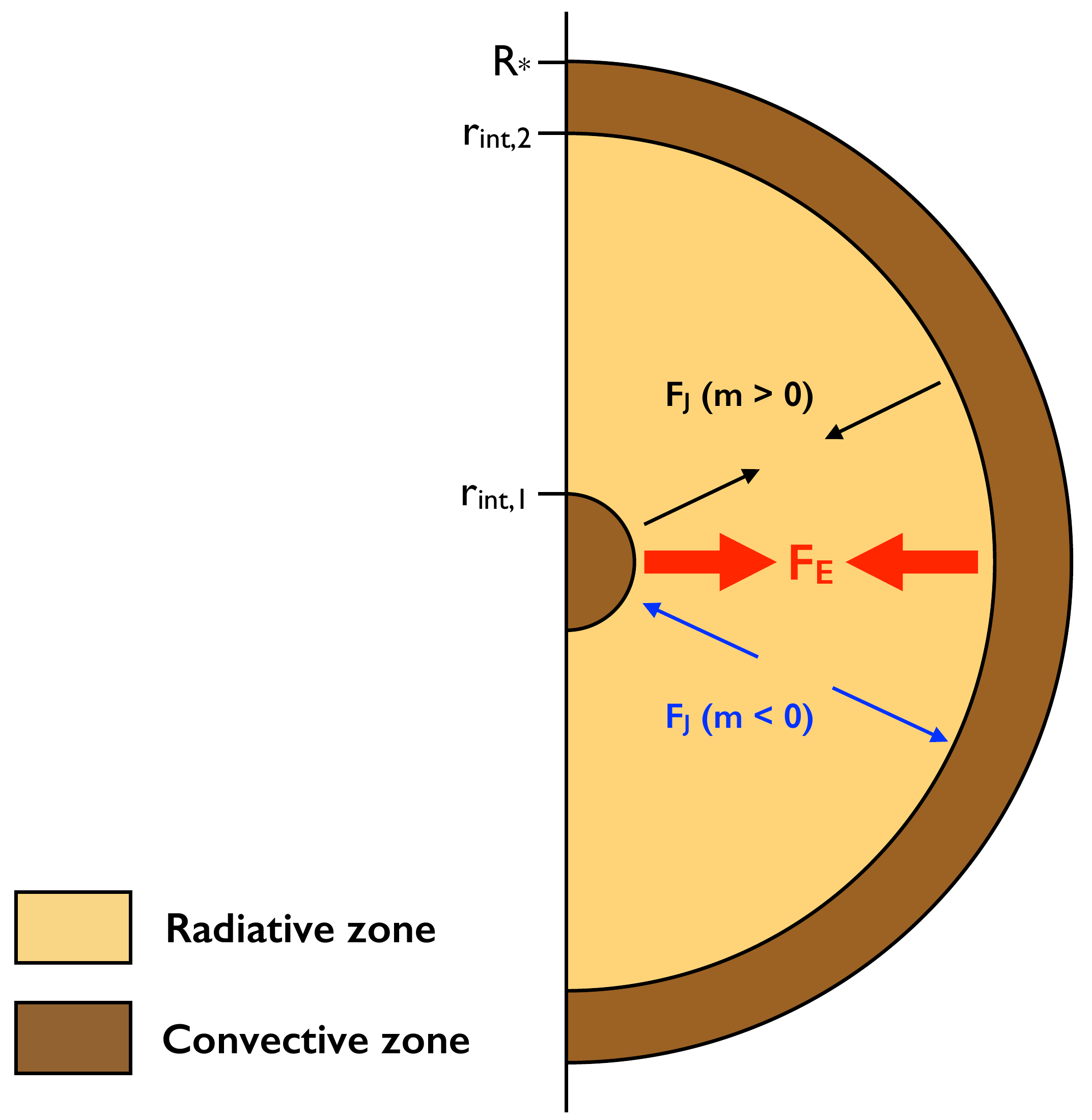}
  \caption{\label{Fstruct} Configurations of the radiative and convective spherical shells for the case of a tri-layer structure. In brown: convective layer. In orange: radiative layer. The red arrows represent the energy fluxes ($F_E$), the black arrows correspond to the angular momentum fluxes ($F_J$) for prograde waves ($m>0$), and the blue arrows stand for the angular momentum fluxes in the case of retrograde waves ($m<0$).}
\end{figure}

As shown in Fig. \ref{Fstruct}, we consider a tri-layer structure with a convective core delimited by the interface located at \(r=r_\text{int,1}\), an intermediate radiative layer and a convective envelope between \(r=r_\text{int,2}\) and \(r=R_\star\). Therefore, inside the radiative zone, tidal gravity wave will propagate inwards and outwards from the interfaces. We assume that all the energy carried by a gravity wave from an interface is dissipated in the radiative zone before reaching the other interface. This way, the outward and inward gravity waves are decoupled. The corresponding values of \(r_\text{int},\ r_\text{rad},\ r_\text{conv}\) and \(\epsilon\) in the case of an outward and an inward energy transport are presented in Table \ref{tab:valuesF}.\\

\begin{table}[!h]
\centering 
      \caption{\label{tab:valuesF} Values of \(r_\text{int},\ r_\text{rad},\ r_\text{conv}\) and \(\epsilon\) in the case of a tri-layer structure.}
      \begin{tabu}{ccccc}
            \hline
            \noalign{\smallskip}
            \text{Configuration} &$r_\text{int}$&$r_\text{rad}$&$r_\text{conv}$&$\epsilon$ \\
            \noalign{\smallskip}
            \hline
            \noalign{\smallskip}
            Inward energy transport &$r_\text{int,2}$& $r_\text{int,1}$&$R_\star$&1\\
            Outward energy transport &$r_\text{int,1}$&$r_\text{int,2}$&0&-1\\
            \noalign{\smallskip}
            \hline
         \end{tabu}
   \end{table}
Far from the interfaces in the radiative zone, the radial displacement of the inward and outward gravity waves can be written in the following WKBJ form
\begin{equation}
\xi_{r,\text{in}} = C_{W,\text{in}}\ \rho_0^{-\frac{1}{2}}r^{-2}{k_r}^{-\frac{1}{2}}\exp\left[i(\tau_W-\tau_0)\right],
\end{equation}
\begin{equation}
\xi_{r,\text{out}} = C_{W,\text{out}}\ \rho_0^{-\frac{1}{2}}r^{-2}{k_r}^{-\frac{1}{2}}\exp\left[-i(\tau_W-\tau_0)\right].
\end{equation}
Each contribution transports angular momentum within the radiative zone, quantified by the corresponding luminosity
\begin{equation}
L_{J,\text{in/out}} =-\epsilon\frac{m}{2}\frac{\omega^2}{l(l+1)}|C_{W,\text{in/out}}|^2.
\end{equation}

As we consider the net torque applied on the radiative zone as a whole, we only focus on the transport of angular momentum at the convective-radiative interfaces (we refer the reader to Appendix E for more details). Since both inward and outward prograde (retrograde) gravity waves transport energy towards the inside (outside) of the radiative layer, the two contributions act in concert and lead to the same change in the rotation of the radiative zone. The total tidal torque then leads to the following tidal dissipation:
\begin{equation}
\begin{split}
&|\Im (k_2)| = \frac{3^{-\frac{1}{3}}\Gamma^2\left(\frac{1}{3}\right)}{4\pi}m \left[l(l+1)\right]^{-\frac{4}{3}}\frac{\omega^\frac{8}{3}}{n^4}\frac{GM_\star^2}{m_p^2R_\star^5}\times\\
&\left(\rho_0(r_\text{int,1})r_\text{int,1}\left|\frac{dN^2}{d\ln r}\right|_{r_\text{int,1}}^{-\frac{1}{3}}\!\!\!\mathcal{F}_\text{out}^2+\rho_0(r_\text{int,2})r_\text{int,2}\left|\frac{dN^2}{d\ln r}\right|_{r_\text{int,2}}^{-\frac{1}{3}}\!\!\!\mathcal{F}_\text{in}^2\right),
\end{split}
\end{equation}
with
\begin{equation}
\begin{split}
\mathcal{F}_\text{out} &=\int_0^{r_\text{int,1}}{\left[\left(\frac{r^2\varphi_T}{g_0}\right)'' - \frac{l(l+1)}{r^2}\left(\frac{r^2\varphi_T}{g_0}\right)\right]\frac{X_{1,\text{out}}}{X_{1,\text{out}}(r_\text{int})}}dr,\\
\mathcal{F}_\text{in} &=\int_{r_\text{int,2}}^{R_\star}{\left[\left(\frac{r^2\varphi_T}{g_0}\right)'' - \frac{l(l+1)}{r^2}\left(\frac{r^2\varphi_T}{g_0}\right)\right]\frac{X_{1,\text{in}}}{X_{1,\text{in}}(r_\text{int})}}dr.
\end{split}
\end{equation}
From the boundary conditions we considered in \S 4 and 5.1, the functions \(X_{1,\text{in}}\) and \(X_{1,\text{out}}\) are solutions of the following Cauchy problems:
\begin{equation}
\begin{cases}
\displaystyle X_{1,\text{out}}''-\frac{\partial_r\rho_0}{\rho_0}X_{1,\text{out}}'-\frac{l(l+1)}{r^2}X_{1,\text{out}}= 0\\
X_{1,\text{out}}(0) = X'_{1,\text{out}}(0) = 0,\\
\end{cases}
\end{equation}
\begin{equation}
\begin{cases}
\displaystyle X_{1,\text{in}}''-\frac{\partial_r\rho_0}{\rho_0}X_{1,\text{in}}'-\frac{l(l+1)}{r^2}X_{1,\text{in}}= 0\\
X_{1,\text{in}}(R_\star) = X'_{1,\text{in}}(R_\star) = 0.\\
\end{cases}
\end{equation}
\begin{figure*}[h]
    \begin{center}
    \includegraphics[scale=0.37]{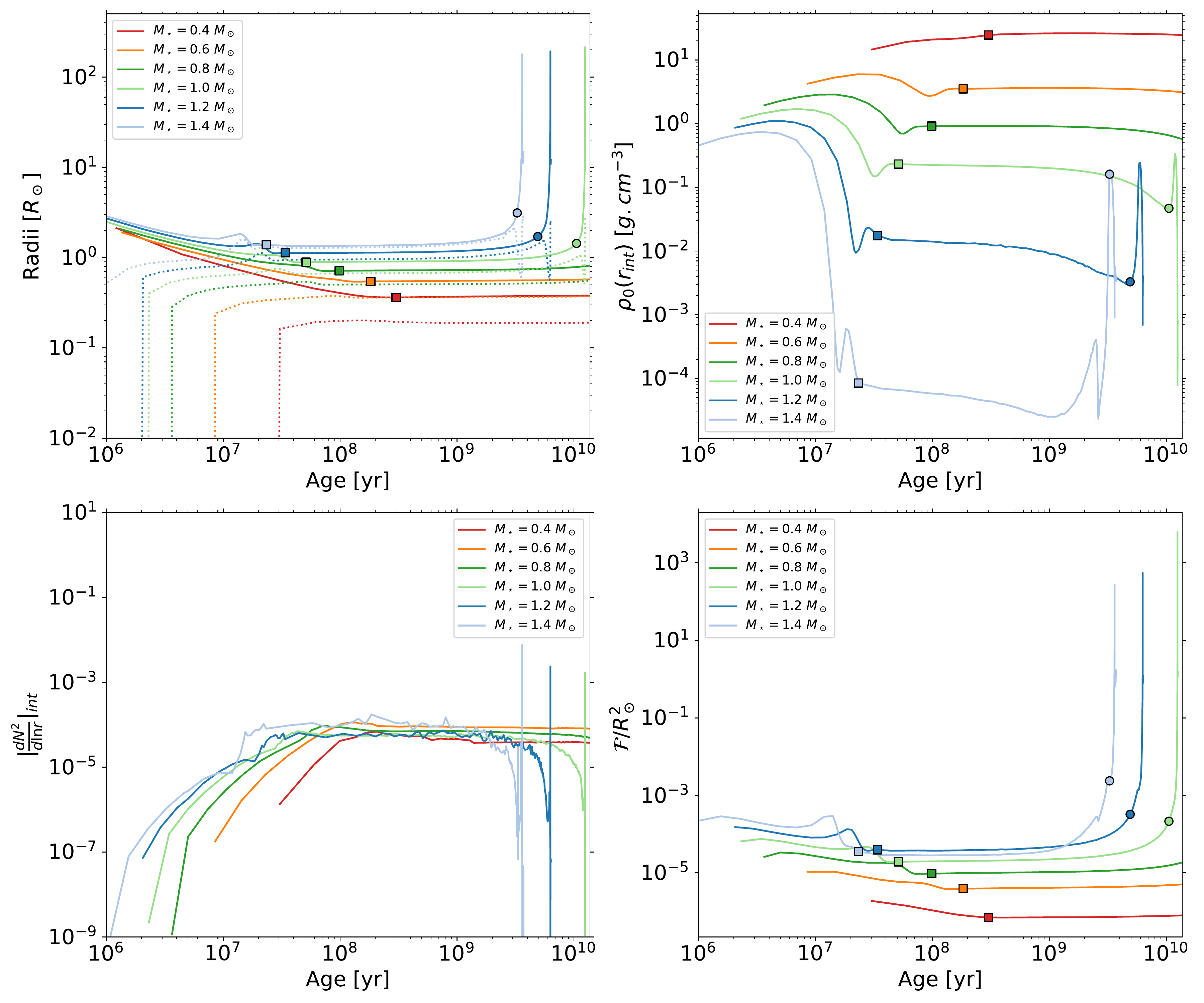}
    \end{center}
\caption{\label{diagnostic} Evolution of the stellar radius (top-left, plain line), the radius of the radiative zone (top-left, dotted line), the density at the interface core-envelope (top-right), the gradient of the square of the Brunt-Väisälä frequency at the interface (botton-left) and the forcing term \(\mathcal{F}\) (bottom-right) as a function of the age of the system, for stellar masses (\(M_\star\)) between 0.4 and 1.4 \(M_\odot\), with \(P_\text{orb} = 1\ \text{d}\), typical value for a hot Jupiter. The colored squares correspond to the ZAMS in each model, and the colored circles to the TAMS.}
\end{figure*}

\section{Variation of tidal dissipation of low-mass stars along their evolution}
\subsection{Physical insight and the \citet{goldreich} approach}
We assumed that the totality of the energy flux carried by internal gravity waves is dissipated before the waves undergo any reflection. Therefore tidal dissipation is directly linked to the efficiency of the excitation of the waves. Furthermore, in the radiative zone, near the interface, the forced oscillations follow the inhomogeneous Airy equation
\begin{equation}
\begin{split}
\frac{d^2\psi}{d\eta^2}+v^2\eta\psi = v^2\eta Z,
\end{split}
\end{equation}
with
\begin{equation}
v^2 =\frac{l(l+1)}{r_\text{int}^2\omega^2}\left|\frac{dN^2}{dr}\right|_{r_\text{int}}
\end{equation}
and
\begin{equation}
\eta =r_\text{int}-r.
\end{equation}
One can therefore introduce a characteristic length \(\lambda\) of variation of the gravity waves in the radial direction \citep{goodman,kushnir}, defined as
\begin{equation}
\lambda = v^{-\frac{2}{3}} = \omega^\frac{2}{3}\left[l(l+1)\right]^{-\frac{1}{3}}\left|\frac{dN^2}{d\ln  r}\right|^{-\frac{1}{3}}_{r_\text{int}}r_\text{int}.
\end{equation}

Then, by following \citet{goldreich}, for a given forcing frequency \(\omega\), the group velocity \(v_g\) of the gravity waves can be estimated as \(v_g \sim \lambda\omega\). Furthermore, since gravity waves are excited by the equilibrium tide, the characteristic velocity \(u\) of the fluid can be assessed as \(u \sim \xi_\text{eq,int}\omega\), where \(\xi_\text{eq,int} \sim -\frac{\varphi_T(r_\text{int})}{g_\text{0,int}}\) \citep{zahn75} is the radial displacement linked to the equilibrium tide estimated at the interface. The energy luminosity \(L_E\) can therefore be expressed as
\begin{equation}
\begin{split}
L_E &\sim \rho(r_\text{int})u^2 v_g r_\text{int}^2\\
&=\rho(r_\text{int})\lambda\omega^3r_\text{int}^2\left(\frac{\varphi_T(r_\text{int})}{g_\text{0,int}}\right)^2.\\
\end{split}
\end{equation}
By introducing \(\varphi_T(r_\text{int}) = r_\text{int}^2 \Psi\) and \(\omega^2_\text{dyn,int} = g_\text{0,int}/r_\text{int}\) we obtain
\begin{equation}
L_E \sim \omega^\frac{11}{3}\rho(r_\text{int})r_\text{int}^5\left|\frac{dN^2}{d\ln  r}\right|^{-\frac{1}{3}}_{r_\text{int}}\left(\frac{\Psi}{\omega^2_\text{dyn,int}}\right)^2,
\end{equation}
which is similar to the \citet{goodman} formulation, as already pointed out by \citet{kushnir}, as well as our prescription. The tidal dissipation then becomes
\begin{equation}
\begin{split}
|\Im (k_2)| & \propto \frac{GM_\star^2}{m_p^2R_\star^5}\frac{1}{n^4}\frac{L_E}{\omega}\\
&\propto \frac{\omega^\frac{8}{3}}{n^4}\frac{GM_\star^2}{m_p^2R_\star^5}\rho_0(r_\text{int})r_\text{int}\left|\frac{dN^2}{d\ln r}\right|_{r_\text{int}}^{-\frac{1}{3}} \left(\frac{\Psi}{\omega^2_\text{dyn,int}}\right)^2.
\end{split}
\end{equation}
Such an approach allows us to better understand the influence of the relevant physical parameters on tidal dissipation. Indeed, for a given star-planet system, a higher density at the interface will increase the energy carried by the waves and therefore the tidal dissipation. A similar effect can be obtained for higher values of \(r_\text{int}\) and smoother slopes of the Brunt-Väisälä profile, which enhance the group velocity.

\subsection{Grid of models for low-mass stars}
To study the influence of stellar structure and evolution on tidal dissipation in the radiative zone of low-mass stars, we rely on grids computed with the 1D stellar evolution code STAREVOL \citep[see][for an extensive description]{siess,lagarde,amard19} for masses ranging from 0.4 to 1.4 \(M_\odot\) at a solar metallicity Z = 0.0134 \citep{asplund}. We study all phases from the PMS to the top of the RGB.\\ 

As shown in Fig. \ref{diagnostic} (top-left panel), the star, completely convective at the beginning of the evolution, is in contraction during the PMS. A radiative core grows after the Hayashi phase and a mass transfer occurs from the convective envelope to the radiative zone, leading to an increase then a decrease of the density at the core-envelope interface during the PMS (top-right panel). At the zero-age main sequence (ZAMS), the stellar radius, the size of the radiative core, and the density at the core-envelope interface reach a value which remains almost constant during the MS. There, the thinner convective envelope of more massive stars lead to smaller densities at the interface. During the Sub-Giant phase, because of the expansion of the envelope and the contraction of the core, low-mass stars are characterized by larger radii, thicker convective layers, and larger densities at the interface compared to their MS counterparts. 
\begin{figure}[!h]
    \centering
    \includegraphics[scale=0.34]{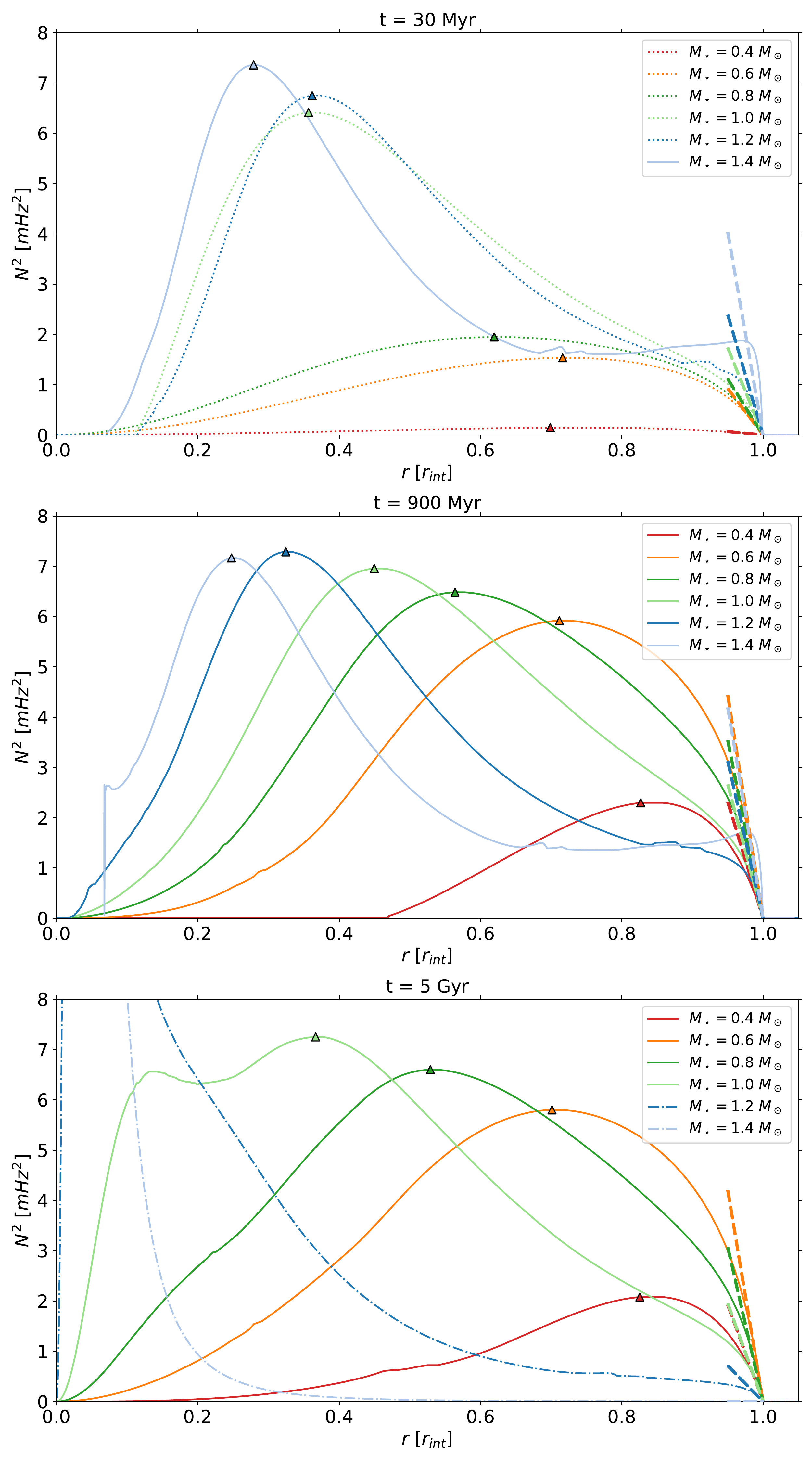}
\caption{\label{Nprofile} Brunt-Väisälä frequency profiles as a function of stellar radius, normalized to the radius of its radiative zone, at 30 Myr (top panel), 900 Myr (middle panel) and 5 Gyr (bottom panel). The dotted profiles are situated in the PMS, the plain lines in the MS, and the dashed-dotted lines in the post-MS phase. The colours correspond to the same stellar masses as in Fig. \ref{diagnostic}. The dashed lines correspond to the tangents to the \(N^2\) profiles at the interface. The triangles indicate the position of the maximal Brunt-Väisälä frequency.}
\end{figure}

A key contribution to assess tidal dissipation in low-mass stars is the forcing term
\begin{equation}
\displaystyle\mathcal{F}=\int_{r_\text{int}}^{R_\star}{6\left(1-\frac{\rho}{\bar\rho}\right)\left[\frac{1}{r}\left(\frac{r^2\varphi_T}{g_0}\right)'-\frac{2}{r^2}\left(\frac{r^2\varphi_T}{g_0}\right)\right]\frac{X_1}{X_1(r_\text{int})}}dr.
\end{equation}
At a given age, such a quantity is computed by relying on the density radial profiles provided by STAREVOL in order to obtain the solution \(X(r)\) in the convective envelope. Below the stellar surface, the density profile is replaced by a polytropic model, whose polytropic index varies with the extent of the convective zone, to ensure a singularity at \(r=R_\star\) (for more details about the numerical implementation of surface boundary conditions, we refer the reader to Eq. \eqref{eqn:boundary} in Appendix C). The differential equations are solved by relying on a fourth-order Runge–Kutta method. 
This way, if we neglect the changes in stellar structure, Eq. \eqref{eqn:Fscaling} leads to \(\mathcal{F} \propto R_\star^5 M_\star^{-1}\), for a given star-planet system (which allows us to keep the tidal potential constant). According to this scaling law, the forcing term presents a time evolution similar to the stellar radius (see the bottom-right panel in Fig. \ref{diagnostic}). Furthermore, if we consider a mass-radius relationship during the MS, the \(\mathcal{F}\) term tends to be higher for more massive stars. However, taking an inhomogeneous distribution of mass in the stellar interior as well as variations in stellar structure into account leads to a more complex behavior of the forcing term \(\mathcal{F}\). Indeed, for \(M_\star = 1.4\ M_\odot\),  \(\mathcal{F}\) reaches higher values than for the \(1.2\ M_\odot\) evolution at the beginning of the PMS and during the SG phase. However, at the end of the PMS and during the MS, a reduced contribution of the forcing term can be seen for \(M_\star = 1.4\ M_\odot\).

\begin{figure*}[!h]
    \begin{center}
    \includegraphics[scale=0.39]{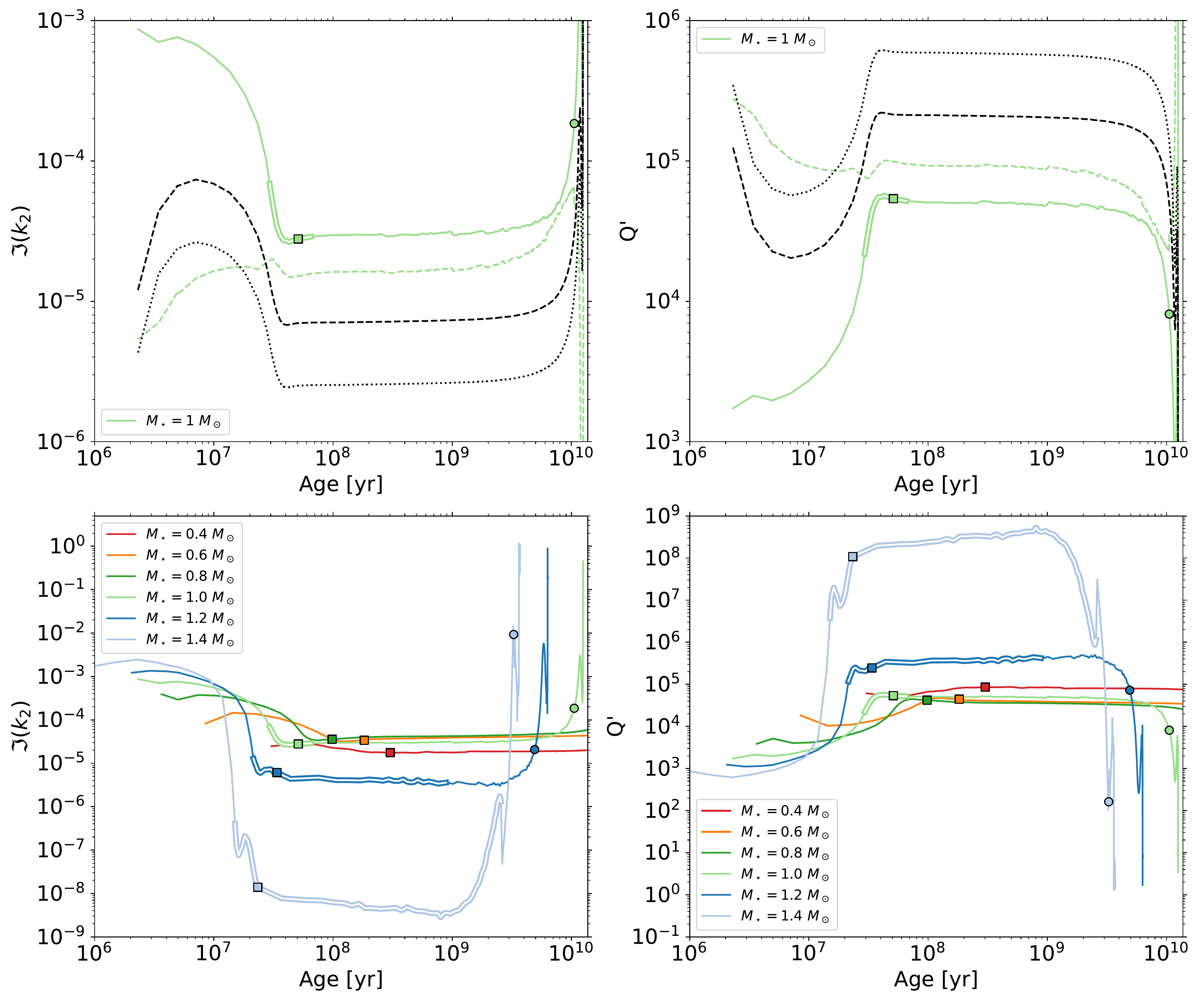}
    \end{center}
\caption{\label{tidaldiss} \textbf{Top-left}: Evolution of the tidal dissipation as a function of time for \(M_\star = 1\ M_\odot\) and \(P_\text{orb} = 1\ \text{d}\). \textbf{Top-right}: Time evolution of the corresponding tidal quality factor. In green: prescription developed in our work. Dashed green: Simplified prescription for a thin convective envelope. Black dashed line: \citet{barker10} prescription. Black dotted line: \citet{goodman} prescription. \textbf{Bottom-left}: Evolution of the tidal dissipation as a function of time for stellar masses (\(M_\star\)) between 0.4 and 1.4 \(M_\odot\), with \(P_\text{orb} = 1\ \text{d}\). \textbf{Bottom-right}: Time evolution of the corresponding tidal quality factor. The colored squares correspond to the ZAMS in each model, and the colored circles to the TAMS. Double lines indicate the presence of a convective core. In such a configuration, wave braking is unlikely to occur.}
\end{figure*}

Despite a noisy profile inherent to the numerical treatment of the stellar interior, we can see on the bottom-left panel in Fig. \ref{diagnostic} that the gradient of the squared Brunt-Väisälä frequency increases during the PMS. It then reaches a stationary value on the MS which does not depend on the stellar mass. This way, its main effect is to enhance tidal dissipation in young systems. Such behavior can be explained by considering the evolution of the Brunt-Väisäla profiles for different stellar masses. As shown in Fig. \ref{Nprofile}, at all ages the Brunt-Väisälä frequency drops at the center as well as at the interface between the convective and the radiative zones. Moreover, during PMS (see the dotted lines in Fig. \ref{Nprofile}), the frequency at a given radius increases due to the contraction of the star, which locally increases the gravity \citep{charbonnel}. This leads to higher maximum values of \(N^2\) during the early evolution Furthermore, since 
\begin{equation}
\displaystyle\left.\frac{dN^2}{d\ln r}\right|_\text{int} = \left.\frac{dN^2}{d\left(\frac{r}{r_\text{int}}\right)}\right|_\text{int},
\end{equation}
it also explains the increase of the gradient of \(N^2\) at the interface during the PMS. During the MS (see the plain lines in Fig. \ref{Nprofile}), as the structure of the star is stabilized, the position of the maximum frequency in the radiative zone approaches the center of the evolving star, marginally changing its value and the general shape of the profile. Thus \(dN^2/d\ln r\) and \(N_\text{max}\) remain approximately constant during this phase. The dependence in stellar mass of these two quantities is then essentially explained by the relative PMS and MS lifetimes of the different types of stars. Ideed, the more massive the star, the faster its PMS evolution, and thus the earlier \(dN^2/d\ln r\) and \(N_\text{max}\) values reach the MS plateau. During the Sub-Giant phase and the RGB (see the dashed-dotted lines in Fig. \ref{Nprofile}), as the maximum value of \(N^2\) increases and gets closer to the center of the star \citep{fuller}, the gradient \(d N^2/d\ln r\) at the convective-radiative interface decreases, which emphasizes tidal dissipation in the radiative zone.

\subsection{Time evolution of tidal dissipation as a function of stellar mass and evolutionary stage}

The evolution of our tidal dissipation prescription (see the green curve in the top panels of Fig. \ref{tidaldiss}), for a planet orbiting a non-rotating star with a period \(P_\text{orb} = 1\ \text{d}\), is similar to the \citet{barker10} (black dashed line in the top panels of Fig. \ref{tidaldiss}) and the \citet{goodman} prescriptions (black dotted line in the top panels of Fig. \ref{tidaldiss}). The discrepancies between those estimates of tidal dissipation come from the parametrization of \(dN^2/d\ln r\) and \(d\xi_r/dr\) at the core-envelope interface, which are sensitive to the hypothesis made in a given stellar evolution model. The simplified version of our prescription, derived in \S 4.2.3 (see the green dashed curve in the top panels of Fig. \ref{tidaldiss}), is only relevant during the MS of higher mass stars, where the thin convective layer approximation remains valid.

As shown in the bottom panels in Fig. \ref{tidaldiss}, tidal dissipation increases at the beginning of the PMS towards its maximum value, due to the formation of a radiative core. Such a maximum increases with stellar mass and is reached earlier for more massive stars, due to their shorter lifetime. Then, the energy transported by IGW is reduced by the decrease of the density at the interface due to mass transfer from the envelope to the core, and to a lesser extent by an increase of the gradient of the Brunt-Väisälä frequency. Their excitation is then inhibited and tidal dissipation decreases to reach an almost constant value during the MS. During this evolutionary stage, G-type and K-type stars (\(M_\star = \{0.6-1\}\ M_\odot\)), present a similar dissipation, as already pointed out by \citet{barker20}. In the case of F-type stars (\(M_\star = \{1.2-1.4\}\ M_\odot\)), tidal dissipation decreases by around four orders of magnitude as the stellar mass increases and the convective envelope becomes thinner. It is important to note that for these stars, the dissipation of progressive waves should not be the dominating contribution to tidal friction in the stellar radiative zone, as standing modes may undergo a more efficient dissipation, enhanced through resonance locking \citep{fuller17}. For M-type stars (here \(M_\star = 0.4\ M_\odot\)), the smaller extension of the radiative core leads to a weaker tidal dissipation. At the end of the MS phase, as the hydrogen is mostly consumed, the core becomes isothermal and contracts. At the same time, the rest of the burning hydrogen migrates to form a shell around the helium core. This leads to an inflation of the convective envelope. Hence, as the density at the interface increases, the tidal dissipation becomes stronger during the Sub-Giant phase and the RGB.

Such an evolution of tidal dissipation as a function of stellar mass and stellar age is expected when $\omega < \omega_c$, when an interaction with a co-rotation layer occurs or when wave braking develops within the radiative core. For this last mechanism, the mass of the companion has to be greater than a critical value $M_\text{cr}$ (we refer the reader to Appendix D, as well as \citealt{barker10,barker11, barker20} for more details). In particular, during the sub-giant phase and the RGB, more massive stars require smaller planetary masses to initiate wave breaking. Hence, super-Earths and hot Neptunes are likely to trigger such a process during these phases of evolution, as the convective envelope of the star is thicker \citep[see Appendix D and][]{barker20}.

\subsection{Dynamical tide in the convective and the radiative zones of low-mass stars: influence of structural and rotational evolution}

\subsubsection{Spin evolution}

A consistent treatment of the tidal dissipation in the convective zone requires to take the evolution of stellar rotation into account. To this end, we rely on STAREVOL evolutionary tracks adapted from \citet{amard19}. We compute stellar models of rotating stars for a range of initial masses between 0.4 and 1.4 \(M_\odot\) at solar metallicity. Star–disk interaction is taken into account at the beginning of the PMS by assuming a constant surface rotation rate during the disk’s lifetime, set by the observations. Over the whole mass range, we selected the fast rotators as calibrated by \citet{galletbouvier15} with a 3 days initial rotation period and a 2.5 Myr disk lifetime. This way, we consider an upper bound of the tidal dissipation through inertial waves.

\begin{figure}[!h]
    \begin{center}
    \includegraphics[scale=0.29]{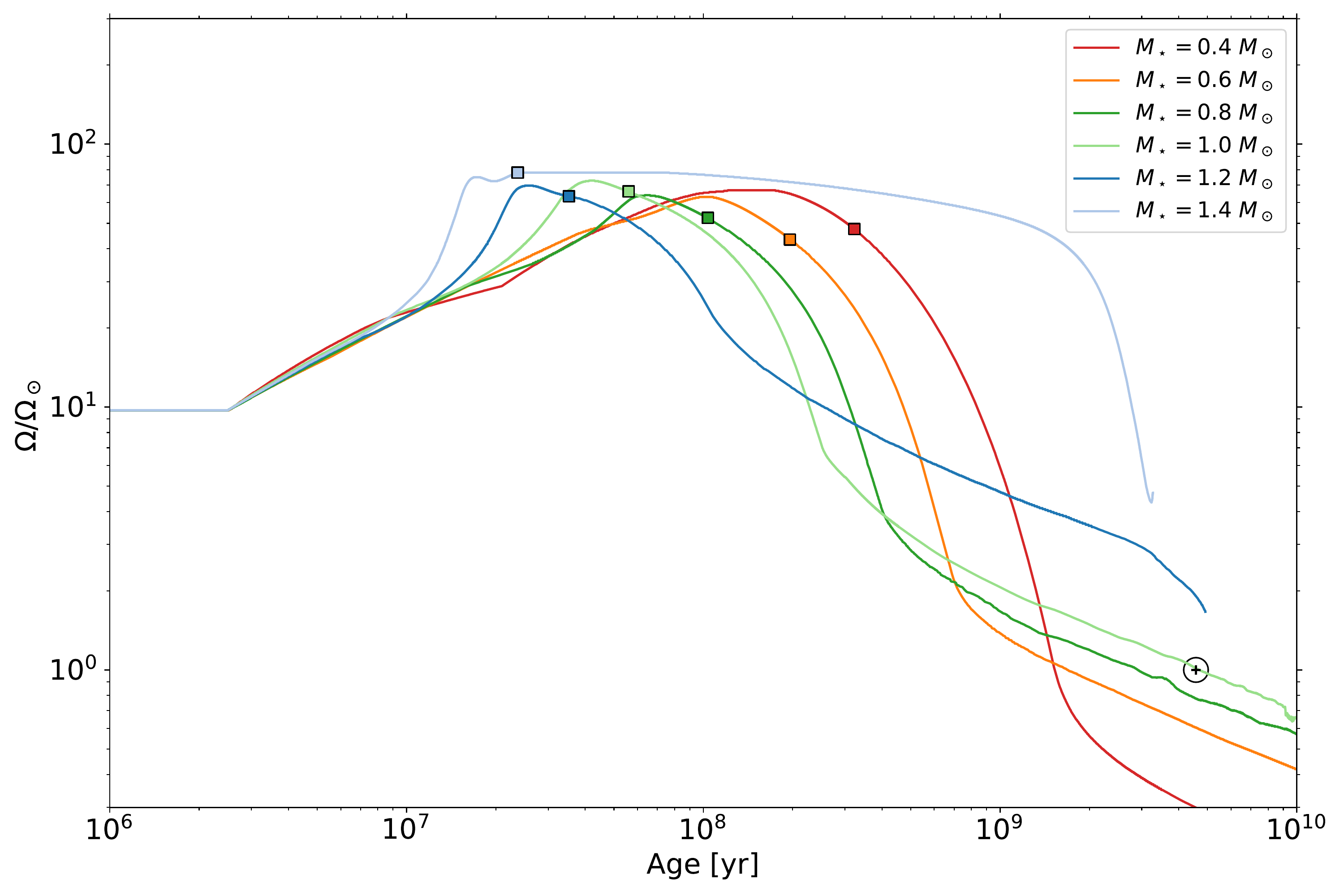}
    \end{center}
\caption{\label{evolrot} Evolution of the surface rotation rate (scaled to the present Sun rotation rate) for stars with a mass ranging from 0.4 to 1.4 solar masses \citep{amard19}. The solar rotation rate at solar age is represented by a black circle. The colored squares correspond to the ZAMS in each model.}
\end{figure}
As seen in Fig. \ref{evolrot}, after the dissipation of the disk, the star spins up during the PMS due to its contraction. Then, during the MS, the magnetized stellar winds carry away angular momentum from the star, leading to its spin-down \citep{skumanich,kawaler,matt15}. As the evolution of the most massive stars is faster than the one of the least massive, the decrease of their rotation rate at the beginning of the main sequence occurs at lower ages. Furthermore, F-stars have a smaller outer convective zone, leading a to less efficient stellar dynamo. Hence, they are less braked during the MS and thus remain fast rotators during most of their life. Therefore, lower-mass stars reach a slower rotation rate at the solar age than their higher-mass counterparts.

\subsubsection{Tidal dissipation in the convective zone}

We now aim to compare the dissipation of the dynamical tide in the radiative and the convective zones of low-mass stars along their evolution. To this end we rely on the formalism of \citet{ogilvie13} and \citet{mathis15} to assess tidal dissipation in the convective zone through inertial waves. In this framework, we assume the same simplified bi-layer structure as in \S 5.3. The stellar convective envelope is assumed to be in solid-body rotation with angular velocity \(\Omega_\star\). Furthermore, centrifugal effect is neglected by assuming moderate rotation i.e. \(\Omega_\star^2/\omega_\text{dyn}^2 \ll 1\). In the case of a coplanar and circular star–planet system, the frequency-averaged tidal dissipation is given by
\begin{equation}
\begin{split}
\left<\Im(k_2)\right>_\text{IW}^\text{CZ} &= \int_{-\infty}^{+\infty}\Im\left(k_2\right)\frac{d\omega}{\omega} = \frac{100\pi}{63}\left(\frac{\Omega_\star}{\omega_\text{dyn}}\right)^2\left(\frac{\alpha^5}{1-\alpha^5}\right)(1-\gamma)^2\\
&\times(1-\alpha)^4\left(1+2\alpha+3\alpha^2+\frac{3}{2}\alpha^3\right)^2\left[1+\left(\frac{1-\gamma}{\gamma}\right)\alpha^3\right]\\
&\times \left[1+\frac{3}{2}\gamma+\frac{5}{2\gamma}\!\left(1+\frac{1}{2}\gamma-\frac{3}{2}\gamma^2\right)\alpha^3-\frac{9}{4}(1-\gamma)\alpha^5\right]^{-2}\!\!\!\!\!\!, 
\end{split}
\end{equation}
with \(\displaystyle\alpha = \frac{R_r}{R_\star}\), \(\displaystyle\beta = \frac{M_r}{M_\star}\) and \(\displaystyle\gamma = \frac{\rho_c}{\rho_r} = \frac{\alpha^3(1-\beta)}{\beta(1-\alpha^3)}\), \(R_r\) being the radius of the radiative zone and \(M_r\) its mass. The effective calculation of the tidal dissipation related to the excitation of inertial modes leads to a strongly frequency-dependent resonant spectrum, which is highly sensitive to the friction in the stellar medium \citep{savonije97,ogilvie13}. A frequency average then provides a reasonable estimate likely to over- or underestimate the effective dissipation at a given frequency. Moreover, the assumption of a bi-layer structure for the star is reasonable for stellar masses lesser than \(1.2\ M_\odot\), as the \citet{mathis15} prescription underestimates tidal dissipation by a factor less than 2 compared to the \citet{ogilvie13} formulation applied to a more realistic stratified structure of the star \citep{barker20}. However, in the case of F-type stars, the two-layer model may generally underpredict the dissipation by at least one order of magnitude.

Furthermore, the dissipation of the equilibrium tide in the convective zone can be estimated as \citep{remus}
\begin{equation}
\left[\Im(k_2)\right]_\text{eq}^\text{CZ} = 4\pi\frac{2088}{35}\frac{R_\star^4}{GM_\star^2}\left|\omega\int_{\alpha}^1 x^8\rho_\text{CZ}\nu_t\ dx\right|,
\end{equation}
where \(\nu_t\) is the turbulent viscosity in the
convection zone, \(\rho_\text{CZ}\) the density in the convection zone and \(x = r/R_\star\) the normalized radial coordinate. Following \citet{duguid}, the turbulent viscosity strength can be assessed from the convective turnover time \(t_c\) as
\begin{equation}
\nu_t = \nu_c l_c F(\omega),   
\end{equation}
where \(\nu_c\) is the typical convective velocity, \(l_c\) is the mixing
length, and
\begin{equation}
F(\omega) = 
\begin{cases}
5,\ |\omega|t_c < 10^{-2}\\
\displaystyle \frac{1}{2}\left(|\omega| t_c\right)^{-\frac{1}{2}},\ |\omega|t_c \in  [10^{-2},5]\\
\displaystyle \frac{25}{\sqrt{20}}\left(|\omega| t_c\right)^{-2},\ |\omega|t_c > 5.
\end{cases}
\end{equation}
Such a prescription accounts for the results of the latest numerical simulations dealing with the interaction between turbulent convection and tidal flows. In particular, they provide strong evidence in favor of the \citet{goldreich} frequency-dependence at high frequencies, as the energetically-dominant modes of the convection contribute the most to the effective viscosity \citep[for a more in-depth discussion, we refer the reader to][]{barker20}. As we only aim here to provide an order of magnitude of tidal dissipation, we assume a constant density in the convective zone, equal to \(\rho_\text{CZ} = 3M_\star(1 - \beta)/4\pi R_\star^3 (1-\alpha^3)\).
Furthermore, we approximate the mixing length \(l_c\) by its maximum, given by the depth of the convective zone \((1 - \alpha)R_\star\). This leads to
\begin{equation}
\left[\Im(k_2)\right]_\text{eq}^\text{CZ} = \frac{696}{35}|t_c \omega|\frac{R_\star}{GM_\star}\nu_c^2 (1-\beta)\frac{1-\alpha^9}{1-\alpha^3} F(\omega).
\end{equation}
Given the stellar luminosity \(L_\star\) as well as the rotation period \(P_\text{rot}\), one can estimate the convective velocity \(\nu_c\) and convective turnover time \(t_c\) from the \citet{mathis16} formulation, based on the mixing-length theory for a rotating body \citep[e.g.][]{stevenson,augustson}:
\begin{equation}
\nu_c = \nu_{c,0}
\begin{cases}
\displaystyle\left(1-\frac{1}{242 Ro^2}\right),\ Ro>0.25\\
\displaystyle 1.5\ Ro^\frac{1}{5},\ Ro<0.25,\\
\end{cases}
\end{equation}
where \(Ro = P_\text{rot}\nu_{c,0}/l_c\) is the convective Rossby number and \(\displaystyle\nu_{c,0} = (L_\star/(\rho_\text{CZ}R_\star^2))^{1/3}\) is the convective velocity from the standard mixing-length
theory, without rotation. For both prescriptions, we rely on STAREVOL to compute the stellar mass, radius, rotation and luminosity as well as to assess the \(\alpha,\beta\) coefficients. Furthermore, we have chosen here the so-called conventional equilibrium tide \citep{zahn66,remus}, defined as the hydrostatic response of the star to the tidal potential, which use in convective zone is discussed in favor of a non-wavelike equilibrium tide \citep{goodman,terquem, ogilvie14}. However, as presented in \citet{barker20}, the first formulation differs from the second by a factor of 2-3. Such a discrepancy is not a deal-breaker in our study since we consider both orders of magnitudes and upper bounds of tidal dissipation.

\subsubsection{Comparison with tidal dissipation in the convective envelope along stellar evolution}

\begin{figure}[!h]
    \begin{center}
    \includegraphics[scale=0.35]{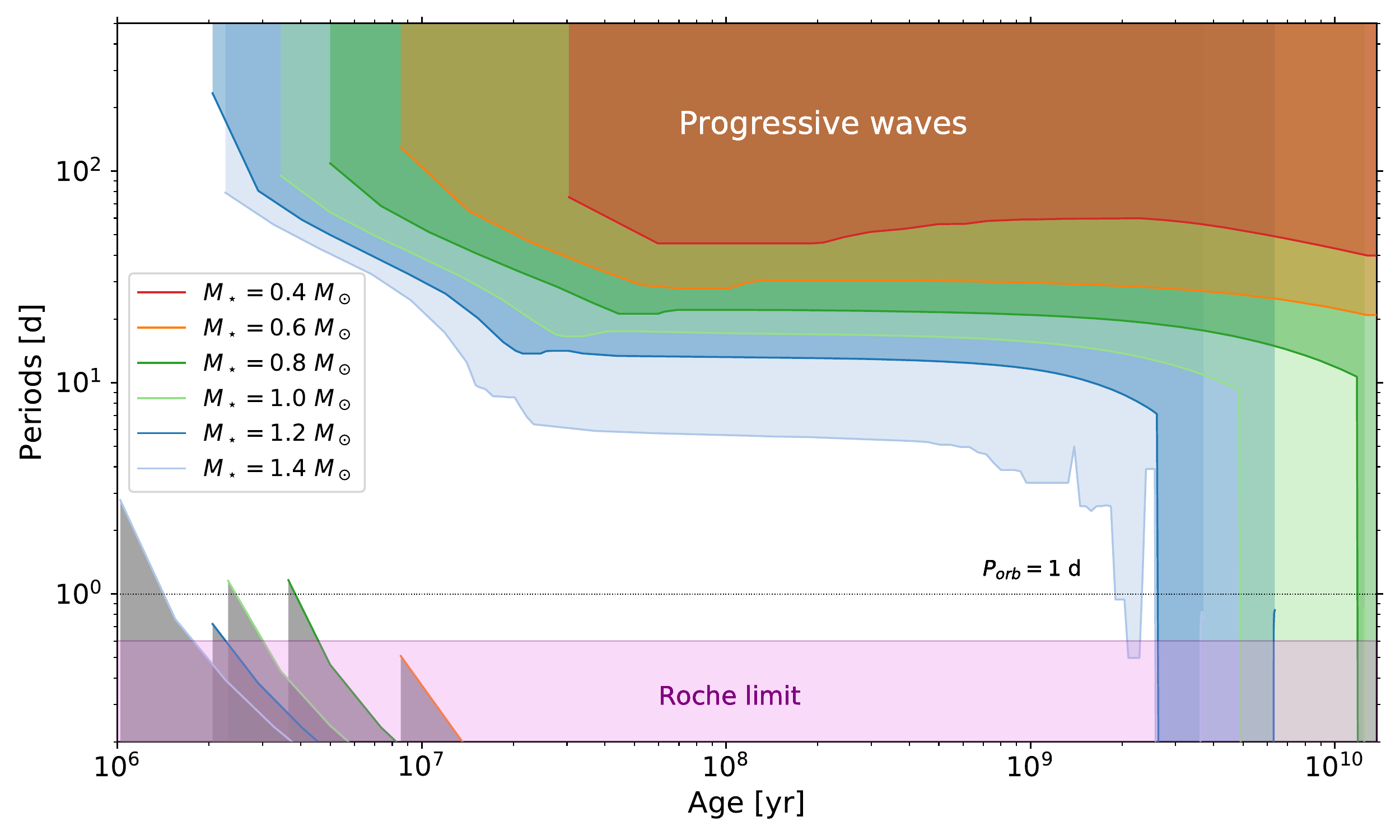}
    \end{center}
\caption{\label{periods} Characteristic periods as a function of the age of the system, for stellar masses between 0.4 and 1.4 Msun. In dark gray: region corresponding to tidal frequencies greater than \(N_\text{max}\), for which no propagation of gravity waves is allowed. The coloured areas correspond to orbital periods likely to excite progressive waves (with frequencies \(\omega < \omega_c\)) for each stellar mass. In purple: typical values corresponding to the Roche limit. The black dashed line corresponds to an orbital period of 1 day.}
\end{figure}

We now aim to compare tidal dissipation in the stellar radiative and convective zones throughout stellar evolution. To do so, we first need to study the frequencies available to tidal gravity waves as a function of stellar mass and stellar age. For tidal frequencies higher the maximal Brunt-Väisälä frequency \(N_\text{max}\), the propagation of internal gravity waves is not allowed (the corresponding periods are represented as gray areas in Fig. \ref{periods}). However, such a case corresponds in general to orbital periods below the Roche limit \citep[in violet in the same figure ; for the calculation of such a limit we refer the reader to][]{benbakoura}. This implies that a planet, after several megayears of evolution, is likely to excite gravity waves throughout the evolution of the star for all the stellar masses considered. 

Fig. \ref{periods} also represents the orbital period corresponding to the cut-off frequency \(\omega_c\). Beyond this critical period, tidal gravity waves are entirely dissipated through radiative damping before undergoing any reflection (see the colored regions in Fig. \ref{periods}). As stratification and thermal diffusivity are more important for the most massive and the oldest stars, close-in planets are more likely to excite progressive waves in the radiative zone of those kind of stars. Thus, during the RGB, even planets located at the Roche limit are able to excite progressive waves. During the PMS and MS of the least massive stars, a planet excites gravity waves that cannot be entirely dissipated through radiative damping. In the absence of other ways of dissipation, g-modes are then formed. However, for sufficiently massive planets, wave breaking may occur at the center of the star or, in the presence of differential rotation, gravity waves may interact with a critical layer. In this case, close-in planets may generate efficient tidal dissipation in the stellar radiative zone.\\
\begin{figure}[!h]
    \begin{center}
    \includegraphics[scale=0.35]{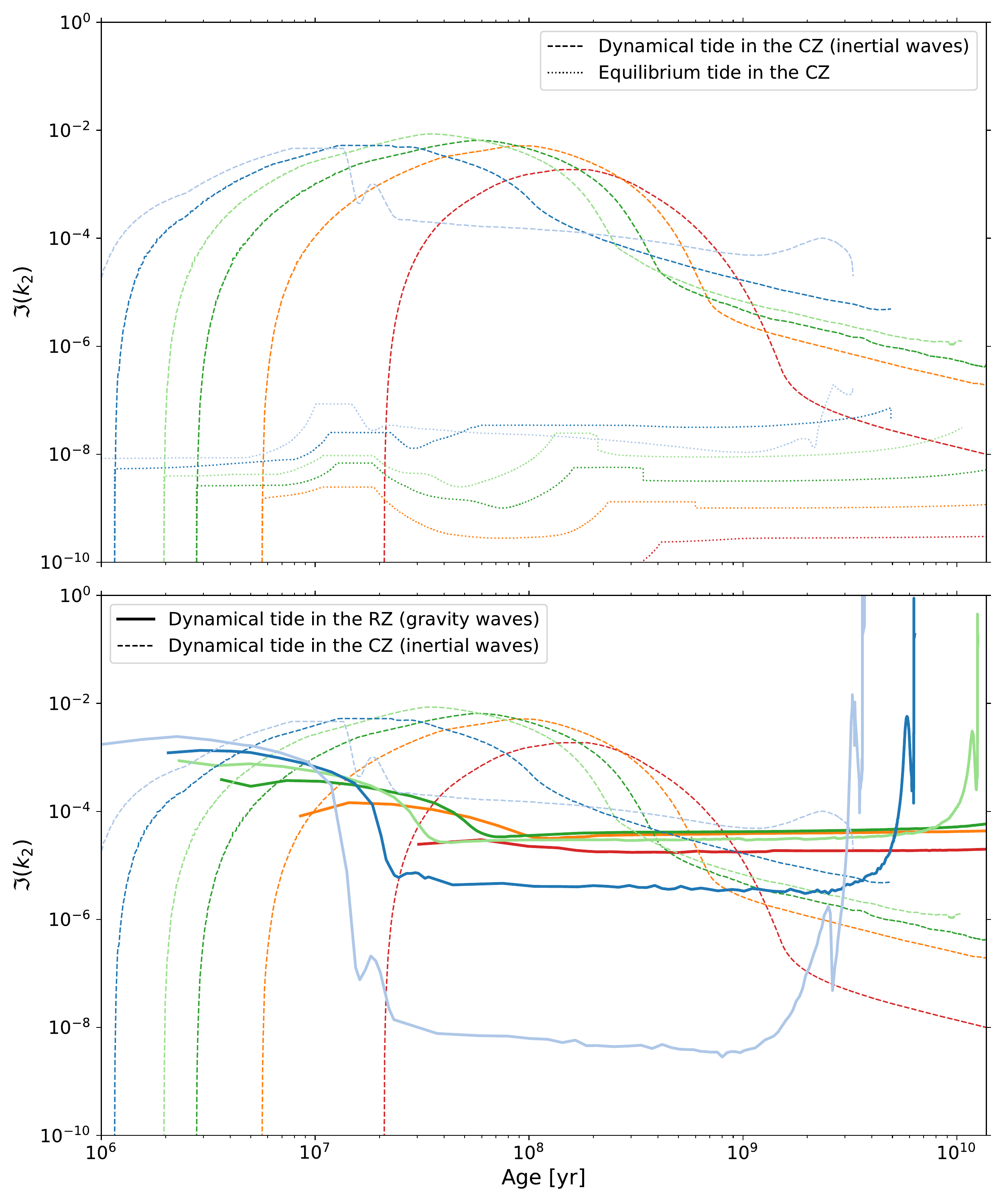}
    \end{center}
\caption{\label{inert} \textbf{Top:} Evolution of the dissipation of the dynamical tide through inertial waves (dashed lines) and of the equilibrium tide in the convective zone (dotted lines). \textbf{Bottom:} 
 evolution of the dissipation of the dynamical tide through gravity waves (solid lines) and of the dynamical tide through inertial waves (dashed lines). Stellar mass ranges between 0.4 and 1.4 \(M_\odot\). The colours correspond to the same masses as in Fig. \ref{periods}.}
\end{figure}

To evaluate an order of magnitude of the contribution of the dissipation of tidal gravity waves, one may be tempted to rely on the method introduced by \citet{ogilvie13,mathis15,gallet17} for tidal inertial waves. They compute a frequency average that would evaluate the capacity of a given stellar structure to dissipate the tidal wave, assuming that it is excited in an impulsive way and dissipated after a finite time by a dissipation mechanism. We would then have
\begin{equation}\label{eqn:average}
\left<\Im(k_2)\right>_\text{IGW}^\text{RZ} = \int_{0}^{\omega_c}\Im\left(k_2\right)\frac{d\omega}{\omega},
\end{equation}
which for solar-type stars gives
\begin{equation}
\begin{split}
\left<\Im(k_2)\right>_\text{IGW}^\text{RZ} = \frac{3^{\frac{2}{3}}\Gamma^2\left(\displaystyle\frac{1}{3}\right)}{32\pi}|m|&\left[l(l+1)\right]^{-\frac{4}{3}}\frac{GM_\star^2}{R_\star^5}\omega_c^\frac{8}{3}\\
&\times\ \rho_0(r_\text{int})r_\text{int}\left|\frac{dN^2}{d\ln r}\right|_{r_\text{int}}^{-\frac{1}{3}}\!\frac{\mathcal{F}^2}{m_p^2 n^4}.
\end{split}
\end{equation}
We stop the frequency average at $\omega_c$, since we are focusing only on progressive waves. 

However, the frequency-averaged method is meaningful in the case of a strongly resonant and erratic dissipation. For instance, in the case of tidal inertial waves, the amplitude of the resonant dissipation depends strongly on the strength of the turbulent friction applied by convection \citep{ogilvie04,ogilvie07,auclair15,mathis16}. However, we cannot use such an approach if the dissipation varies with a power-law of the tidal frequency, as it is the case for progressive gravity waves (the same holds for the equilibrium tide). In such a configuration, the frequency-averaged dissipation would lead to a lower bound which is not representative. We thus choose here an effective evaluation for a typical hot-Jupiter system with a period of 1 day.

Fig. \ref{inert} shows the evolution of the dissipation of the equilibrium tide in the convective zone (dashed lines) as well as of the dynamical tide through inertial waves (dashed lines) and gravity waves (solid lines), for \(M_\star = {0.4-1.4} M_\odot\). The dissipation of the equilbrium tide in the radiative zone is neglected here, as its efficiency is much lower than the other contributions \citep{zahn66,zahn77}. Moreover, as we consider fast rotators, the dissipation through inertial waves that we calculate here acts as an upper bound. While the tidal dissipation through gravity waves is comparable to its counterpart in the convective zone at the beginning of PMS, the mass transfer from the convective to the radiative zone decreases the efficiency of the dissipation in the stellar core, as presented in \S 5.3. Thus, during most of the PMS and the beginning of the MS, when the rotation of the star is at its highest, the dissipation of the dynamical tide in the convective zone dominates. The dissipation of the equilibrium tide is then lower by several orders of magnitude than the other two contributions. 

During the MS, as magnetic braking substantially decreases the stellar rotation rate, the tidal dissipation through inertial waves loses efficiency to the benefit of the dissipation in the core. Indeed, the latter remains constant during the MS due to the small changes in stellar structure. During the most advanced phases, the equilibrium tide becomes comparable to the contribution of inertial waves. As in the case of the most massive stars, the choice of a two-layer model leads to an overestimate of the dissipation of the dynamical tide in the convective zone by at least an order of magnitude \citep{barker20}, the equilibrium tide is likely to become the dominant contribution within the stellar envelope. Such a scenario is consistent with the evolution of eccentricities observed for red-giant binaries observed by the \textit{Kepler} mission \citep{beck18}. However, due to the core contraction and the expansion of the envelope during the SG phase and the RGB, tidal dissipation through gravity waves becomes largely dominant. This may have strong impact on the secular evolution of star-planet systems and binaries during the most advanced phases \citep{schlaufman,essick,weinberg,vick20}. When considering slower rotators (with a rotation period between 5 and 10 days), the tidal dissipation though inertial waves is less efficient, as $\Im[k_2] \propto \Omega_\star^2$. The dissipation through gravity waves then dominates the evolution of the system during longer phases, at the beginning of the PMS and at the end of MS.

\subsection{Influence of stellar metallicity}

Stellar metallicity may affect secular evolution of star-planet systems and henceforth the orbital period distribution of hot Jupiters \citep[e.g.][]{gonzalez,santos03}. In order to assess its influence on tidal dissipation through gravity waves, we focus on a 1 \(M_\odot\) star with three different metallicities, that is: \(Z=0.004,\ Z=0.0134=Z_\odot,\text{ and } Z=0.0255\), as in \citet{bolmont17}. When the metallicity of the star increases, its opacity evolves in the same way \citep{kippenhahn}. This has the effect of decreasing the stellar luminosity as well as the effective temperature, which subsequently increases its overall lifetime. Furthermore, the radiative zone loses in extension, which makes the density at the convective-radiative interface higher at a given evolutionary phase.

\begin{figure}[!h]
    \begin{center}
    \includegraphics[scale=0.34]{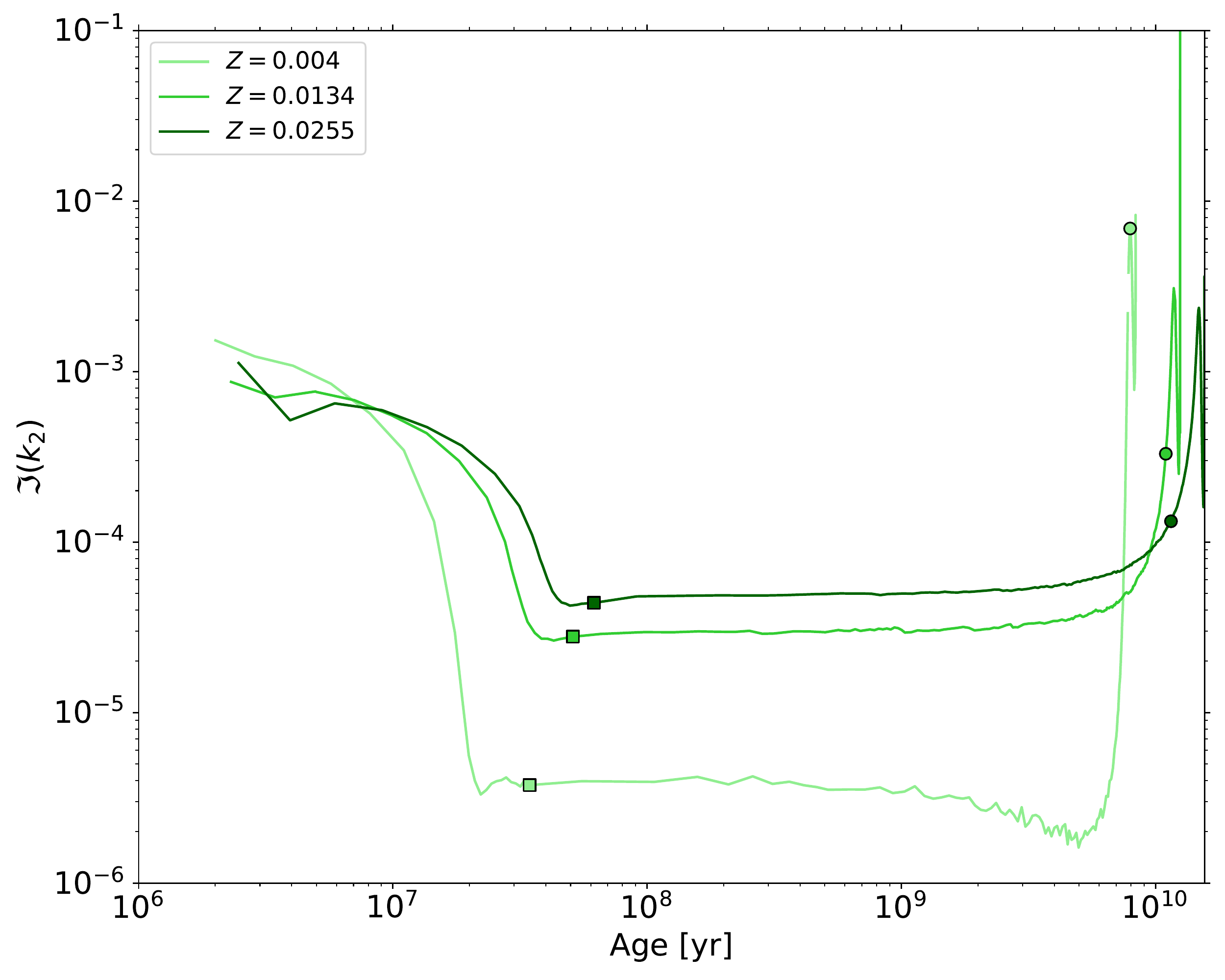}
    \end{center}
\caption{\label{Met} Evolution of the tidal dissipation through gravity waves as a function of time for three different stellar metallicities, that is: \(Z=0.004,\ Z=0.0134=Z_\odot,\text{ and } Z=0.0255\), with \(M_\star = 1\ M_\odot\) and \(P_\text{orb} = 1\ \text{d}\). The colored squares correspond to the ZAMS in each model, and the colored circles to the TAMS.}
\end{figure}

As seen in Fig. \ref{Met}, tidal dissipation through gravity waves evolves in the same way regardless of the stellar metallicity considered. Furthermore, dissipation increases along with metallicity. Such a behavior may lead to a discrepancy of around one order of magnitude near the ZAMS between the solar metallicity case (\(Z = 0.0134\)) and the metal-poor case (\(Z = 0.004\)). At the beginning of the evolution, one can notice than metal-poor stars undergo a stronger dissipation than for metal-rich stars. Indeed, the higher the metallicity, the later the radiative core is formed. Then, as the radiative zone reaches its maximal extension at the end of the PMS, the opposite behavior is observed and dissipation increases with metallicity. At the end of evolution, as metal-poor stars reach the Sub-Giant phase and the RGB before their metal-rich counterparts, the observed trend reverses again and the dissipation of metal-poor stars becomes the most important.

\subsection{Tidal dissipation in massive solar-type stars: bi-layer vs. tri-layer structure}

\begin{figure}[!h]
   \begin{center}
      \includegraphics[scale=0.35]{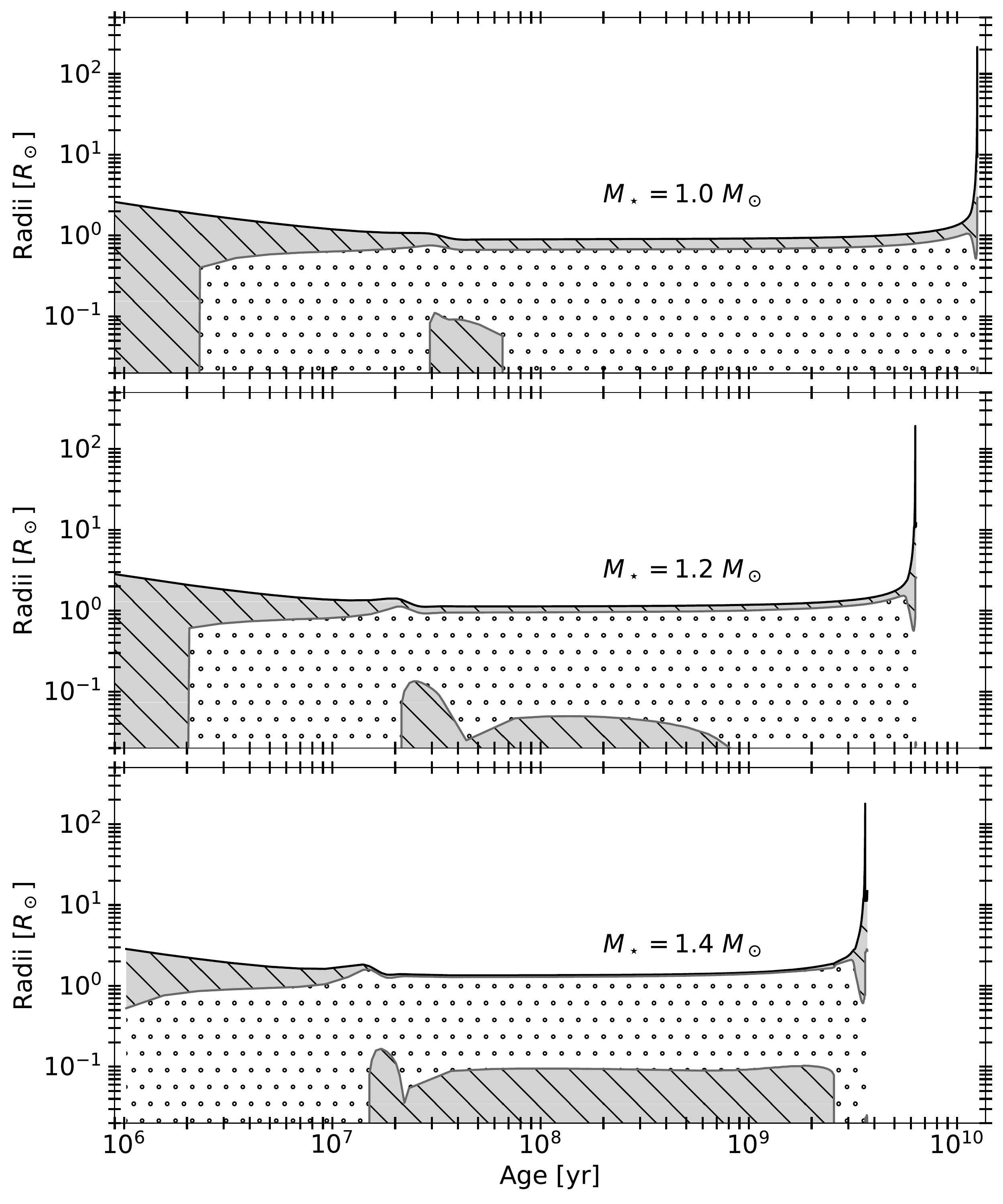}
   \end{center}
\caption{\label{struct_massive} Kippenhahn diagram showing the evolution of the internal structure of the non-rotating star for \(M_\star = 1\ M_\odot\) (top), \(M_\star = 1.2\ M_\odot\) (middle) and \(M_\star = 1.4\ M_\odot\) (bottom) from the PMS up to the end of the MS. The upper line represents the stellar radius in solar radii. The hatched grey areas depict convective regions, while the dotted areas stand for radiative regions.}
\end{figure}

We now focus on the STAREVOL models with stellar masses ranging from 1 to 1.4 \(M_\odot\). As shown in Fig. \ref{struct_massive}, all models form a convective core near the Zero-Age Main Sequence (ZAMS). However, for \(M_\star = 1.4\ M_\odot\), the star keeps its core during the entire Main Sequence while for \(M_\star = 1.2\ M_\odot\) it disappears at around 1 Gyr.  For \(M_\star = 1\ M_\odot\) the convective core stays only for 40 Myr.

\begin{figure}[!h]
   \begin{center}
      \includegraphics[scale=0.35]{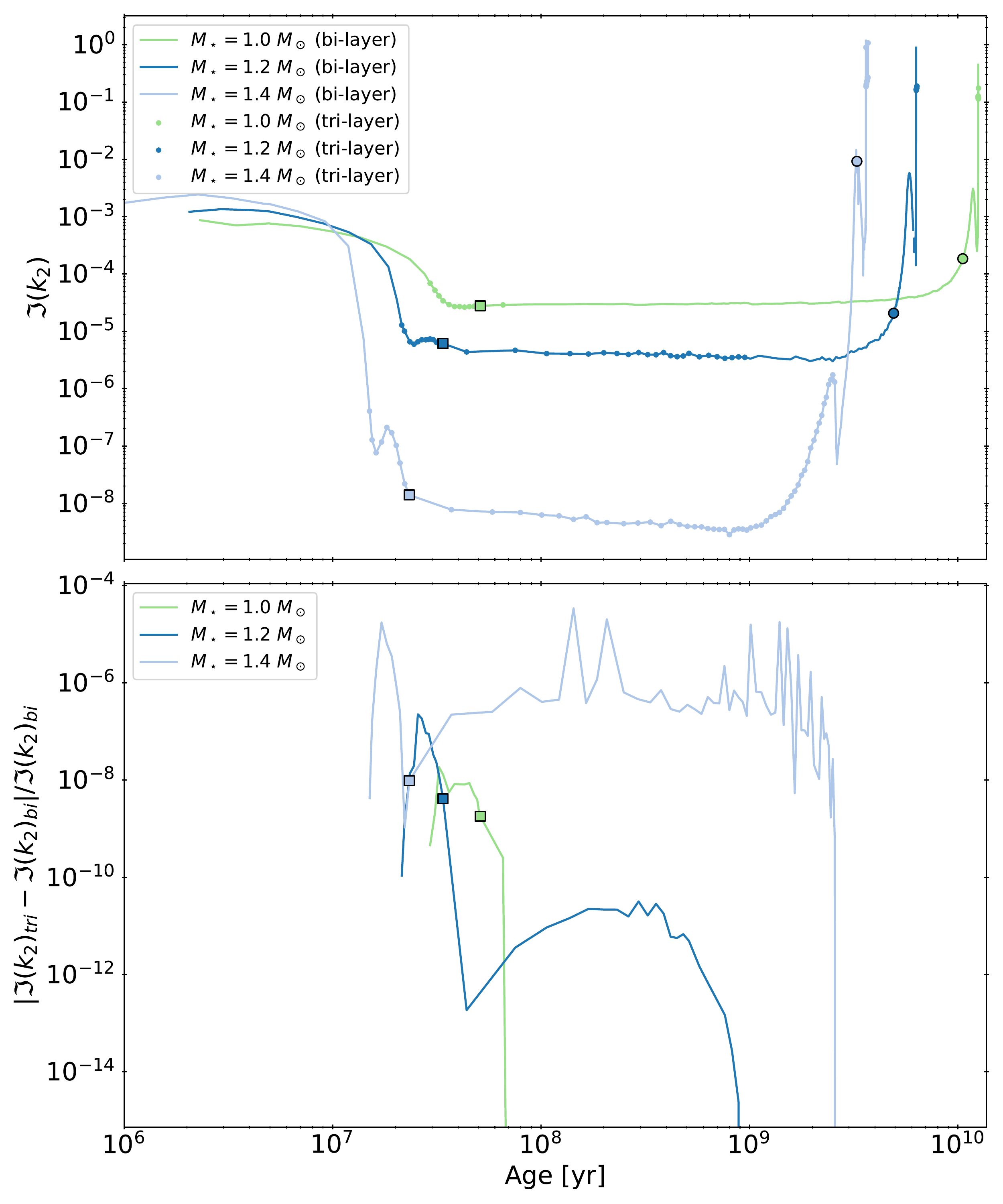}
   \end{center}
\caption{\label{trilayer} Top: Evolution of the tidal dissipation in the case of bi-layer (blue curves) and a tri-layer (blue dots) structure as a function of time for stellar masses \(M_\star = \{1-1.4\}\ M_\odot\), with \(P_\text{orb} = 1\ \text{d}\). Bottom: Time evolution of the relative difference of tidal dissipations derived from bi-layer and tri-layer structures. The colored squares correspond to the ZAMS in each model, and the colored circles to the TAMS.}
\end{figure}

In this configuration, as we see in Fig. \ref{trilayer}, the contribution of a convective core to the total tidal dissipation is negligible compared to the contribution of the outer thin convective envelope we derive for a bi-layer structure. Indeed, despite a higher density at lower radii, the low extent of the core tends to reduce tidal dissipation through outward gravity waves, especially with the aid of the forcing term \(\mathcal{F}\). Such a contribution involves a \(\left(1-\frac{\rho}{\bar\rho}\right)\) factor, with \(\bar{\rho}\) the mean density inside a sphere of radius \(r\), which in a core configuration quantifies the inhomogeneity of the mass distribution near the center. A tiny convective core, coupled with a flat mass distribution near the center from spherical symmetry, then leads to a weak dissipation.

\section{Conclusions and discussions}

In this paper, we provide a general formalism to assess tidal dissipation in stellar radiative zones in all type of stars. We focused on the angular momentum flux transported by progressive tidal gravity waves, which are more likely to affect the secular evolution of the considered binary or planetary system \citep{goodman,terquem}. Such an approach allows us to gather the founding work of \citet{zahn75}, \citet{goldreich}, \citet{goodman}, and \citet{barker10} prescriptions, among others, into a unique flexible framework applicable to all type of stars and planetary systems. In the case of low-mass stars, we investigate the influence of stellar structure and evolution on tidal dissipation through gravity waves. We show that, for a given star-planet system, tidal dissipation reaches a maximum value on the PMS for all the stellar masses we considered. This behavior is the result of the expansion of the radiative zone, allowing the propagation of higher-frequency waves, as well as the subsequent decrease of stellar density at the convective-radiative interface, which cuts off the energy transported by the tidal gravity waves. Then, as stellar structure stabilizes, tidal dissipation  evolves to a stationary value during the MS, which is maximal for K-type stars and decreases by several orders of magnitude for F-type stars, due to their thinner convective envelope. We also find that during most of the PMS and the beginning of the MS, as the star rotates rapidly, the dissipation of the dynamical tide in the convective zone dominates the other contributions for fast rotators. Then, as magnetic braking spins down the star on the MS, tidal gravity waves dissipation becomes the largest contribution. We confirm that stellar metallicity has also a significant influence on tidal dissipation. Indeed, the dissipation is two orders of magnitude larger in a metal-rich star (\(Z=0.0255\)) than in a metal-poor one (\(Z = 0.0040\)). However, we find that at a given age, the dissipation is more important in low-metallicity stars during the PMS, SGB and RGB because of their quicker evolution. Finally, we show that the contribution of a convective core for F-type stars is negligible compared to the tidal dissipation derived by assuming a bi-layer structure. Indeed, as tidal dissipation is enhanced by density inhomogeneities in the convective zone (see Eq. \eqref{eqn:Finhom} for more details), a spherical convective core presents a geometry and mass distribution unfavorable to the excitation and dissipation of tidal gravity waves.\\
 
We find that a massive star structure and a solar-type star structure lead to similar prescriptions regarding tidal dissipation. However, such a symmetry is made possible by assuming a polytropic behavior of stellar matter at the stellar surface, which leads to a vanishing density there. A more detailed study of the boundary conditions at the surface, initiated in Appendix C, is required to assess the robustness of such an analogy.

We also consider in this paper a linear approximation of \(N^2\) near the convective-radiative interfaces. Taking more complex \(N^2\) profiles into consideration \citep{lecoanet} may change the frequency dependency of the induced tidal torque and may be in better agreement with 3D numerical simulations of internal wave breaking \citep{barker11,ivanov}. Nonetheless, according to \citet{barker11}, an improved prescription may deviate from ours by a factor of 2 at most, meaning that our model provides a suitable order of magnitude of the dissipation of tidal gravity waves to follow the secular evolution of a star-planet system.

Taking dissipation processes \citep{zahn97, barker10} as well as angular momentum transport in stellar interiors into account (see \citet{mathis13} for a review) is also among the perspectives of this work. In this context, one needs to factor in stellar rotation \citep{ogilvie04}. Regarding our formulation, adding the Coriolis acceleration may directly alter the forcing term \(\mathcal{F}\), since in the convective zone gravito-inertial waves are no more evanescent in a sub-inertial regime (where \(|\omega| <2\Omega_\star\)). Such an effect is expected to increase tidal dissipation in the radiative zone \citep{rocca87,rocca89,ogilvie07,witte02}. In the same propagation regime, low-frequency waves in the radiative zone are trapped near the equatorial plane, leading to a geometry significantly different from what we have considered in this work \citep{rieutord97, dintrans99}. Such a change is likely to modify the tidal dissipation accordingly. In addition, since we find that the dissipation through IGW is maximal during the radiative core formation, when the star is spinning up, a progressive trapping of gravito-inertial waves in the sub-inertial regime may occur. Stellar rotation also increases the radial wavenumber of the gravito-inertial waves, leading a stronger radiative damping. These are therefore deposited closer to their excitation region than in the non-rotating case \citep{pantillon07,mathis08}.

Differential rotation on tides should also be taken into account, as it
affects propagation of gravito-inertial waves, leading to a large variety of resonant cavities as well as chaotic zones \citep{mathis09,prat18}. It also allows the deposit of angular momentum in critical layers and therefore interactions between waves and mean flows \citep{goldreich,alvan13,astoul}. A strong differential rotation may set up during the PMS due to stellar contraction \citep{charbonnel,hypolite,gouhier}, when the tidal dissipation through gravity waves is maximal. Such an effect can therefore have a significant influence on the evolution of binary and planetary systems. The presence of a magnetic field may also affect tidal dissipation, modifying the propagation and the damping of tidal gravity waves \citep{MathisDB11,MathisDB12}.

Despite those limitations, a consistent formulation of the dynamical tide in stellar radiative zones provides an avenue to an exhaustive study of the fate of star-planet systems. In this spirit, comparing systematically tidal dissipation in radiative and convective zones \citep{mathis15,gallet17} requires a secular evolution model to track the complex evolution of the tidal forcing frequency \citep{benbakoura}. Indeed, we have compared in our work the capacity of the star to dissipate progressive tidal gravity waves in the radiative zone and inertial modes tidally excited in the convective zone. In a more realistic model, we need to estimate the passing of a gravito-inertial wave from one zone to another as well as possible reflections. Such an undertaking then requires to take into account the complex temporal evolution of the tidal frequency related to the orbital dynamics of a given system.

In the case of exoplanetary systems for example, a secular evolution model with a formalism that models tidal effects coherently in stellar convective and radiative zones would allow us to study the migration of nearby planets and their lifetime throughout the stellar evolution, thus reconciling the results of recent studies \citep[e.g.][]{barker10,guillot,barker20,bolmont16,gallet17,bolmont17,benbakoura}. Such an approach would ultimately allow us to take into account other star-planet interactions and to explain the observed statistical distributions \citep{MMA13}. The implementation of such secular evolution models is of primary importance for analyzing data of upcoming space missions such as PLATO
\citep{rauer}.

\begin{acknowledgements}
We would like to thank the referee, Jim Fuller, as well as Allan-Sacha Brun and Antoine Strugarek for helpful comments and suggestions regarding our work. J.A. and S.M. acknowledge the PLATO CNES funding at CEA/IRFU/DAp. J.A. acknowledges funding from the European Union’s Horizon-2020 research and innovation programme (Grant Agreement no. 776403 ExoplANETS-A). S.M. and J.A. acknowledge funding by the European Research Council through the ERC grant SPIRE 647383. L.A. acknowledges funding from the European Reseach Council AWESoMeStars 682393.
\end{acknowledgements}

\newpage
\appendix
\section{Computation of a particular solution in the radiative zone near the interface}
The goal of this section is to compute a particular solution of Eq. \eqref{eqn:rad} in the radiative zone, near the interface. In such a configuration, Eq. \eqref{eqn:rad} becomes
\begin{equation}\label{eqn:Airyinhom}
\frac{d^2\psi}{d\eta^2}+v^2\eta\psi = v^2\eta Z.
\end{equation}
For a given function \(f\), we will write in this section \(\frac{df}{d\eta} = f'\). We choose \(\psi_1(\eta) = \text{Ai}[v^{\frac{2}{3}}(-\eta)]\) and \(\psi_2(\eta) = \text{Bi}[v^{\frac{2}{3}}(-\eta)]\) as the two basis solutions of the corresponding homogeneous equation. Their Wronskian \(\Lambda_A\) can be written as
\begin{equation}
\Lambda_A = \psi_1\psi_2' - \psi_2\psi_1' = -\frac{v^{\frac{2}{3}}}{\pi}.
\end{equation}
Then a particular solution of the inhomogeneous Airy equation, vanishing at the interface, can be expressed as
\begin{equation}
\begin{split}
\psi_\text{p}(\eta) = &-\left(\int_0^\eta{\Lambda_A^{-1}v^2\eta Z\psi_2(\eta) d\eta}\right)\psi_1(\eta)\\
&+\left(\int_0^\eta{\Lambda_A^{-1}v^2\eta Z\psi_1(\eta) d\eta}\right)\psi_2(\eta).
\end{split}
\end{equation}
From Eq. \eqref{eqn:Airyinhom} we obtain
\begin{equation}
\frac{\psi_\text{p}(\eta)}{v^{-\frac{2}{3}}\pi} = \left(\int_0^\eta{\!\!\!-Z\psi_2''(\eta) d\eta}\right)\psi_1(\eta)+\left(\int_0^\eta{Z\psi_1''(\eta) d\eta}\right)\psi_2(\eta),
\end{equation}
which leads to
\begin{equation}
\begin{split}
\frac{\psi_\text{p}(\eta)}{v^{-\frac{2}{3}}\pi}= &-Z(\eta)\Lambda_A+Z(0)\left[\psi_2'(0)\psi_1(\eta) - \psi_1'(0)\psi_2(\eta)\right]\\
&+\left[Z'\psi_2\right]_0^\eta \psi_1 - \left[Z'\psi_1\right]_0^\eta \psi_2 + \mathcal{I},
\end{split}
\end{equation}
with \(\mathcal{I} = -\left(\int_0^\eta{Z''\psi_2(\eta) d\eta}\right)\psi_1+\left(\int_0^\eta{Z''\psi_1(\eta) d\eta}\right)\psi_2\). Such a term will be neglected from now on by assuming that the equilibrium tide varies slowly compared to the dynamical tide in the radiative zone. Since \(\psi_1'(0) = -v^{\frac{2}{3}}\text{Ai}'(0)\) and \(\psi_2'(0) = -v^{\frac{2}{3}}\text{Bi}'(0)\), we obtain:
\begin{equation}
\begin{split}
\psi_\text{p}(\eta) = &Z(\eta)-Z(0)\pi\left[\text{Bi}'(0)\psi_1(\eta) - \text{Ai}'(0)\psi_2(\eta)\right]\\
&-v^{-\frac{2}{3}}\pi Z'(0)\left[\text{Bi}(0)\psi_1(\eta)-\text{Ai}(0)\psi_2(\eta)\right].
\end{split}
\end{equation}
Knowing that \(\pi = [\text{Ai}(0)\text{Bi}'(0)-\text{Bi}(0)\text{Ai}'(0)]^{-1} = \frac{1}{2}3^{\frac{3}{2}}\Gamma\left(\frac{4}{3}\right)\Gamma\left(\frac{2}{3}\right)\) the particular solution becomes
\begin{equation}
\begin{split}
\psi_\text{p}(\eta) = Z(\eta)+\left(\frac{\tau}{2}\right)^{\frac{1}{3}}\left[\alpha_\text{rad,p}J_{\frac{1}{3}}(\tau)+\beta_\text{rad,p}J_{-\frac{1}{3}}(\tau)\right],
\end{split}
\end{equation}
where \(\displaystyle \alpha_\text{rad,p} = -\frac{dZ}{d\eta}(0)\left(\frac{v}{3}\right)^{-\frac{2}{3}}\Gamma\left(\frac{4}{3}\right)\) and \(\beta_\text{rad,p} = -Z(0)\Gamma\left(\frac{2}{3}\right)\).

\newpage
\section{Forcing term and radial displacement for a low-mass star}
We aim in this section to link the forcing term to the radial displacement linked to the dynamical tide, which is comparing formulations from \citet{zahn75} and \citet{goodman}. One can express the derivative of the radial displacement at the interface linked to the dynamical tide with the notations we used in Sect. 2: 
\begin{equation}
\left.\partial_r{\xi_r^\text{dyn}}\right|_\text{int} = \partial_r\left[{\rho_0^{-\frac{1}{2}}r^{-2}(\rho_0^{-\frac{1}{2}}X-Z)}\right]_\text{int}.
\end{equation}
By expressing the radial displacement \(\xi_r\) in the (\(S_+,S_-\)) basis, knowing that \(\mathcal{T}_1 =0\) we obtain from Eqs. \eqref{eqn:homrel}, \eqref{eqn:coeffpart} and \eqref{eqn:intK0}:
\begin{equation}
\begin{split}
&\left.\partial_r{\xi_r^\text{dyn}}\right|_\text{int} = \left.\partial_r\left(\rho_0^{-\frac{1}{2}}r^{-2}\right)\right|_\text{int}\left[\frac{\beta_\text{conv}+\beta_\text{conv,p}}{\Gamma\left(\frac{2}{3}\right)}-Z(0)\right]\\
&+\rho_0^{-\frac{1}{2}}(r_\text{int})r_\text{int}^{-2}\left[\frac{\beta_\text{conv}+\beta_\text{conv,p}}{\Gamma(\frac{2}{3})}\frac{\frac{d}{dr}(\rho_0^{-\frac{1}{2}}X_h)_\text{int}}{(\rho_0^{-\frac{1}{2}}X_h)_\text{int}}+\mathcal{T}-\partial_r Z(0)\right]
\end{split}
\end{equation}
with 
\begin{equation}
\begin{split}
\mathcal{T}=&\rho_0^{\frac{1}{2}}(r_\text{int})r_\text{int}^2\left\{-\frac{\rho_0'(r_\text{int})}{\rho_0(r_\text{int})}-\frac{\left(\rho_0^{-\frac{1}{2}}Z\right)_\text{int}'}{\left(\rho_0^{-\frac{1}{2}}Z\right)_\text{int}}+\frac{X_1'(r_\text{int})}{X_1(r_\text{int})}\right\}\left(\frac{\varphi_T}{g}\right)_\text{int}\\
&+\rho_0^{\frac{1}{2}}(r_\text{int})\mathcal{F}.
\end{split}
\end{equation}
Since \(C_2 = 0\), the homogeneous solution \(X_h\) in the convective zone is proportional to the basis solution \(X_1\), which leads to 
\begin{equation}
\partial_r{\xi_r^\text{dyn}} =\left.\frac{\partial_r(\rho_0^{-1}r^{-2}X_h)}{\rho_0^{-1}r^{-2}X_h}\right|_\text{int}\xi_r^\text{dyn}(r_\text{int})+r_\text{int}^{-2}\mathcal{F},
\end{equation}
where \(\displaystyle \xi_r^\text{dyn}(r_\text{int})=\rho_0^{-\frac{1}{2}}(r_\text{int})r_\text{int}^{-2}\left(\frac{\beta_\text{conv}+\beta_\text{conv,p}}{\Gamma\left(\frac{2}{3}\right)}-Z(0)\right)\).
By assuming slow variations of the density and the radius compared to the characteristic length of variation of the gravity waves in the radial direction \(\lambda = v^{-\frac{2}{3}} = \omega^\frac{2}{3}\left[l(l+1)\right]^{-\frac{1}{3}}\left|\frac{dN^2}{d\ln  r}\right|^{-\frac{1}{3}}_{r_\text{int}}r_\text{int}\) \citep{goodman,kushnir}, we obtain from Eq. \eqref{eqn:homrel}
\begin{equation}
\left.\partial_r{\xi_r^\text{dyn}}\right|_\text{int}\sim \frac{\Gamma(\frac{2}{3})}{3^{\frac{2}{3}}\Gamma(\frac{4}{3})}\frac{\alpha_\text{conv}}{\beta_\text{conv}}\frac{\xi_r^\text{dyn}(r_\text{int})}{\lambda}+r_\text{int}^{-2}\mathcal{F}.
\end{equation}
Since \(\xi_r^\text{dyn}(r_\text{int}) \sim \lambda \left.\partial_r\xi_r^\text{dyn}\right|_\text{int}\) \citep{goodman} and \(\alpha_\text{conv}\ll\beta_\text{conv}\), we have
\begin{equation}
\left.\partial_r\xi_r^\text{dyn}\right|_\text{int}\sim r_\text{int}^{-2}\mathcal{F},
\end{equation}
which ensures the equivalence between the formulations from \citet{zahn75} and \citet{goodman}.
\newpage
\section{Stress-free condition at the stellar surface}
\begin{figure}[!h]
   \begin{center}\label{ratio}
      \includegraphics[scale=0.55]{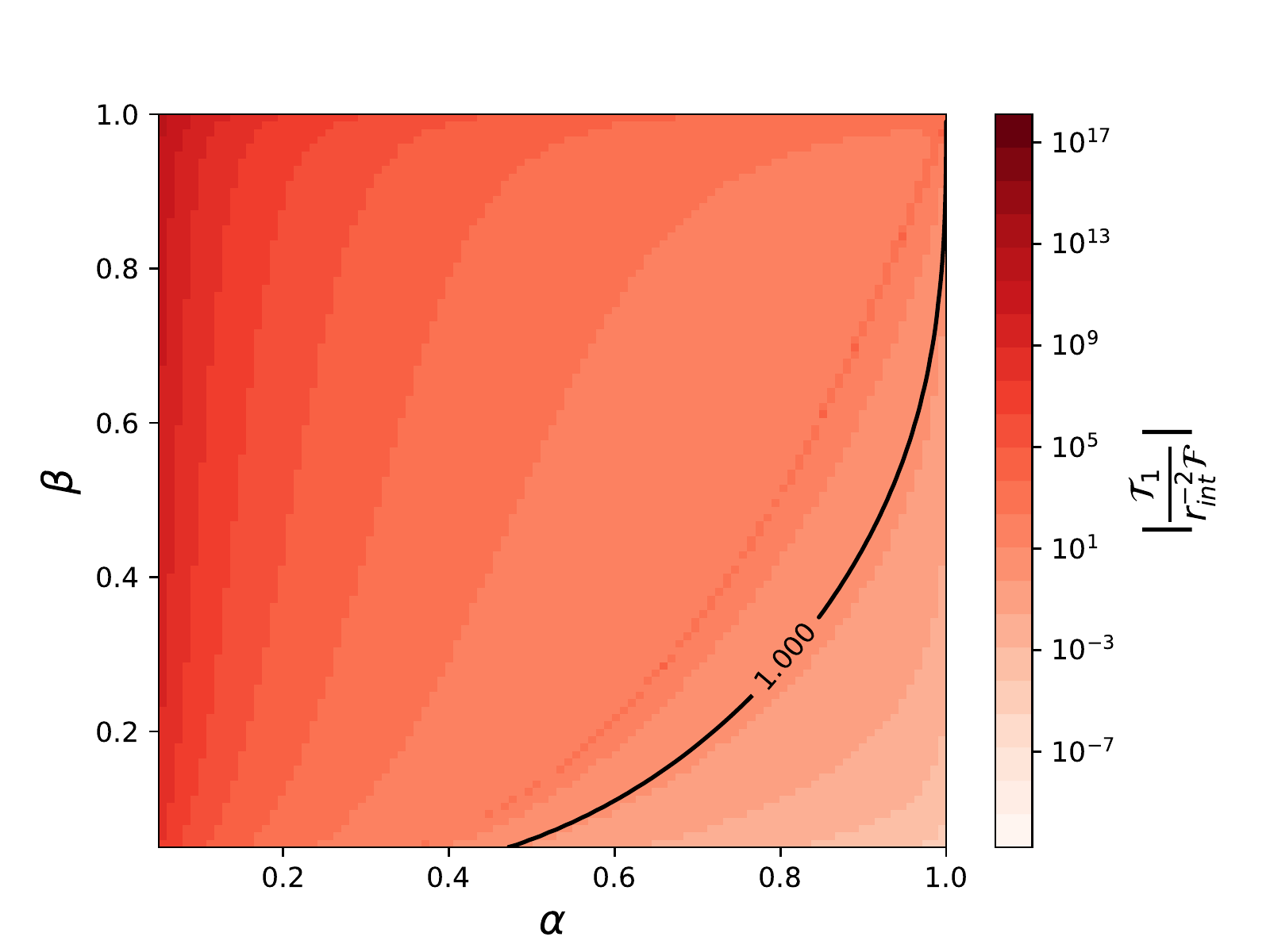}
   \end{center}
\caption{Value of \(\mathcal{T}_1/r_\text{int}^{-2}\mathcal{F}\) for all the possible values of \(\alpha = R_r/R_\star\) and \(\beta = M_r/M_\star\). In black: values of \(\alpha\) and \(\beta\) for which the ratio is equal to 1.}
\end{figure}

The goal of this section is to investigate the consequences on tidal dissipation of low-mass stars of a stress-free boundary condition coupled with a nonzero density at the stellar surface. In this case, the Lagrangian perturbation of pressure vanishes at \(r=R_\star\):
\begin{equation}
\delta P = \rho_0(R_\star)\left[y-g_0(R_\star)\xi_r(R_\star)\right] = 0.
\end{equation}
We will consider the case where \(\rho_0(R_\star) >0\) and \(\frac{\rho_0'(R_\star)}{\rho_0(R_\star)}\) is the only large parameter involved, to account for the density behavior near the photosphere.
From Eqs. \eqref{eqn:press} in the anelastic approximation, we obtain
\begin{equation}
\xi_r(R_\star) = -\frac{\varphi_T(R_\star)}{g_0(R_\star)}+\frac{\omega^2}{\omega_\text{dyn}^2}\frac{1}{l(l+1)R_\star}\partial_r(r^2 \xi_r).
\end{equation}
By assuming low frequency waves, \textit{i.e.} \(\omega\ll\omega_\text{dyn}\), the stress-free condition becomes
\begin{equation}
X(R_\star) = -\rho_0(R_\star)R_\star^2\frac{\varphi_T(R_\star)}{g_0(R_\star)},
\end{equation}
which accounts for the equilibrium tide. In the convective zone, we choose the basis solution \(X_1\) as the solution of the homogeneous equation associated to Eq. \eqref{eqn:conv} verifying \(X_1(R_\star) = X(R_\star)\) and \(X_1'(R_\star) = X'(R_\star)\). This way, we fix \(C_2 = \mathcal{T}_0 = \mathcal{T}_2 = 0\). As in section 5.1, near the surface, \(X_1\) is a solution of the following equation:
\begin{equation}
X''-\frac{\rho_0'}{\rho_0}X'=0,
\end{equation}
which leads to \(X_1' \propto \rho_0\). We now impose that \(X'(R_\star) = X_1'(R_\star) = R_\star^2\rho_0(R_\star)\).\\
\\
If we assume that the interface between convective and radiative zone is close to the stellar surface, one can assume that
\begin{equation}\label{eqn:boundary}
\begin{split}
X_1'(r) &= R_\star^2\left[\rho_0(R_\star) + \rho_0'(R_\star)(r-R_\star)\right]\\
X_1(r) &= X_1(R_\star)\!+\! R_\star^2\!\left[\rho_0(R_\star)(r-R_\star)\! +\! \frac{1}{2}\rho_0'(R_\star)(r-R_\star)^2\!\right].
\end{split}
\end{equation}
Such expressions are adopted as a boundary condition in our numerical treatment of tidal dissipation to deal with the surface singularity in the case of solar-type stars. Furthermore, if we assume that \(\displaystyle\frac{\rho_0(R_\star)}{\rho_0'(R_\star)}\ll R_\star^2(\alpha-1)^2\) we obtain
\begin{equation}
\displaystyle\frac{X_1'(R_\star)}{X_1(r_\text{int})}=\frac{1}{-\frac{\varphi_T(R_\star)}{g(R_\star)}+R_\star(\alpha-1)+\frac{1}{2}\frac{\rho_0'(R_\star)}{\rho_0(R_\star)}R_\star^2(\alpha-1)^2}\ll 1,
\end{equation}
\begin{equation}
\displaystyle\frac{X_1(R_\star)}{X_1(r_\text{int})}=\frac{-\frac{\varphi_T(R_\star)}{g(R_\star)}}{-\frac{\varphi_T(R_\star)}{g(R_\star)}+R_\star(\alpha-1)+\frac{1}{2}\frac{\rho_0'(R_\star)}{\rho_0(R_\star)}R_\star^2(\alpha-1)^2}\ll 1.
\end{equation}
The \(\mathcal{T}_1\) term then becomes
\begin{equation}
\begin{split}
\mathcal{T}_1 = &\frac{\varphi_T(R_\star)}{g(R_\star)}\alpha^{-2}\frac{\rho_0'(R_\star)}{\rho_0(R_\star)}\frac{X_1(R_\star)}{X_1(r_\text{int})}\\
&\sim -2\left(\frac{\varphi_T(R_\star)}{g(R_\star)}\right)^2\alpha^{-2}(1-\alpha)^{-2}R_\star^{-2}.
\end{split}
\end{equation}
Furthermore, in the case of a thin convective layer, from Eq. \eqref{eqn:Fscaling} we have 
\begin{equation}
r_\text{int}^{-2}\mathcal{F} = 3\frac{1-\gamma}{1-\alpha}\frac{\alpha^5}{\beta}\left(\frac{2\alpha}{3}-1\right) \alpha^{-2}\frac{\varphi_T(R_\star)}{g(R_\star)}R_\star^{-1}.
\end{equation}
Then we obtain
\begin{equation}
\frac{\mathcal{T}_1}{r_\text{int}^{-2}\mathcal{F}} = -\frac{2}{3}\frac{\beta}{\alpha^5(1-\gamma)(1-\alpha)\left(\displaystyle\frac{2\alpha}{3}-1\right)} \frac{\varphi_T(R_\star)}{g(R_\star)}R_\star^{-1}.
\end{equation}
Since \(\gamma = \frac{\alpha^3(1-\beta)}{\beta(1-\alpha^3)}\) we have
\begin{equation}
\frac{\mathcal{T}_1}{r_\text{int}^{-2}\mathcal{F}} = -\frac{2}{3}\frac{\beta^2(1+\alpha+\alpha^2)}{\alpha^5\left(\displaystyle\frac{2\alpha}{3}-1\right)(\beta-\alpha^3)} \frac{\varphi_T(R_\star)}{g(R_\star)}R_\star^{-1}.
\end{equation}
The values of such a ratio for all the possible values of \(\alpha = R_r/R_\star\) and \(\beta = M_r/M_\star\) are represented in Fig. \ref{ratio}. As we assumed that the convective zone is sufficiently thin to linearize the density profile, only values of \(\alpha\) close to 1 are relevant in this analysis. Therefore, we can assume that the \(\mathcal{T}_1\) term marginally affects our prescription for tidal dissipation in the radiative zone in the case of a surface density sufficiently weak. A more detailed study on surface boundary conditions is left for future work.

\newpage
\section{A wave breaking criterion for solar-type stars}
\begin{figure}[!h]
   \begin{center}\label{Mcrit}
      \includegraphics[scale=0.55]{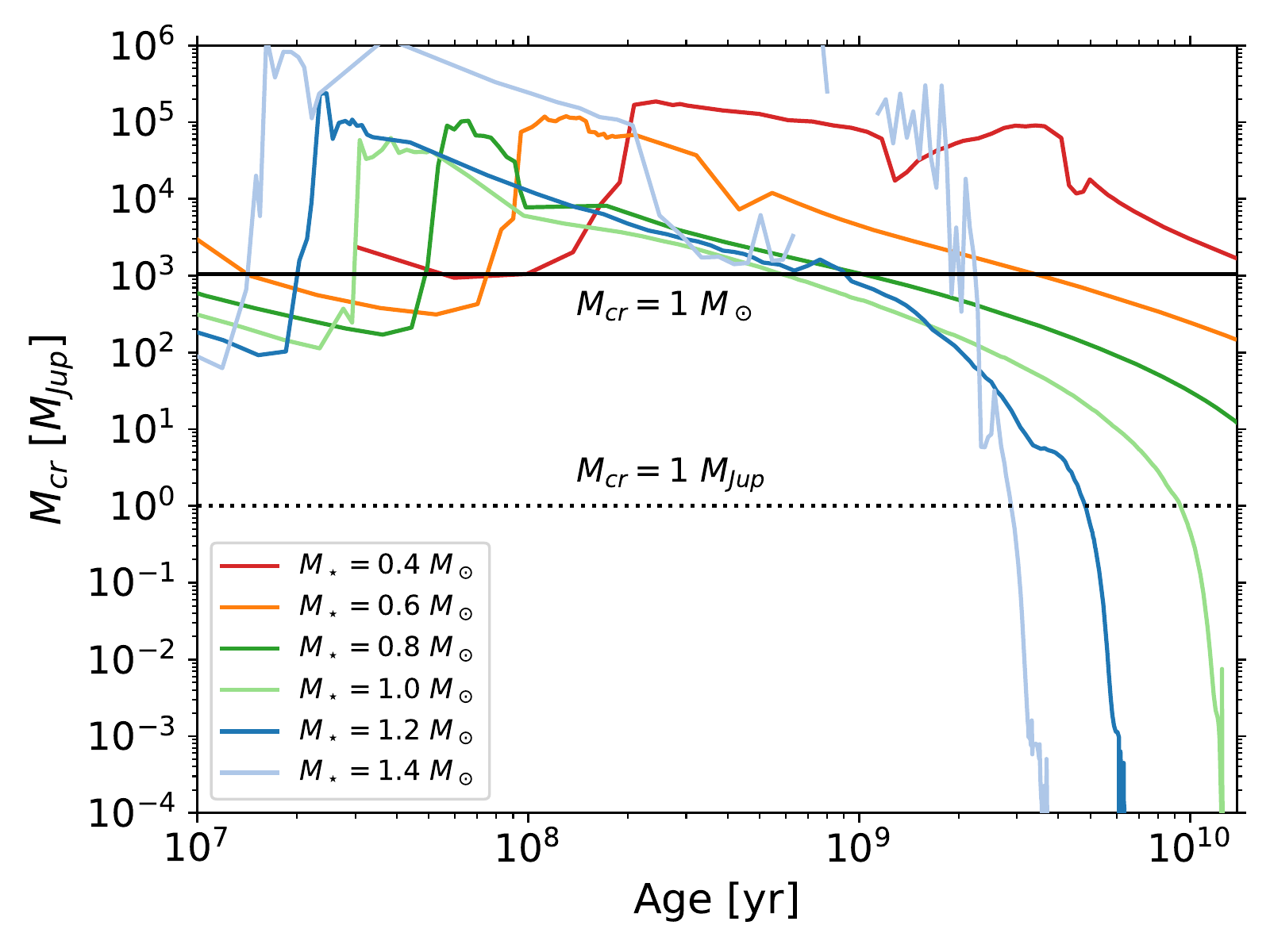}
   \end{center}
\caption{Evolution of the critical planetary mass $M_\text{cr}$ as a function of the age of the system, for stellar masses (\(M_\star\)) between 0.4 and 1.4 \(M_\odot\). Wave braking may occur for planetary masses higher than $M_\text{cr}$.}
\end{figure}

To provide a general criterion for wave braking in the case of solar-type stars, we rely on the non-linearity factor $\epsilon_{nl}$, which is the ratio of the amplitude of the radial displacement to the radial wavelength \citep{press,barker10, barker20}
\begin{equation}
\epsilon_{nl}= |k_r \xi_r|.
\end{equation}
Non-linearities become important when $\epsilon_{nl} \geq 1$. In particular, wave braking is likely to occur. One can assess this non-linearity factor through the energy luminosity of the tidal gravity wave. Indeed, from Eq. \eqref{eqn:energylum}, we have
\begin{equation}
|L_E| = \frac{\omega^3}{2l(l+1)}|C_{W}|^2.
\end{equation}
Furthermore, by defining \(\xi_r^\text{dyn}\) the radial displacement linked to the dynamical tide, we obtain in the WKBJ approximation:
\begin{equation}
\xi_r^\text{dyn} (r) = \rho_0^{-\frac{1}{2}}r^{-2}C_{W}\frac{1}{\sqrt{k_r}}e^{\epsilon i(\tau_W-\tau_0)},
\end{equation}
where \(k_r \approx \sqrt{\frac{N^2}{\omega^2}\frac{l(l+1)}{r^2}}\) is the radial wavenumber. This leads to
\begin{equation}
|C_{W}|^2 = \rho_0 r^3 \frac{N}{\omega}\sqrt{l(l+1)}\left|\xi_r^\text{dyn}\right|^2.
\end{equation}
The energy luminosity then becomes:
\begin{equation}
|L_E| = \frac{N\omega^2 \rho_0 r^3}{2\sqrt{l(l+1)}}\left|\xi_r^\text{dyn}\right|^2.
\end{equation}
Hence, one can assess $\epsilon_{nl}$ as
\begin{equation}
\epsilon_{nl} = \sqrt{\frac{2[l(l+1)]^\frac{3}{2}N|L_E|}{\rho_0r^5\omega^4}},
\end{equation}
 which is similar to the expression obtained in Eq. (53) in \citet{barker20}.
Furthermore, in the case of a solar-type star, we find that the energy luminosity $L_E$ is equal to 
\begin{equation}
L_E =-\frac{3^\frac{2}{3}\Gamma^2\left(\displaystyle\frac{1}{3}\right)}{8\pi}\omega^\frac{11}{3}\left[l(l+1)\right]^{-\frac{4}{3}}\rho_0(r_\text{int})r_\text{int}\left|\frac{dN^2}{d\ln r}\right|_{r_\text{int}}^{-\frac{1}{3}}\mathcal{F}^2.
\end{equation}
with 
\begin{equation}
\mathcal{F} = \int_{r_\text{int}}^{R_\star}{\left[\left(\frac{r^2\varphi_T}{g_0}\right)'' - \frac{l(l+1)}{r^2}\left(\frac{r^2\varphi_T}{g_0}\right)\right]\frac{X_1}{X_1(r_\text{int})}}dr.
\end{equation}
As $L_E \propto \mathcal{F}^2$, we introduce $|L_E| = L_{E,0}\ m_p^2 n^4 \omega^\frac{11}{3}$, where $L_{E,0}$ is independent of the tidal frequency and planetary properties, $m_p$ is the planetary mass and $n$ the mean motion of its orbit. Hence, wave braking may occur if $\epsilon_{nl} \geq 1$, which leads to
\begin{equation}
\frac{2[l(l+1)]^\frac{3}{2}NL_{E,0}m_p^2n^4}{\rho_0r^5}\omega^{-\frac{1}{3}} \geq 1
\end{equation}
In the absence of stellar rotation, we have $\omega = 2n$. 
Furthermore, near the stellar center, the radial profile of the Brunt-Väisälä is approximately linear. We then assume that $N \approx Cr$,  where $C \approx 8 \times 10^{-11}\ \text{m}^{-1}.\text{s}^{-1}$ for the current Sun \citep{barker20}. Such a quantity is here estimated at a given stellar mass and stellar age by relying on STAREVOL grids. Following \citet{goodman}, one can estimate the location $r_\text{inner}$ of the inner turning point, defined as $N=\omega$, as follows:
\begin{equation}
r_\text{inner} = \frac{2n}{C}.
\end{equation}
Then, estimating the non-linearity factor at the inner turning point gives
\begin{equation}
\frac{2^{-\frac{10}{3}}[l(l+1)]^\frac{3}{2}C^5 L_{E,0}}{\rho_0(r_\text{inner})}m_p^2n^{-\frac{1}{3}} \geq 1
\end{equation}
which, by introducing the orbital period $P_\text{orb}$, gives the following criterion on the planetary mass
\begin{equation}
m_p \geq (2\pi)^{\frac{1}{6}}\frac{2^{\frac{5}{3}}\rho_0(r_\text{inner})}{[l(l+1)]^\frac{3}{4}C^\frac{5}{2} L_{E,0}^\frac{1}{2}}P_\text{orb}^{-\frac{1}{6}} \equiv M_{cr}.
\end{equation}
We then find a similar result as the \citet{barker10} criterion (see also \citealt{barker11,barker20}), which is based on an overturning of the stratification. Such a condition depends weakly on the orbital period (a decrease by one order of magnitude in $P_\text{orb}$ lead to an increase of $M_\text{cr}$ of around 31.9 \%). If a given planet has a mass higher than $M_\text{cr}$, then wave breaking may occur in the star, and the tidal quality factor is expected to behave according the results of our work. Otherwise, other dissipation processes, like radiative damping in the case of progressive waves or critical layers, may lead to a similar tidal dissipation. We present in Fig. \ref{Mcrit} the evolution of $M_\text{cr}$ as a function of the age of the system, for stellar masses (\(M_\star\)) between 0.4 and 1.4 \(M_\odot\). As already pointed out in \citet{barker20}, for stellar masses larger than 0.9 \(M_\odot\), the critical planetary mass may fall below 1 Jupiter mass for all ages greater than 10 Gyr. It then allows super-Earths and hot Neptunes to trigger wave braking in their host stars during the sub-giant phase and the RGB.

\section{Angular momentum transport and tidal torque}
The goal of this section is to clarify the relationship between the angular momentum transport and the net torque applied to the radiative zone. To this end, we consider a radiative zone between $r=r_0$ and $r=r_1 > r_0$.
The equation for the transport of angular momentum, horizontally averaged and focusing only on waves, is given by \citet{mathis09}:
\begin{equation}
\rho_0\frac{d}{dt}\left(r^2\int_{\theta=0}^{\theta=\pi}\sin^3\theta\ \Omega d\theta\right) = -\frac{1}{2\pi r^2}\partial_r\left(r^2 \int_{\theta=0}^{\theta=\pi}F_J\sin \theta d\theta\right),
\end{equation}
where $\Omega$ is the angular velocity of the radiative zone and $F_J$ is the radial component of flux of angular momentum transported by gravity waves' Reynolds stresses, whose expression is given in Eq. \eqref{eqn:angmomflux}. By integrating along the radial and latitudinal directions, this leads to 
\begin{equation}
\frac{dJ_{RZ}}{dt} = - \int_{r_0}^{r_1}(\partial_r L_J) dr,
\end{equation}
where $\displaystyle J_{RZ} = \rho_0\int_{r=r_0}^{r=r_1}\int_{\theta=0}^{\theta = \pi}\int_{\varphi=0}^{\varphi = 2\pi} r^4\sin^3\theta\ \Omega\  d\varphi d\theta dr$ is the total angular momentum of the radiative zone and $\displaystyle L_J = 2\pi \int_{\theta=0}^{\theta=\pi} r^2F_J\sin \theta d\theta$ is the luminosity of angular momentum. Hence we obtain
\begin{equation}
\frac{dJ_{RZ}}{dt} = L_J(r_0) - L_J(r_1).
\end{equation}
In the case of an inward energy transport ($\epsilon = 1$, corresponding to the configuration of solar-type stars), tidal gravity waves are excited at $r=r_1$ and are totally dissipated before reaching the radius $r=r_0$. Hence we obtain
\begin{equation}
\frac{dJ_{RZ}}{dt} = - L_J(r_1).
\end{equation}
In the case of an outward energy transport ($\epsilon = -1$, corresponding to the massive and intermediate-mass stars' configuration), tidal gravity waves are excited at $r=r_0$ and are totally dissipated before reaching the radius $r=r_1$. In this configuration, we have
\begin{equation}
\frac{dJ_{RZ}}{dt} = L_J(r_0).
\end{equation}
Hence, one can compute the torque $T$ applied to the whole radiative zone with a single expression:
\begin{equation}
T = -\epsilon L_{J,\text{exc}},
\end{equation}
where $L_{J,\text{exc}}$ is the luminosity of angular momentum estimated at the region of excitation of tidal gravity waves.

\end{document}